\documentclass[aps,prb,10pt,twocolumn,floatfix,superscriptaddress]{revtex4-2}
\usepackage[utf8]{inputenc}
\usepackage[T1]{fontenc}
\usepackage[english]{babel}
\usepackage{microtype}
\usepackage{mathtools}
\usepackage[varbb,varg,otfmath]{newtx} 
\usepackage{bm,dsfont,eucal}
\AtBeginDocument{\renewcommand{\vec}[1]{\bm{#1}}}
\newcommand{\pdagger}{\phantom{\dagger}}
\usepackage[dvipsnames,svgnames,x11names]{xcolor}
\usepackage{hyperref}
\hypersetup{ 
    colorlinks=true, linkcolor=NavyBlue, urlcolor=NavyBlue, citecolor=NavyBlue
}
\usepackage{physics,braket,siunitx,slashed}
\usepackage[capitalise]{cleveref}
\usepackage{enumitem} 

\begin{document}
\title{Magnetism in the Dilute Electron Gas of Rhombohedral Multilayer Graphene}

\author{Tobias M.~R.~Wolf}
\affiliation{Department of Physics, University of Texas at Austin, Austin, Texas 78712, USA}
\author{Nemin Wei}
\affiliation{Department of Physics, Yale University, New Haven, CT 06520, USA}
\author{Haoxin Zhou}
\affiliation{Department of Physics, University of California, Berkeley, California 94720, USA}
\affiliation{Department of Electrical Engineering and Computer Sciences,
University of California, Berkeley, California 94720, USA}
\affiliation{Materials Sciences Division, Lawrence Berkeley National Laboratory, Berkeley, California 94720, USA}
\author{Chunli Huang}
\affiliation{Department of Physics and Astronomy, University of Kentucky, Lexington, Kentucky 40506-0055, USA}
\date{\today}

\begin{abstract}
Lightly-doped rhombohedral multilayer graphene has recently emerged as one of the most promising material platforms for exploring electronic phases driven by strong Coulomb interactions and non-trivial band topology.  This review highlights recent advancements in experimental techniques that deepen our understanding of the electronic properties of these systems, especially through the application of weak-field magnetic oscillations for studying phase transitions and Fermiology. 
Theoretically, we advocate modeling these systems using an electron gas framework, influenced primarily by two major energy scales: the long-range Coulomb potential and band energy. The interplay between these energies drives transitions between paramagnetic and ferromagnetic states, while smaller energy scales like spin-orbit coupling and sublattice-valley-dependent interactions at the atomic lattice scale shape the (magnetic anisotropic energy) differences between distinct symmetry-broken states.
We provide first-principles estimates of lattice-scale coupling constants for Bernal bilayer graphene under strong displacement field, identifying the on-site inter-valley scattering repulsion, with a strength of $g_{\perp \perp}=269\text{meV nm}^2$ as the most significant short-range interaction. The mean-field phase diagram is analyzed and compared with experimental phase diagrams. New results on spin and valley paramagnons are presented, highlighting enhanced paramagnetic susceptibility at finite wavevectors and predicting valley and spin density-wave instabilities. Additionally, we examine the unique magnetic properties of orbital magnetism in quarter-metal states and their associated low-lying ferromagnon excitations. The interplay between superconductivity and magnetism, particularly under the influence of spin-orbit coupling, is critically assessed. The review concludes with a summary of key findings and potential directions for future research.
\end{abstract}
\maketitle
\tableofcontents
\newpage 

\section{Scope of the Review}
The profound understanding of emerging phases from fundamentally simple microscopic constituents, like the pressure-temperature phase diagram of liquid $^3$He, has significantly advanced the field of condensed matter physics. This review article explores the electronic phases of matter emerging from a dilute concentration of electrons in the $\pi$ bands of multilayer graphene. Despite the $\pi$ bands being primarily composed of rather simple $2p_z$ atomic orbitals, the electrons in these bands can exhibit a surprisingly rich variety of electronic phases as the electronic density of states becomes large.

The phases identified so far in the experiments include metallic ferromagnets with spontaneous spin and orbital moments, intervally-coherent metals, and nematic metals that reduce the $C_3$ rotational symmetry of the lattice, metals with non-linear IV characteristics, spin-singlet superconductivity, magnetic-field induced superconductivity, nearly half-metal ferromagnets, and nearly quarter-metal ferromagnets. Remarkably, all these phases emerge through simple adjustments of the electric field and carrier density via electrical gates, highlighting the quintessence of emergent properties.
We anticipate that this list will expand over time. The aim of this review is to provide an experimental overview of well-established findings and to discuss the theoretical minimal models that describe interaction effects at various energy scales.

This review is structured as follows: First, we explore the suite of experimental methods that have led to the discovery of these fascinating phases and examine the advanced techniques that have facilitated their detection. We then provide an in-depth review of the initial experimental reports on Bernal bilayer and rhombohedral trilayer graphene. Then, we discuss how to model interaction effects in these dilute electron systems. We then examine the mean-field phase diagram for Bernal bilayer and rhombohedral trilayer graphene, highlighting the valley-doublets and their energy crossings. Subsequently,  we introduce new results from time-dependent Hartree-Fock theory that study spin and paramagnons in Bernal bilayer graphene, which predict spin-density and valley-density wave instabilities.  We also  provide estimates for the short-range, layer-valley dependent interactions, which are important for resolving magnetic anisotropic energies. Next, we examine the influence of adjacent tungsten diselenide layers on the bilayer graphene phase diagram. The review concludes with a discussion on the criteria to observe superconductivity and provides a simple outlook on future research directions.

\section{Experiments}
\label{sec:exp}
\subsection{Suite of Experimental Techniques}
The motivation for these studies was partly driven by the discovery of twisted bilayer graphene, the use of electric fields, boron nitride encapsulation, and the synthesis of metastable ABC-stacked materials. We will touch upon the most advanced methods, such as the use of rhombohedral AFM tips, and briefly discuss the new phases that have been discovered.

The rapid development of fabrication, measurement, and data analysis techniques in recent years makes it possible to study the delicate correlated physics in ultra-clean two-dimensional material heterostructures. In this section, we review the suite of techniques that were exploited in various studies on rhombohedral multilayer graphene.

\subsubsection{Sample Fabrication}
Multilayer graphene samples are typically prepared by mechanically exfoliating bulk graphite crystals.
The flakes are then deposited on a silicon substrate with a thermal oxide layer of appropriate thickness to enhance optical contrast, allowing the identification of the number of carbon layers using optical microscopy for few-layer graphene \cite{blake_making_2007, teo_the_2008}.
For thicker flakes, atomic force microscopy is commonly employed to estimate the layer number \cite{shi_electronic_2020}.
Once the samples are prepared and the layer numbers are identified, the samples with rhombohedral stacking order need to be screened.
In multilayer graphene with more than two layers, multiple stacking orders can exist, distinguished by the relative positions of the carbon atoms from different layers. For instance, in trilayer graphene, there are two distinct stacking orders: Bernal (ABA) stacking and Rhombohedral (ABC) stacking \cite{guinea_electronic_2006, aoki_dependence_2007}.
For multilayer graphene with more than three layers, the stacking orders become more complex, including Bernal stacking, rhombohedral stacking, and a mix of the two.
In mechanically exfoliated flakes, different stacking orders usually coexist and form domains \cite{ju_topological_2015}.

While samples with different stacking orders cannot be distinguished with conventional optical microscopes, their different electronic band structures lead to different infrared and Raman spectra and can therefore be identified with these spectroscopic techniques.
Two techniques have been widely applied to identify the stacking orders: scanning near-field infrared microscopy \cite{ju_topological_2015} and scanning Raman spectroscopy \cite{lui_imaging_2011, cong_raman_2011, wu_raman_2018}.
Scanning near-field infrared microscopy is an imaging technique that combines the high spatial resolution of scanning probe microscopy with the chemical specificity of infrared spectroscopy\cite{vicent_scanning_2021}.
It works by illuminating a sharp metallic or dielectric tip with infrared light. This tip is brought very closely to the sample surface, typically within a few nanometers.
The electric field of infrared light focused onto the tip, hence sample, scatters and generates a near-field signal, which contains information about the local optical properties of the sample.
The sub-100nm spatial resolution allows it to not only identify domains with different stacking orders but also resolve one-dimensional AB-BA domain walls in Bernal-stacked bilayer graphene\cite{ju_topological_2015}.
Scanning Raman spectroscopy is an alternative technique that can be used to identify the stacking order.
In such setups, a laser beam is shed onto the sample on a scanning stage, with the position, intensity, and/or shapes of the G-band and 2D-band being measured to distinguish the stacking orders~\cite{lui_imaging_2011, cong_raman_2011, wu_raman_2018}.
Compared to scanning near-field infrared microscopy, the far-field nature of the Raman measurement limits the spatial resolution to the same order of the laser wavelength, but facilities more rapid sample screening~\cite{zhou_half_2021}, which is convenient for the fabrication process.

After the rhombohedrally stacked flakes or domains are identified, the samples are used to create Van der Waals heterostructures.
Despite the rhombohedral stacking order naturally occurring in graphite and mechanically exfoliated thin flakes, it is mechanically meta-stable~\cite{latychevskaia_stacking_2019}.
Mechanical and thermal disturbances during the fabrication process may lead to stacking relaxation.
Several techniques have been developed to reduce the probability of the relaxation.
One is to cut the flakes so that the uniform rhombohedral domains are physically isolated from the original flakes that contain different stacking orders.
This prevents the lattice relaxation caused by the domain wall movement.
During the cutting process, a metal-coated atomic force microscope tip operating in contact mode scratches over the sample, while an a.c. voltage is applied to the tip apex. This setup can induce an electrochemical reaction between the graphene and water in the air and oxidize the carbon layer under the tip. The oxidized carbon layer was subsequently removed mechanically by the tip, generating a gap~\cite{giesbers_nanolithography_2008, masubuchi_fabrication_2009, neubeck_scanning_2010, kumar_a_2010, biro_nanopattering_2010}. In-situ identification and isolation of the rhombohedral domains can be conveniently achieved by combining this setup with near-field infrared microscopy.

After the rhombohedral multilayer graphene flakes are prepared, conventional dry-transferred processes are applied to fabricate the Van der Waals heterostructures~\cite{wang_one_2013}
. To reduce the chance of lattice relaxation, the mechanical manipulation of the rhombohedral multilayer should be minimized. In this context, a sample flipping technique has been proved to be helpful~\cite{zeng_high_2019, zhou_half_2021}.

\subsubsection{Sample Characterization}

\label{sec:Sample Characterization}

\begin{figure*}[t]
\hspace*{-3.5cm}
\includegraphics[width=1.4\linewidth]{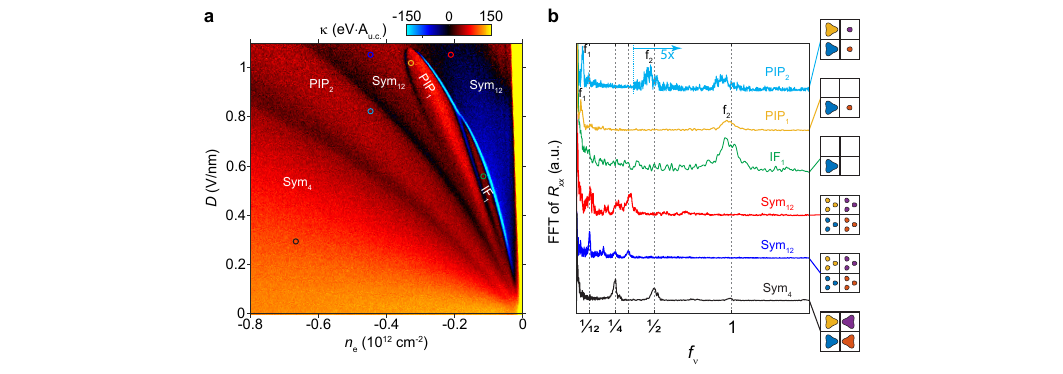}
\caption{
Phase Diagram of Bernal Bilayer Graphene. Panel a) shows how the inverse electronic compressibility  $\kappa$ varies with electron density $n_e$ and displacement field $D$. Panel b) shows the frequency of the Shubnikov-de Haas magnetic oscillations $R_{xx}$ v.s.~ $1/B$ at selected points within the phase diagram, revealing the number of Fermi surfaces and their respective enclosed areas. The insets provide schematics of the Fermi sea' topology, with four squares depicting the Fermiology of the four spin-valley flavors.
}
\label{fig:bbg_kappa_phase_diagram}
\end{figure*}

\begin{table*}[t]
	\centering
    \begin{ruledtabular}
	\begin{tabular}{ m{5.7cm} | m{3.6cm} | m{8cm} }
		\textbf{Metals}                                 & \textbf{Fundamental Oscillation Frequencies $f_\nu$} & \textbf{Remarks}                                                                                                                                            \\
		\hline
		Symmetric-12                                    & $1/12$                                               &
		A paramagnetic metal with 12 Fermi pockets, resulting from the presence of 4 spin-valley flavors. Each flavor hosts 3 Fermi pockets due to strong trigonal warping. This state typically occurs at lower densities near neutrality, where bandstructure details are important.                \\
		\hline
		Symmetric-4                                     & $1/4$                                                & A paramagnetic metal with 4 large Fermi pockets, occurring at higher densities compared to the Symmetric-12 state. It distributes carriers equally across the 4 spin-valley flavors. \\
		\hline
		Half-Metal                                      & $1/2$                                                & A ferromagnetic state where only two out of the four possible spin-valley flavors are occupied.
		\\
		\hline
		Quarter-Metal                                   & $1$                                                  & A ferromagnetic state where only one out of the four possible spin-valley flavors are occupied.                                                                                      \\
		\hline
		Almost-Half-Metal Ferromagnet (PIP$_2$)      & $1/2-\delta\;,\; \delta$                             & A ferromagnetic state that exhibits slight deviations from perfect half-metallicity due to the presence of small Fermi pockets with a fractional volume $\delta$.                    \\
		\hline
		Almost-Quarter-Metal Ferromagnet   (PIP$_1$) & $ 1-\delta\;,\; \delta$                              & A ferromagnetic state that exhibits slight deviations from perfect quarter-metallicity due to the presence of small Fermi pockets with a fractional volume $\delta$.                 \\
	\end{tabular}
    \end{ruledtabular}
	\caption{Different types of metals revealed by quantum oscillations. The four flavors refers to the combination of valley and spin degrees of freedom intrinsic to graphene.
		Note $\delta$ reflects the fraction of minority flavor, which can manifest as multiple small Fermi pockets.The nature of the order parameter for the symmetry-broken phases, whether Ising, XY, or Heisenberg-like, is determined by smaller energy scales, which will be discussed later. The nature of the order parameter for these generalized ferromagnets—whether Ising, XY, or Heisenberg-like—is influenced by subtle energy scales that are smaller compared to the dominant exchange and band energies. 
		PIP stands for partially-iosospin-polarized. }
	\label{table:terminologies}
\end{table*}

Rhombohedral graphene heterostructures has been studied with various cryogenic electronic measurement techniques including charge transport measuremen , quantum capacitance measurement , scanning nano superconducting interference device microscopy and optical spectroscopy. In this review, we discuss in detail on transport and capacitance measurement techniques.

Electronic transport has been widely used as a probe for material properties due to its relative simple setup and adaptability to extreme environments. Since sample resistance can be measured with very small currents, transport measurements do not suffer significantly from external heating. This allows for extremely low electron temperatures in the sample by cooling it down with dilution refrigerators and applying a series of low-pass electronic filters to reduce thermal noise. In Ref. \cite{zhou_superconductivity_2021} and\cite{zhou_isospin_2021}, superconducting phases with critical temperature below 100mK were identified. In addition, due to the low carrier density and low disorder nature of the rhombohedral graphene heterostructures, clean Shubnikov-de Haas (SdH) oscillations can be observed at low magnetic fields. 
 These oscillations can be used to probe the isospin degeneracy and Fermi surface topology \cite{varlet2014anomalous,seiler2024probing} of the system.
In these measurement, the longitudinal resistance $R_{xx}$ is measurement as a function of the externally applied perpendicular magnetic field $B_\perp$. In the low-field regime before the Landau level is formed, $R_{xx}$ oscillates periodically as a function of $1/B_\perp$, the oscillation frequency is $f=\frac{h}{2\pi e} A_{FS}$, where $A_{FS}$ is the area enclosed by the Fermi contour in the $k$-space. By performing a discrete Fourier transform and mapping out the oscillation frequency as a function of carrier density and electrical displacement perpendicular to the sample, many phase transitions originating from isospin symmetry breaking and changes in the Fermi surface topology can be clearly identified~\cite{zhou_half_2021, zhou_isospin_2021, zhang2023enhanced}. To obtained high-resolution SdH spectrum, proper sampling of the magnetic field is essential. In Ref.~\cite{zhou_half_2021}, a uniform sampling in $1/B_\perp$ in the high-field regime followed by uniform sampling in $B_\perp$ is used. In addition, the application of a window function, such as the Hanning Window, for the discrete Fourier transform may also be considered.

The limitation of charge transport measurement is that the resistance of the sample can have various intrinsic and extrinsic origins, making it challenging to directly compare the results to theoretical models. On the other hand, electrical measurements of thermodynamic quantities, such as electronic compressibility and chemical potential, would provide more insights of the systems.
Quantum capacitance measurement~\cite{eisenstein_negative_1992, eisenstein_compressibility_1994} is one of the approach to access the thermodynamic properties of the system. Compared to resistance, the quantum capacitance has a relative simple physical origin, making it easier to fit to theoretical models. Probing the capacitance of mesoscopic samples like the Van der Waals heterostructures is more challenging than probing the resistance. The main reason is that the parasitic capacitance of the electrodes and measurement setup is usually orders of magnitude larger than the sample capacitance. In recent experiments\cite{zibrov_even_2018},a capacitance bridge circuit including a high electron mobility transistor (HEMT) is used to decouple the sample from the measurement system, allowing the probing of the capacitance values among the sample and metallic gates. In Ref.~\cite{zhou_half_2021} and \cite{de2022cascade}, the penetration field capacitance -- capacitance between the top and bottom gates are measured to extract the inverse electronic compressibility $\kappa = \partial \mu / \partial n$. The result can be directly fitted to a theoretical model to identify the symmetry breaking phases and anisotropic interactions. Apart from capacitance measurement, the chemical potential of the system can be also be directly probe by adding a sensing layer in the heterostructure\cite{yang_experimental_2021, han_chemical_2023}. By measuring the dependence of chemical potential on environmental parameters such as temperature and magnetic field, additional thermodynamic properties such as entropy and magnetization can be derived.

Apart from electrical measurements, other techniques have also been used to study rhombohedral graphene systems. In Ref~\cite{kerelsky_moireless_2021}, scanning tunneling microscopy (STM) is used to study ABCA-stacked four-layer graphene domains. One major limitation for STMs is that the sample has to be open-faced such that the probe can directly access the sample surface. Therefore, the dual-gate control of carrier density and displacement field needs extra sensing layer and can be complicated. Scanning superconducting quantum interference device (SQUID) is another scanning probe setup that can directly sense the local magnetic field. In Ref~\cite{arp_intervalley_2023}, the local magnetization of rhombohedral trilayer graphene is studied with this technique. Optical and photocurrent measurement are also common techniques to study condensed matter systems. In Ref~\cite{yang_spectroscopy_2022}, photocurrent spectroscopy is used to study the correlated phase transition in rhombohedral trilayer graphene / hexagonal boron nitride moire superlattice. In Ref.~\cite{holleis_isospin_2024}, flavor polarization of rhombohedral trilayer graphene is optically detected by probing the exciton spectra of a ${\rm WSe_2}$ sensing layer.

\subsection{Bernal Bilayer Graphene}
\label{sec:exp_bilayer}

Let us begin our discussion with the experimental data for Bernal bilayer graphene as presented in Ref.~\cite{zhou_isospin_2021}. \cref{fig:bbg_kappa_phase_diagram} illustrates how the thermodynamic inverse compressibility, defined as the infinitesimal change of chemical potential with respect to density, $\kappa\equiv \partial \mu /\partial n_e $, varies as a function of density $n_e$ and $D$, at very low temperature of $T\approx 20$mK. The most striking features in the data are the sharp blue lines and several darker regions identified by $\kappa$ minima, which fan out from the origin where both $D$ and $n_e$ are small. These features, particularly the prominent blue lines, strongly suggest a potential phase transition between two phases with distinct chemical potentials  $\mu=\partial E/\partial n_e$. To confirm this hypothesis, the dependencies of longitudinal resistance and inverse compressibility on a weak perpendicular magnetic field were studied, revealing oscillatory behaviour at low temperatures—a characteristic commonly observed in metals.
The results for $R_{xx}$ are shown in \cref{fig:bbg_kappa_phase_diagram}b), indeed reveal that the oscillation frequencies change across different parts of the phase diagram. The $x$-axis of  \cref{fig:bbg_kappa_phase_diagram}b) is the ratio of the oscillation frequency normalized by the electron density $|n_e|$ times flux quanta  $\phi_0=h/e$:
\begin{equation} \label{eq:qo_frequency}
	f_{\nu}=\frac{F_{\nu}}{\phi_0|n_e|}.
\end{equation}
In the semiclassical limit, where the spacing between Landau levels is small compared to the Fermi energy, electrons exhibit phase coherence and orbit the Fermi surface with an enclosed area $A_{\mathrm{FS},\nu}$ at the fundamental frequency $F_{\nu} = \hbar A_{\mathrm{FS},\nu} / (2\pi e)$.
This parameter, $f_{\nu}$, thus provides insights into the number and relative areas of occupied Fermi surfaces, serving as an important tool for detecting Fermi surface reconstructions often associated with magnetic phase transitions. In the paramagnetic state, typically, four Fermi surfaces are expected, correlating with the two spin and two valley degrees of freedom, which we refer to as flavor degrees of freedom in this article. 
In symmetry-broken phases, some flavors may not be occupied, and their characteristics can vary. For example, wavefunctions from opposite valleys might mix.

In scenarios where either electron-like or hole-like Fermi surfaces are present, each Fermi surface corresponds to a quantum oscillation frequency, $f_{\nu}$, with all frequencies summing to unity:
\begin{equation}
    \sum_{\nu=1}^{k} f_{\nu} = 1,
\end{equation}
where $k$ represents the number of Fermi surfaces.  For example, a typical paramagnetic state with a single simple Fermi surface per flavor will exhibit four frequencies of $0.25$ each. Conversely, a symmetry-broken phase with a single, simply-connected Fermi surface in one flavor will show a singular frequency of $1$, indicative of strong ferromagnetism, termed a quarter metal.
 If a symmetry-broken phase hosts two equally occupied Fermi surfaces, each enclosing half the total electron density $|n_e|$, it results in two frequencies of $0.5$ -- a half-metal. When additional carriers are introduced into a half-metal across a maximum threshold, tiny new Fermi surfaces typically emerge, leading to more complex frequency distributions in what is known as an Partially Isospin Polarized (PIP) phase, yet the sum of all frequencies remains equal to 1. From the frequencies, the PIP phases resembles an almost-half metal ferromagnet. For an annular Fermi sea, where both electron-like and hole-like Fermi surfaces exist, the frequency sum rule must be adjusted so that hole-like and electron-like frequencies contribute with opposite signs in the sum-rule. The paramagnetic and generalized ferromagnetic metals identified in Ref.~\cite{zhou_isospin_2021} are catalogued in Table.~\ref{table:terminologies}.

Similar phases have been reported in the quantum oscillations of inverse compressibility by Barrera et.~al.~in Ref.~\cite{de2022cascade}. Additionally, they noted that these symmetry-broken phases exhibit enhanced layer polarization.
While magnetic oscillations clearly identify spontaneous symmetry breaking, pinpointing the order parameter is more challenging. This usually requires a multifaceted experimental approach, including the study of the anomalous Hall effect, the influence of in-plane magnetic fields on phase boundaries, and local probes to reveal local magnetization. Comparing the experimental phase diagram with theoretical predictions is also beneficial. For instance, band structure calculations shown \cref{fig:bands_params_bilayer} reveal an enhanced density-of-states region in the electron density versus displacement field space, where the topology of the Fermi surface changes from three distinct pockets to a single, simply connected Fermi surface. While this enhanced density of states due to van-Hove singularities is important in driving these phase transition, it does not correspond straightforwardly to a particular line observed in the experimental data. Moreover, the Hartree-Fock phase diagram, calculated at a high dielectric screening constant $\epsilon=38$, \cref{fig:phase_diagram_bilayer} reproduces certain generic features of the phase diagram. We will delve into this comparison in greater detail later. For now, let us explore how other experimental data illuminate the nature of these phase transitions.

\begin{figure*}[t]
\centering
\includegraphics[width=1.0\linewidth]{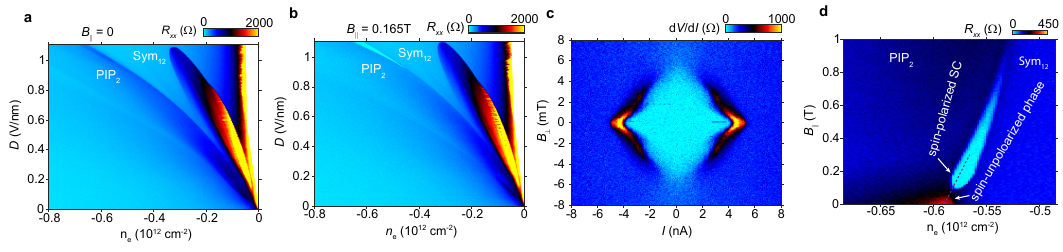}
\caption{
Magnetic Field-Induced Superconductivity in Bernal Bilayer Graphene. Panel (a) shows the resistance map without a magnetic field, and panel (b) shows the same map under an in-plane magnetic field of  $B_{\parallel}=0.165$T. Panel (c) shows the critical current versus out-of-plane magnetic field; note that the critical out-of-plane magnetic field is almost two orders of magnitude smaller than the in-plane field shown in panel b). $B_{\parallel}$ vs $n_e$ for a fixed displacement field
$D=1.02 $V/nm.
}
\label{fig:n_D_BBG_in_plane_field}
\end{figure*}

\cref{fig:n_D_BBG_in_plane_field} shows the change of the $n_e-D$ phase diagram under an applied in-plane magnetic field.
Surprisingly, the region between the nearly ferromagnetic half-metal state (PIP$_2$) and the paramagnetic state Sym$_{12}$ transitions into a superconductor.  This superconductivity is distinguished not only by a significant drop in resistance but also by a critical current that varies with the perpendicular magnetic field, as shown in Figure ~\ref{fig:n_D_BBG_in_plane_field}c). These observations confirm a genuine phase transition to a superconducting state, rather than a simple increase in metallic conductivity.

\begin{figure*}[t]
\centering
\includegraphics[width=0.98\linewidth]{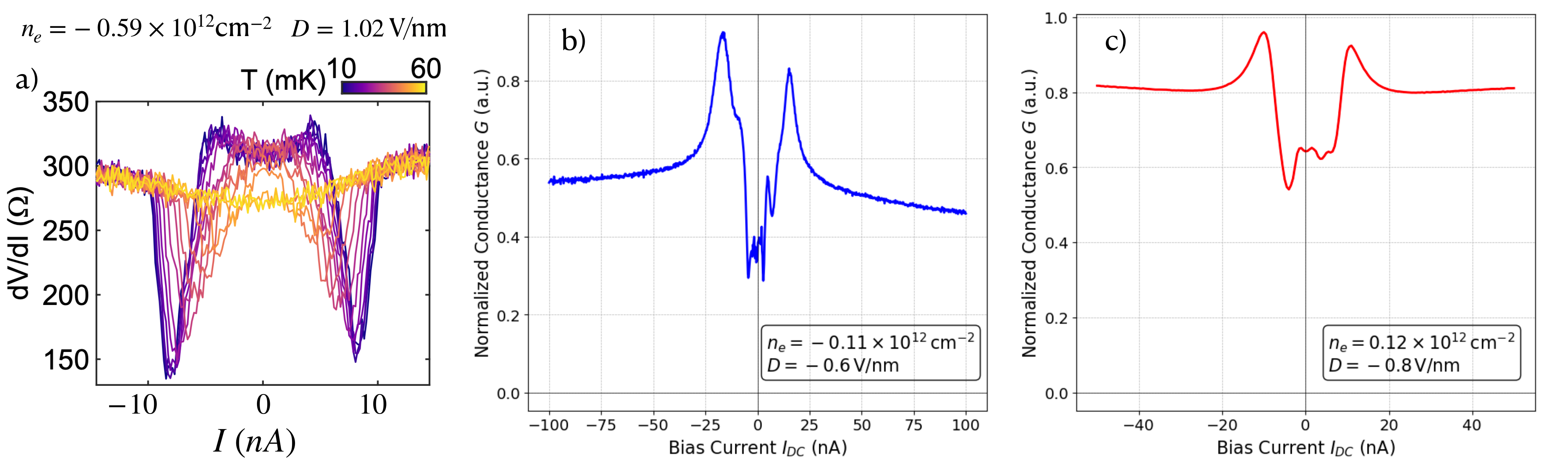}
\caption{%
Non-Ohmic IV transport characteristics in Bernal bilayer graphene across different densities and displacement fields. Panel (a) from Zhou.~et.al \cite{zhou_isospin_2021} uses four-terminal measurements to report differential resistance. This $(n_e,D)$ coordinate correspons to the ``normal-state'' of the field-induced superconductor, as shown in \cref{fig:n_D_BBG_in_plane_field}. Panels (b) and (c) from Seiler et.al.~\cite{seiler2022quantum,seiler2023interaction} use two-terminal measurements to report conductance, demonstrate similar non-Ohmic IV transport characteristics. Our theoretical study supports these experimental observations by identifying spin and valley density wave instabilities in the symmetric-12 paramagnetic state, see Figs.~\ref{fig:intervalley} and \ref{fig:intravalley}.
}
\label{fig:nonlinear-IV}
\end{figure*}

Fig.~~\ref{fig:n_D_BBG_in_plane_field}d) shows that superconductivity extends to a finite $B_{\parallel}$, but fades out at high $B_{\parallel}$.  If this superconductor is a spin-triplet superconductor, the mechanism limiting the critical in-plane field remains to be determined. Is the primary factor orbital depairing, or are other unknown influences at play? For instance, the critical out-of-plane magnetic field is very small, approximately $B_{\perp,c} \approx 5\,\mathrm{mT}$. Compared to the field required to stabilize the superconductor, even a slight misalignment between the 2D material plane and the applied field could suppress superconductivity.
The supplementary data from Ref.~\cite{zhou_superconductivity_2021} reveal that the superconducting critical temperature actually increases with the strength of the in-plane field, climbing from 30 mK at $B_{\parallel} = 0.165$ T to 40 mK at $B_{\parallel} = 0.3$ T.  This dependence of superconductivity on the in-plane magnetic field, combined with its proximity to nearly half-metal and paramagnetic state, suggests that a phonon-mediated pairing mechanism is unlikely to be the dominant pairing mechanism. This observation points towards alternative superconducting mechanisms.


The normal state from which this superconductivity emerges is particularly noteworthy. \cref{fig:nonlinear-IV}a)
shows the non-Ohmic behavior of the differential conductance $dI/dV$ vs $I$ across various temperature. By taking the high-temperature resistance at temperature $T=60$mK resistance as the Ohmic baseline ($R_{xx}^0\sim 300\Omega$), it is observed that at low currents and low temperature, the resistance is elevated above $R_{xx}^0$. When a temperature-dependent critical current is exceeded, the resistance falls back to approximately  $R_{xx}^0\sim 300\Omega$.
Notably, during this transition, there is a significant overshoot where the resistance sharply decreases below the baseline before stabilizing. This low-temperature nonlinear I-V characteristic is consistent with the phenomenology of sliding density-wave \cite{monceau2012electronic}.

Seiler et.~al.~\cite{seiler2022quantum} explored two-terminal conductance behavior on the hole-doped side of bilayer graphene within the density range of $-2\times 10^{11}\text{cm}^{-2}<n_e$, a focus area that was not extensively covered in Ref.~\cite{zhou_superconductivity_2021}. They observed a similar non-Ohmic transport behavior as shown in \cref{fig:nonlinear-IV}b).
More importantly, they observed that this low-current, low-temperature enhanced resistive state follows a specific trajectory in the $n_e$-$B_{\perp}$ parameter space where the magnetic flux per carrier equals to -2. This interesting report led Seiler et.~al. to propose the existence of an anomalous Hall Wigner crystal in Bernal bilayer graphene. 
However, their two-terminal measurement setup did not allow for Hall resistance measurements, which could have further substantiated their hypothesis.
More recently, Seiler et.~al.~\cite{seiler2024interaction} reported similar non-Ohmic behavior on the electron-doping side, as shown in \cref{fig:nonlinear-IV}c). Supplementary data in Fig.~S12c) of Ref.~\cite{seiler2023interaction} indicate these nonlinearities vanish as temperature increases from 10 mK to 1K, confirming that these effects are due to genuine many-body interactions. To further understand these data,
this review introduces new results shown in \cref{fig:intervalley,fig:intravalley}, where time-dependent Hartree-Fock theory is used to investigate finite momentum instabilities in the paramagnetic state of bilayer graphene, see Section ~\ref{sec:TDHF} for more details.
It is important to note that not all enhanced resistive states with nonlinear IV characteristics, as shown in \cref{fig:nonlinear-IV}, are associated with superconductivity. Specifically, superconductivity is associated only with the enhanced resistive state that occur between an almost-half-metal ferromagnet and a symmetric-12 metal. This observation suggests that non-linear IV characteristics are not directly correlated with the presence of superconductivity in Bernal bilayer graphene.


\begin{figure*}[t]
\hspace*{-3cm}
\includegraphics[width=1.3\linewidth]{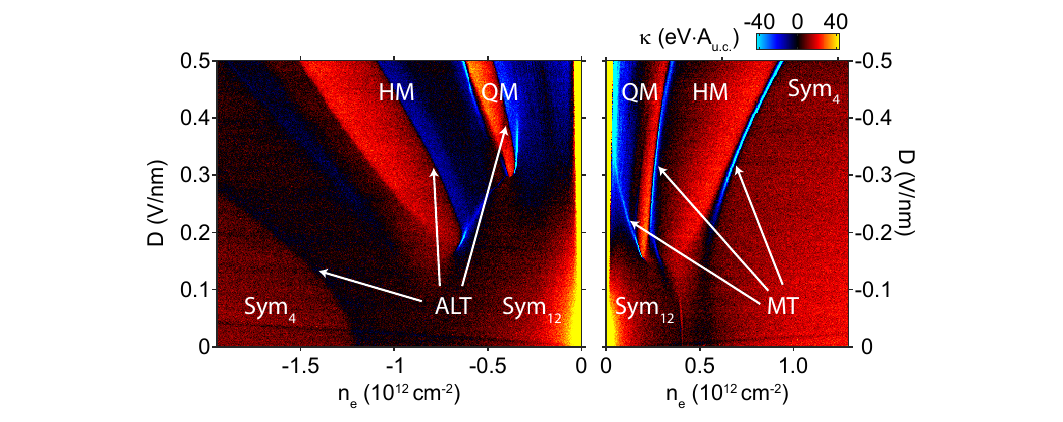}
\caption{
The change in inverse compressibility $\kappa \equiv \partial \mu/\partial n_e$ of rhombohedral trilayer graphene as a function of density and displacement field reveals several Magnetic Transitions (MT) and Annular Lifshitz Transitions (ALT). QM and HM denote symmetry broken quarter-metal and half-metal states, respectively, while Sym$4$ and Sym${12}$ represent two types of fully-symmetric paramagnetic metals, with their properties detailed in Table I. The magnetic transitions on the electron side ($n_e>0$) are distinct and clearly first-order, whereas those on the hole side ($n_e<0$) are less clearly defined.
}
\label{fig:RTG_compressibility}
\end{figure*}

\begin{figure*}[t]
\hspace*{-2.5cm}
\includegraphics[width=1.20\linewidth]{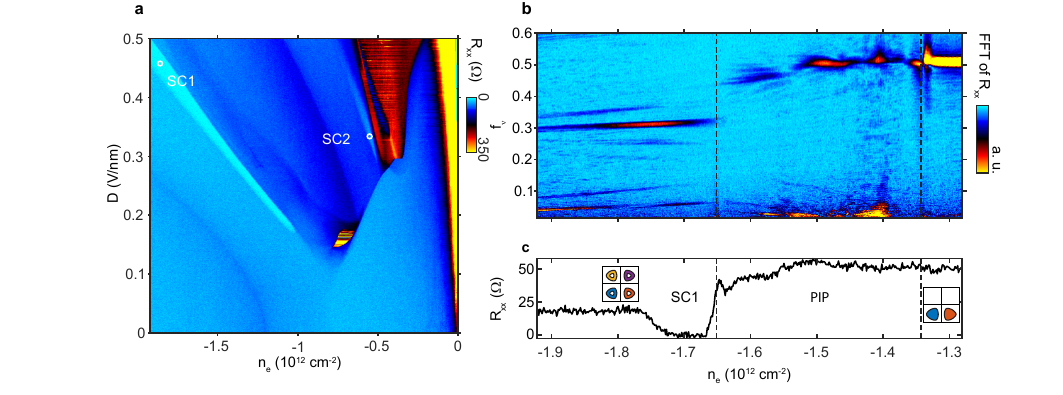}
\caption{%
Rhombohedral trilayer phase diagram. Panel a) shows resistance in the density $n_e$ and displacement field $D$ parameter space. Panel (b) shows the normalized Shubnikov-de Haas oscillation frequencies $f_{\nu}$ vs $n_e$ at $D=0.4$V/nm.
}
\label{fig:RTG_resistance}
\end{figure*}

Recently, Lin et.~al.~\cite{lin2302spontaneous} performed detailed angle-dependent nonlinear transport experiments on a bilayer graphene device with a "sunflower" geometry. In these experiments, an AC current was injected through a specific pair of leads, while the second-harmonic DC voltage response was recorded across others.
By fitting the voltage response with two cosine waves, they observed that the current response of the some metallic phase displays a directional dependence inconsistent with the $C_3$ rotational symmetry expected from the Slonczewski-Weiss-McClure band Hamiltonian.
This observation suggests that interaction effects drive momentum space condensation in these dilute electron systems \cite{jung2015persistent,dong2023isospin,huang2023spin}—a phenomenon where the distribution of electrons in momentum space becomes more compact, allowing them to collective minimize their exchange energy. Specifically, in multilayer graphene, where multiple small Fermi pockets exist, this effect may lead to a consolidation of electrons into specific pockets. This redistribution of electrons in momentum space reduces the original lattice symmetries of graphene ($C_{6v}$), resulting in transport coefficients that exhibit lower symmetry, detectable in transport experiments using devices in Ref.~\cite{lin2302spontaneous}. These experimental data can be used to apply the anisotropic magnetoconductance formula developed by Vafek \cite{vafek2023anisotropic}, which provides a closed-form expression for the electrical potential at any point on the disk when the current source and drain are located along the circumference.
Additionally, by examining the multiplicity of normalized quantum oscillation frequencies in Bernal Bilayer graphene with WSe$_2$, which were found to be inconsistent with multiples of three, Ludwig et al.~\cite{holleis2023ising} also discovered evidence of momentum space condensation, where tiny Fermi pockets merge. We will further expand on this discussion in \cref{sec:SOC} on spin-orbit coupling.



\subsection{Rhombohedral Trilayer Graphene (RTG)}
\label{sec:exp_trilayer}
In this section, we review the electron density ($n_e$) and displacement field ($D$) phase diagram of rhombohedral trilayer graphene (RTG), as characterized by its inverse compressibility $\kappa$ and longitudinal resistance $R_{xx}$. Magnetic oscillations of these quantities not only reveal magnetic phase transitions (flavor-symmetry breaking) in both conduction and valence bands but also Lifshitz transitions that modify the topology of the Fermi sea in the valence band.  Below, we discuss key features properties associated in these transitions.

In the electron-doped regime ($n_e>0$), Fig.\ref{fig:RTG_compressibility} shows that $\kappa$ exhibits three prominent blue lines indicating sudden dips across narrow density windows. These dips suggest energy level crossings between quantum many-body states, i.e.~first-order phase transitions. Indeed, quantum oscillations data reveal that these lines separate metallic phases with different number of Fermi surfaces. The phases identified include the two paramagnetic metals (symmetric-4 and symmetric-12 state) and two generalized ferromagnetic metal (the  half-metal phase and the quarter-metal). See Table.~\ref{table:terminologies} for more discussions on their properties. As $n_e$ decreases towards zero from the electron side, the system undergoes a sequence of phase transitions: it transitions from a paramagnetic state to a half-metal, then to a quarter-metal, and finally reverts to the paramagnetic symmetric 12-metal state.
This sequence of symmetry breaking, often referred to as the "cascade-transition", is primarily due to the collective lowering of exchange energy among the doped charges when their pseudospins are aligned.

On the hole-doped side ($n_e < 0$), in addition to the above mentioned magnetic phase transitions, the behavior of $\kappa$ becomes more complex due to the potential formation of annular shaped Fermi sea within each flavor. The outer and inner contours of these annular Fermi seas correspond to hole-like and electron-like Fermi lines (surfaces), respectively, with the hole-like sections exhibiting higher Fermi velocities than their electron-like counterparts, as illustrated in \cref{fig:bands_params_trilayer}. This transition from a simply connected Fermi sea to an annular Fermi sea is termed an Annular Lifshitz Transition (ALT). During this ALT, $\kappa$ show a steep drop and often changes sign, driven by a significant negative contribution from the exchange energy of the tiny electron-like Fermi surfaces to the chemical potential \cite{huang2023spin}.
The behavior of $\kappa$ thus differs at magnetic phase transitions compared to annular Lifshitz transition lines. At magnetic transitions, such as those observed in the conduction band, $\kappa$ typically shows a dip indicative of a first-order transition. However, when the topology of Fermi sea changes from simply connected to annular type at an ALTs, as found in the valence bands, $\kappa$ undergoes a steep drop and often changes sign. Unlike the ALT on the hole side and the magnetic transition on the electron side, the magnetic transitions on the hole side do not produce a pronounced signature in $\kappa$. However, the quantum oscillation frequency $f_\nu$ does change abruptly, as shown in \cref{fig:RTG_resistance}b, indicating a (weakly) discontinuous phase transition from the symmetric-4 paramagnetic metal with an annular Fermi sea to the nearly half-metal ferromagnet.

At the ALT, there is also noticeable increase in electrical resistance, as shown in \cref{fig:RTG_resistance}. This enhanced resistance is attributable to the newly formed electron-like Fermi surfaces, which offer a high density of states for scattering but has very lower mobility due to their small velocity. 
These features are supported by mean-field theory \cite{huang2023spin}. A particularly prominent feature in \cref{fig:RTG_resistance} is the strong resistance peak at the endpoint of the second ALT, around $(n_e, D) = (-0.6, 0.15)$. This region marks the convergence of the paramagnetic sym$_{12}$ state, a paramagnetic state with an annular Fermi sea and a half-metal phase.

In contrast to bilayer graphene, superconductivity in RTG emerges without the application of an in-plane magnetic field or proximity to transition-metal dichalcogenides layer.
The phase diagram features two distinct superconductivity regions: SC1, situated close to the nearly half-metal PIP$_2$ phase, and SC2, situated close to the nearly quarter-metal PIP$_1$ phase. Notably, SC1 does not exhibit significant Pauli limit violation, whereas SC2 displays strong Pauli limit violation.
Significant theoretical attention has been devoted to explaining SC1. Since it occurs at the boundary of an annular Fermi sea and an almost-half-metal ferromagnet (PIP$_2$), there are theories that focus on attraction due to the shape of the annular Fermi sea \cite{fRG_wei2022,ghazaryan2021unconventional}, and there are theories \cite{chatterjee2022inter} that focus on the attraction mediated by XY-like fluctuations of the inter-valley coherent order in the almost-half-metal ferromagnet \cite{chatterjee2022inter}.  Additionally, some studies explore more conventional phonon-mediated pairing mechanisms \cite{chou2022acoustic}. For a more detailed discussion on these diverse theories of superconductivity, please refer to Ref.~\cite{pantaleon2023superconductivity}.  There is currently no consensus on the microscopic theory of superconductivity in multilayer graphene.
Lesser theoretical attention has been focused on SC2, partly due to the lesser-known ferriomologies surrounding this transition. Recently, Ref.~\cite{arp2024intervalley} carried out a detail study on the nature of the order parameter within the displacement density range where the quarter-metal is the ground state. They found that the quarter metal order parameter can either be Ising-like, termed valley-imbalance, or XY-like termed intervalley coherent state. These states are separated by a weakly first-order phase transition. This transition can be think of magnetic anisotropic transition, where the magnitude of the order parameter remains constant, but its orientation in the extended spin-valley space rotates. Similar to traditional magnets, the magnetic anisotropic energy in multilayer graphene is significantly influenced by spin-orbit coupling. Specifically, in multilayer graphene, Kane-Mele type spin-orbit coupling tends to favor the valley-imbalance state over the intervalley coherent state. This preference arises because the valley and spin can align (or anti-align) straightforwardly in the valley-imbalance state. By studying the shift of the magnetic anisotropic phase transition boundary under an external magnetic field, Ref.~\cite{arp2024intervalley} estimated the intrinsic spin-orbit coupling to be approximately $50\mu$eV, aligning with previous findings.
They observed that on the electron-side, the valley-imbalance quarter-metal appears at lower densities, closer to neutrality, compared to the intervalley coherent quarter-metal. Conversely, on the hole-side, the intervalley coherent state is favored at lower densities. Hartree-Fock studies indicate that their energies are very close. At high displacement fields, valley-imbalance occurs at lower densities, with this trend reversing at smaller displacement fields. We will delve into the details of magnetic anisotropic energy in greater detail in later sections.


\subsection{Summary of Experimental Insights}

Let us provide a concise summary of the key experimental findings in both Bernal bilayer graphene and rhombohedral trilayer graphene. 
A quick glance at the resistance values $R_{xx}$ across the $n_e-D$ phase diagram reveals that they are significantly below the von Klitzing constant $25813\Omega$  - the natural unit of resistance in two dimensions. This suggests that the quasiparticle wavefunctions are extended throughout the 2D material and there are many Landauer conduction channels. 
Additionally, pronounced quantum oscillations in both $\kappa$ and $R_{xx}$ at low temperatures indicate that lightly-doped multilayer graphene behaves like a good metal with coherent quasiparticles and should conform to Fermi liquid theory. Multilayer graphene then distinguishes itself from traditional metals by possessing extended spin and valley degrees of freedom, along with many small Fermi pockets.

Starting with the simplest scenarios in the electron-doped region of RTG, there are three distinct magnetic transitions. The symmetric-4 paramagnetic state transitions to a half-metal, then to a quarter-metal, and finally transitions back to the symmetric-12 paramagnetic state. This sequence of ground state evolution with density agrees well with predictions from  Hartree-Fock theory that accounts for long-range Coulomb repulsion, except for the predicted but experimentally unobserved three-quarter metal state between the half-metal and paramagnetic state.

The next layer of complexity involves the magnetic phase transitions and annular Lifshitz transitions observed on the hole-doped side of RTG. In addition to the quarter-metal and half-metal, there are now almost-half-metal ferromagnet and almost-quarter-metal ferromagnet states, both featuring small Fermi surfaces in the minority flavor. Surprisingly, superconductivity (SC1) occurs at the boundary between the symmetric-4 and almost-half-metal ferromagnet, and another superconducting phase (SC2) appears at the boundary between the half-metal and almost-quarter-metal ferromagnet. These superconducting phases occur at the phase boundaries of different metals.

The most complicated scenario happens on the hole-doped side of Bernal bilayer graphene. Here, the symmetric$_{12}$ state reemerges between the almost-quarter-metal ferromagnet (PIP$_1$) and the almost-half-metal ferromagnet (PIP$_2$), and do not conform to the conventional cascade transition pattern that typically involves depopulating one flavor at a time. Between PIP$_2$ and the symmetric-12 state, the normal state exhibits non-linear IV characteristics, suggesting metal with broken continuous translation symmetry. Upon application of an in-plane magnetic field, it transitions into a field-induced superconductor. Again, the superconducting phase occurs at the phase boundaries of different metals.






\section{Spin and orbital Metallic Ferromagnetism}
In this section, we discuss the minimal theoretical model to study phase transitions in multilayer graphene driven by carrier density and electric displacement field. 
The two-dimensional electron gas (2DEG) and the (extended) Hubbard model, as foundational models in condensed matter physics, offer potential starting points. A pertinent question arises: which model is more suitable for capturing the essential physics behind the phase transitions in multilayer graphene?

To answer this, it is important to note that the phase transitions in the experiment are observed when the average concentration of excess or missing $\pi$-electrons per carbon atom is on the order of $10^{-4}$.
At such low filling fraction, the probability for two electrons to occupy the same carbon atoms is extremely small. This scenario makes it challenging to use an extended Hubbard model, which includes on-site interaction $U$, along with a few neighboring interaction terms ($V$, $V'$, etc.), to accurately describe such dilute systems. 
The Hubbard-$U$, which results from overlapping $\pi$ orbitals, are small compared to the $\pi$-electron bandwidth. 
Consequently, while the extended Hubbard model may yield insights at higher filling fractions, such as near the half-filling of the $\pi$-electron band, its application and conclusions must be approached with caution in the context of low doping in multilayer graphene. 
We note there are ongoing attempts to simulate long-range interactions on real-space lattices with dense momentum meshes \cite{fischer2024spin} to capture tiny Fermi surfaces at the Brillouin zone corners. 
Overcoming these computational challenges could provide an accurate description of interaction effects in these systems. 

Further insights can be gained by recognizing that the dilute concentration of carriers in the $\pi$-band of multilayer graphene aligns well with the situation in the 2DEG. 
At low energies, the electron wavefunction can be expanded around the two high-symmetry, inequivalent corners of the hexagonal 2D Brillouin zone, using ${\bm{k}\!\cdot\!\bm{p}}$ theory \cite{slonczewski1958band}. 
The wavefunction then is a product of a plane wave $e^{i\vec{k} \cdot \vec{r}}$ and a multicomponent spinor $\chi_{n\vec{k}}$:
\begin{equation} \label{eq:wavefunction}
	\psi_{n\vec{k}}(\vec{r}) = \frac{e^{i\vec{k} \cdot \vec{r}}}{\sqrt{\mathcal{A}}} \chi_{n\vec{k}},
\end{equation}
where $\mathcal{A}$ is the sample area, $n$ is a band index, $\vec{k}=(k_x,k_y)$ is the two-dimensional wave-vector relative to the Brillouin zone corner, and $\vec{r}=(x,y)$ are real-space coordinates. 
The spinor $\chi_{n\vec{k}}$ represents the probability amplitude and relative phase for the electron to occupy different carbon sublattices and is $\vec{r}$-independent. 
For $N$-layer graphene, $\chi_{n\vec{k}}$ is a $2N$-dimensional vector. 
To account for contributions from spin and valley degrees of freedom, the dimensionality of the spinor is often enlarged to $8N$: 
\begin{align}
	\chi_{n\vec{k}} = \sum_{\alpha} z_{n\vec{k},
			\alpha} \psi_{\vec{k}\alpha} ,
\end{align}
where $\alpha=(s,\tau,\sigma,l)$ represents the spin $s\in\{\uparrow,\downarrow\}$, valley $\tau\in\{K,K'\}$, sublattice $\sigma\in\{A,B\}$, and layer $l\in\{1,2...,N\}$ degrees of freedom, respectively. Equation \eqref{eq:wavefunction} shows that the spatial dependence of the electron wavefunction in multilayer graphene is that of a simple plane wave, abstracting away the details of the underlying lattice structure. All specific lattice information is encoded in the $\vec{k}$-dependent spinor $\chi_{n\vec{k}}$, and their winding around the valley enriches the electron gas problem in multilayer graphene with topological numbers. When multiple electrons are
introduced, the primary interaction they encounter is the long-range Coulomb repulsion, which depends solely on the electrons' spatial coordinates and is independent of the spinor.
This simple picture provides a robust foundation for understanding the phase transitions observed with dilute doping in multilayer graphene.


In subsequent subsections, we will explore the interaction effects that shape the electronic properties of multilayer graphene. We will begin by examining dominant energy scales: band energy and exchange energy derived from long-range Coulomb interactions. We will then consider lower energy scales, focusing on sublattice- and valley-dependent interactions, as well as spin-orbit coupling. Finally, we will tackle the complexities of correlation energies, which are challenging to estimate accurately.

\begin{figure*}[th]
\centering
\includegraphics[width=0.99\linewidth]{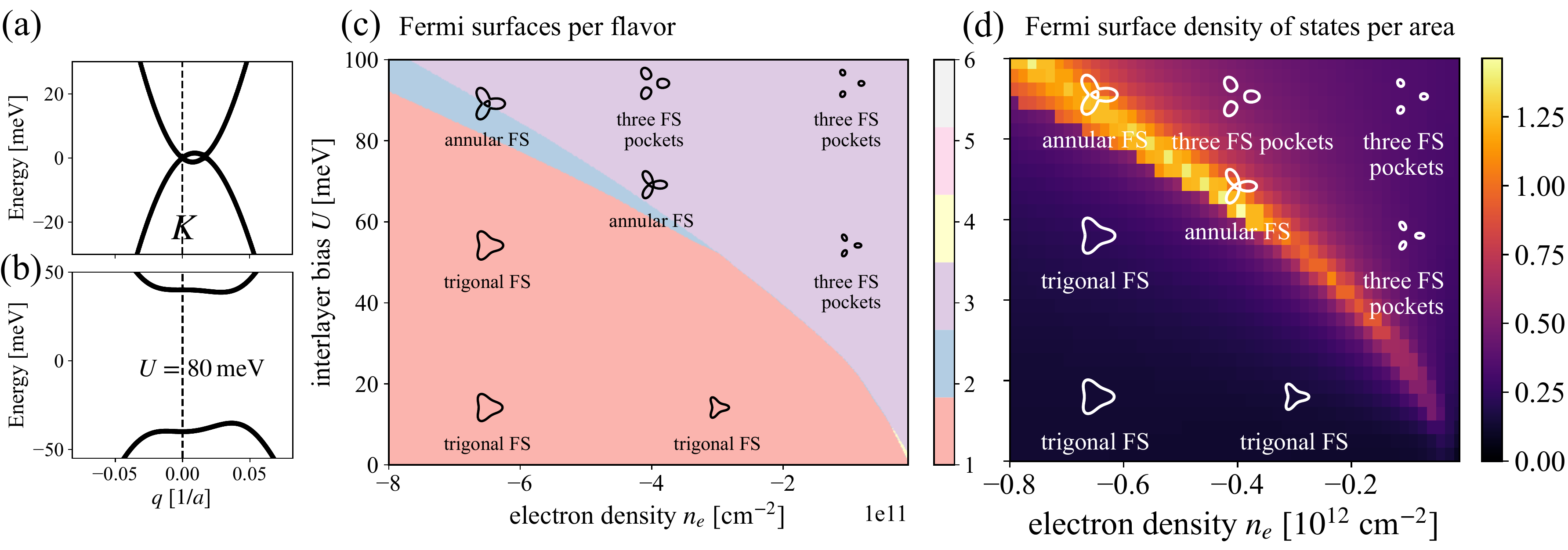}
\caption{%
Band structure model of bernal bilayer graphene in a finite displacement field $U_D$. (a) Band dispersion for $U_D=0$ (top) and $U_D=80$ meV (bottom). (b) Valence-band Fermi sea (FS) topology controlled by displacement field $U_D$ and hole doping $|n_e|$: we have sectors with a single pocket (red), an annular FS (blue), and three small disjoint pockets (purple) per valley flavor. (c) Density of states at the Fermi surface, illustrating how van-Hove discontinuities and singularities appear at Lifshitz transitions [region boundaries in panel (b)]. 
}
\label{fig:bands_params_bilayer}
\end{figure*}

\subsection{Slonczewski-Weiss-McClure
	(SWMc) Model}
\label{sec:SWMc}

The energy dispersion of Bloch waves induced by the crystalline potential of carbon allotrope can be conveniently described by the Slonczewski-Weiss-McClure (SWMc) model \cite{wallace1947band,mcclure1957band,slonczewski1958band}. 
The SWMc model is a tight-binding model orginally designed for for graphite and it consists of six parameters, $\gamma_{i}$, where $i=0,1,\ldots,5$, describe the hopping between different carbon atoms. A comprehensive summary of this model can be found in the review article by Dresselhaus et al \cite{dresselhaus1981intercalation}. 
Monolayer graphene is the fundamental building block of carbon allotrope. 
When considering a few layers of graphene, since their outermost layers are exposed to us, the potential energy differences between layers and sublattices become significant and can be controlled experimentally. 
These energy differences are described by another set of interlayer potential differences, commonly denoted as $\delta$ or $\Delta$. These parameters have been determined through various techniques, including infrared spectroscopy \cite{zhang2008determination}, transport measurements \cite{shi2018tunable,zhou_half_2021,zhou_superconductivity_2021,che2020substrate}. 
More recently, SQUID-on-tip measurements \cite{zhou2023imaging} have found remarkable agreement between experimental data of the measured magnetic field and theoretical calculations of thermodynamic magnetization using Landau levels of the SWMc Hamiltonian. This provides strong support for the SWMc model and yields highly accurate SWMc parameters. We recommend the readers to refer to Extended Data Table 1 of Ref.\cite{zhou2023imaging} for a comprehensive list of SWMC parameters.


For a given set of SWMc parameters, we choose a basis (e.g., A1, B1, A2, B2, etc.) and construct the hopping matrix $\hat{T}_{\vec{k}}$ for the Hamiltonian
\begin{align}
	H_0 = \sum_{\alpha,\alpha'} (\hat{T}_{\vec{k}})_{\alpha \alpha'}^{\pdagger} \,  c_{\vec{k},\alpha}^{\dagger}c_{\vec{k},\alpha'}^{\pdagger} .
\end{align}
Diagonalizing the matrix $\hat{T}_{\vec{k}}$ determines the orientation of the spinors $\chi_{n\vec{k}}$ at each wavevector $\vec{k}$ and band index $n$, as well as the band dispersion $\epsilon_{n\vec{k}}$:
\begin{align}
	\hat{T}_{\vec{k}} \, \chi_{n\vec{k}} = \epsilon_{n\vec{k}} \chi_{n\vec{k}}.
\end{align}
Below, we discuss the two simplest cases of Bernal-stacked (or ``AB-stacked'') bilayer graphene and rhombohedrally stacking (or ``ABC-stacked'') trilayer graphene. Their SWMc hopping matrices are 
\begin{align}
\label{eq:continuum_hamiltonian_AB}
	\hat{T}_{\vec{k}}^{\rm AB} &= \begin{bmatrix}
		                         t_{0}(\vec{k}) + U_1 & t_{12}(\vec{k})                 \\
		                         t_{12}^\dagger(\vec{k})         & t_{0}(\vec{k}) + U_2
	                         \end{bmatrix}_{4\times 4} \!\!\!,
\\
\label{eq:continuum_hamiltonian_ABC}
	\hat{T}_{\vec{k}}^{\rm ABC} &= \begin{bmatrix}
		                          t_{0}(\vec{k}) + U_1 & t_{12}(\vec{k})                 & t_{13}                          \\
		                          t_{12}^\dagger(\vec{k})         & t_{0}(\vec{k}) + U_2 & t_{12}(\vec{k})                 \\
		                          t_{13}^\dagger                  & t_{12}^\dagger(\vec{k})         & t_{0}(\vec{k}) + U_3
	                          \end{bmatrix}_{6\times 6} \!\!\!\!,
\end{align}
where $\vec{k}=(k_x,k_y)$ is the Bloch momentum relative to valley $\pm\vec{K}$. 
Here $U_{l}$ are the electric potentials of each layer $l=1,2,3$ and the $t$ matrices describe the intra-layer and inter-layer tunneling. The two valleys are related by time-reversal symmetry through $T_{\bm{k},-K}=T_{-\bm{k},K}^*$.
The tunneling $t(\vec{k})$ has a relatively simple matrix form when considering the dispersion near valley $K$: the matrix elements for electrons tunneling to different in-plane $(x,y)$ positions acquire a phase through $k_{x} + i k_{y}= k e^{i\theta_{\vec{k}}}$:
\begin{align}
	t_{0}(\vec{k})      & = \begin{bmatrix}
		                    0                    & v_0 k e^{-i\theta_{\vec{k}}} \\
		                    v_0 k e^{i\theta_{\vec{k}}} & 0
	                    \end{bmatrix},   \\
	t_{12}(\vec{k}) & = \begin{bmatrix}
		                    -v_4 k e^{-i\theta_{\vec{k}}} & v_3 k e^{i\theta_{\vec{k}}}  \\
		                    \gamma_1              & -v_4 k e^{-i\theta_{\vec{k}}}
	                    \end{bmatrix}, \quad
	t_{13}           = \begin{bmatrix}
		                    0 & \gamma_2/2 \\
		                    0 & 0
	                    \end{bmatrix},
\end{align}
where the velocity parameters are $v_i=(\sqrt{3}/2)a\gamma_i/\hbar$. We emphasize that $t_{13}$ has no $\vec{k}$-dependence because the $A1$-$B3$ atoms are only vertically displaced in rhombohedral stacking. This configuration creates extra repulsion between electrons in the first and third layers, making it thermodynamically less favorable. In contrast, for Bernal stacking, $t_{13}$ includes the additional phase factor $e^{\pm i\theta_{\vec{k}}}$, causing this matrix element to vanish at the high-symmetry point $K$. This subtle difference significantly impacts the low-energy band dispersion.

The spectrum of $\hat{T}_{\vec{k}}^{\rm AB}$ and $\hat{T}_{\vec{k}}^{\rm ABC}$ has been explored in previous theoretical studies \cite{mccann2013electronic,neto2009electronic}. In this review, we wish to highlight a few significant aspects regarding the eigenspectrum of $\hat{T}_{\vec{k}}$ in multilayer graphene that is important to the density-displacement field driven phase transitions.

\begin{figure*}[th]
\centering
\includegraphics[width=0.90\linewidth]{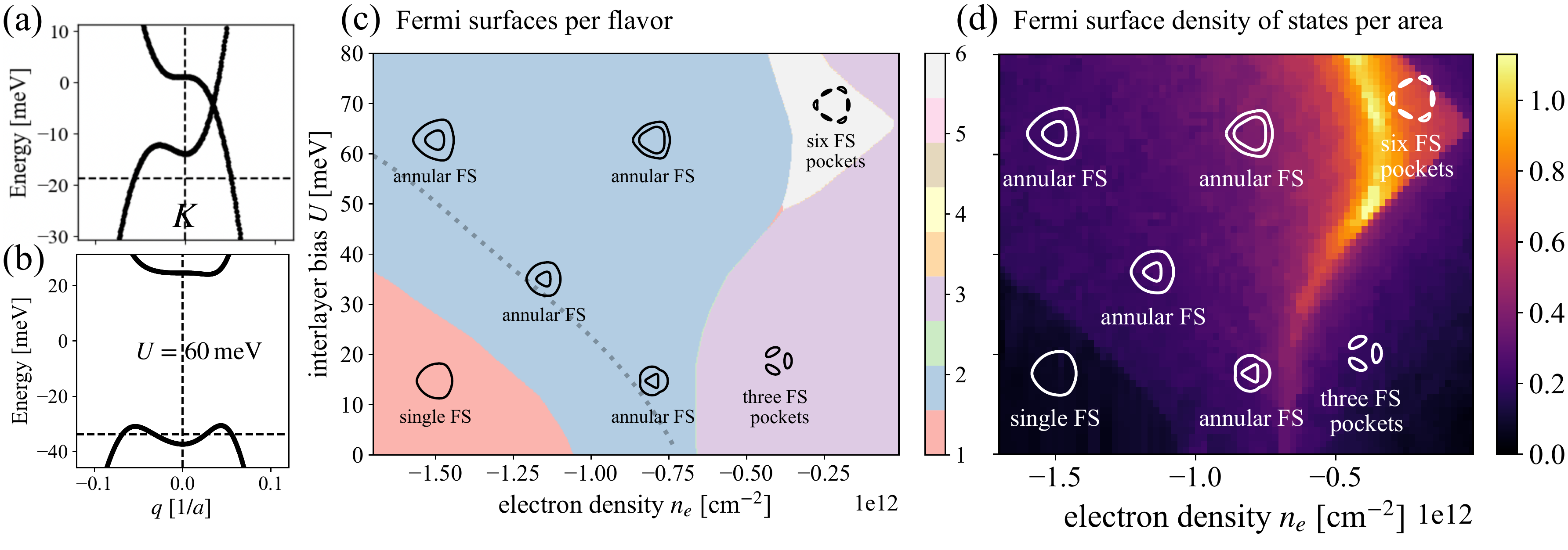}
\caption{%
Band structure model of rhombohedral trilayer graphene in a finite displacement field $U_D$. (a) Band dispersion for $U_D=0$ (top) and $U_D=60\,\mathrm{meV}$ (bottom). (b) Valence-band Fermi sea (FS) topology controlled by displacement field $U_D$ and hole doping $|n_e|$: we have sectors with a single pocket (red), an annular FS (blue), three small pockets (purple) per valley flavor, and six small pockets per flavor (white). (c) Density of states at the Fermi surface, illustrating how van-Hove discontinuities and singularities appear at Lifshitz transitions [region boundaries in panel (b)].
Compared with \cref{fig:bands_params_bilayer}, we note that the annular sector takes up a large part of the parameter space. 
}
\label{fig:bands_params_trilayer}
\end{figure*}

\Cref{fig:bands_params_bilayer,fig:bands_params_trilayer} summarize the band structure, Fermi sea topology, and density-of-states for $\hat{T}_{\vec{k}}^{\rm AB}$ and $\hat{T}_{\vec{k}}^{\rm ABC}$ as functions of hole-density $n_e$ and displacement field potential $U_D$. 
Figures~\ref{fig:bands_params_trilayer}a and b show the band dispersion around the Brillouin zone corner $K$ for rhombohedral trilayer graphene at interlayer potential $U_D=0$ and $U_D=60\,\mathrm{meV}$. Due to the rhombohedral stacking, the band-degenerate points are shifted away from the high-symmetry $K$ point. Because of the $C_3$ symmetry, there are three such points. This shift causes the energy at the $K$ point to exhibit a shallow minimum in the valence band. This feature is significant as it indicates a change in the Fermi sea topology. Additionally, there is a noticeable particle-hole asymmetry: the shallow maximum in the conduction band at the $K$ point is barely visible compared to the local minimum in the valence band. Consequently, we expect a smaller window for an annular Fermi sea in the conduction band.
\Cref{fig:bands_params_trilayer}b) shows that an interlayer potential $U_D$, driven by an applied external electric displacement field, opens a band gap. Figure \ref{fig:bands_params_trilayer}c) illustrates how the Fermi sea topology changes in the $n$-$U_D$ space. Notably, there are two types of Lifshitz transition: an annular Lifshitz transitions occur where new Fermi surfaces appear at $k=0$, and saddle-point Lifshitz transitions where Fermi surfaces merge/appear at $k \neq 0$. As shown in Figure  \ref{fig:bands_params_trilayer}d), the density of states exhibits a discontinuous jump in the former case and diverges logarithmically in the latter.  Importantly, these intriguing Fermi sea topologies, including annular shapes and distinct pockets, are not merely theoretical predictions. They have been identified experimentally through high-resolution magnetic oscillation measurements, as shown in \cref{fig:RTG_resistance}.

\Cref{fig:bands_params_bilayer}a shows the band structure of Bernal-stacked bilayer graphene. Unlike rhombohedral-stacked multilayer graphene, the band-touching point remains situated at the $K$ point. Notably, because $\gamma_3 \neq 0$, there is a small trigonal warping around the band-touching point. Similar to rhombohedral-stacked graphene, the bands become flat near the band edge when an electric displacement field is applied. Figures~\ref{fig:bands_params_bilayer}c and d illustrate the Fermiology and the associated density of states. The regions of enhanced density of states near the Lifshitz transitions differ significantly between Bernal bilayer and rhombohedral trilayer graphene, which can be traced back to their distinct stacking arrangements.

\begin{figure}[h]
\centering
\includegraphics[width=1.0\linewidth]{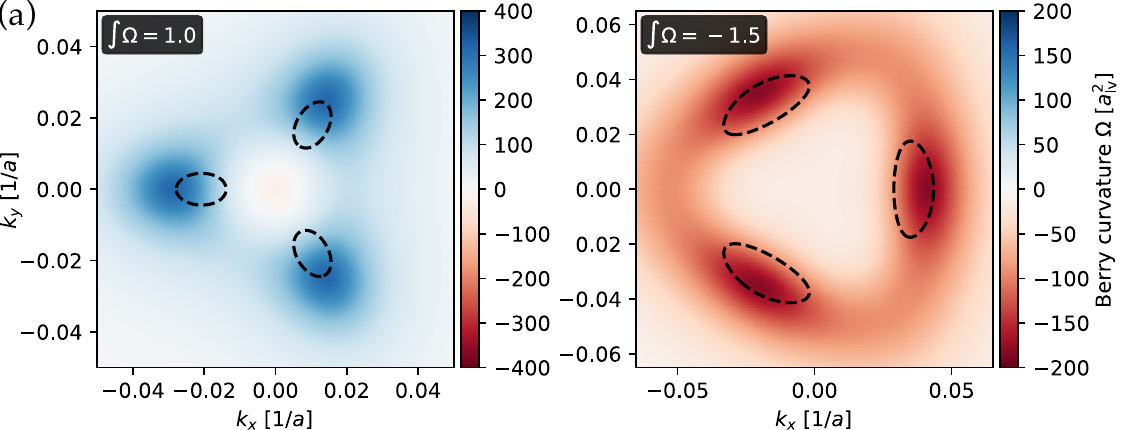}
\caption{%
Berry curvature of the valley-projected valence band at finite displacement fields in (a) Bernal bilayer graphene ($U_D=30$~meV) and (b) rhombohedral trilayer graphene ($U_D=30$~meV). The dashed lines indicate the Fermi surfaces at small hole doping. The integrated Berry curvature yields values $1$ and $3/2$, respectively.
}
\label{fig:bands_berrycurvature}
\end{figure}

Bloch states in (multilayer) graphene exhibit
singularities, where the value of the spinor depends on the direction from which the point is approached: $\lim_{\vec{k}\rightarrow0}\chi_{n\vec{k}}=\chi_{n\vec{\hat{k}}}$. These singularities arise from band-touching points, where the spinor's orientation rapidly rotates around the degeneracy, resembling a vortex in momentum space. The non-trivial winding of the spinor around the high symmetry Brillouin zone corners can be quantified by the gauge-invariant Berry curvature
\begin{align}
	\Omega_{n\vec{k}} = \nabla_{\vec{k}} \times \left( \chi_{n\vec{k}}^{\dagger} i \nabla_{\vec{k}} \chi_{n\vec{k}} \right).
\end{align}
The finite value of $\Omega_{n\vec{k}}$ depends on how the
momentum-space singularity is resolved. When introducing a band gap in multilayer graphene by applying an electric displacement field, $\Omega_{n\vec{k}}$ takes on a finite value. When the topological band is fully occupied, the Berry curvature integrated over the band becomes a topological invariant, the so-called Chern number 
\begin{align}
	C_n = \frac{1}{4\pi}\int  \Omega_{n\vec{k}} d^2 k.
\end{align}
%
\Cref{fig:bands_berrycurvature} shows the Berry curvature distribution for the valley-projected first valence band in Bernal bilayer and rhombohedral trilayer graphene. The valley-projected contribution to the Chern number is ${C_n^K=1}$ and ${C_n^K=3/2}$, respectively, and summing over valleys yields an integer Chern number.

We emphasize that the non-trivial topology of the spinor significantly enriches the electron gas problem in multilayer graphene. Coupled with its high experimental tunability, this electronic system serves as an ideal ``toy model'' for studying interaction effects with a topologically non-trivial band structure. In our opinion, this model strikes a perfect balance of complexity: it features straightforward long-range Coulomb repulsion between the plane-waves alongside a non-trivial topology associated with the spinor.

\subsection{Coulomb Interaction and Hartree-Fock Theory}
\label{sec:hf_meanfield}

As discussed at the beginning of the theory section, the most dominant interaction in the dilute-doping limit of the multilayer graphene electron gas problem is the long-range Coulomb potential, represented by $V(r)=\frac{e^2}{\epsilon r}$ in CGS units.  In second quantized form, the Hamiltonian for this interaction is 
\begin{align}
	H_{\rm s} = & \frac{1}{2}\sum_{\vec{k},\vec{k'},\vec{q}} \sum_{\alpha,\alpha'} V_{q} \, \psi^{\dagger}_{\vec{k}+\vec{q},\alpha}\psi^{\dagger}_{\vec{k'}-\vec{q}, \alpha'}
	\psi_{\vec{k'},\alpha'}\psi_{\vec{k},\alpha}.
\end{align}
$\psi_{\vec{k},\alpha}$ is the annihilation operator of momentum $\vec{k}$ and ``pseudospin'' $\alpha$, where
$\alpha=\{l,\sigma,\tau,s\}$ encapsulates the layer, sublattice, spin, and valley degree of freedom. Here $V_q=\frac{2\pi e^2}{\epsilon q} $ is the Fourier component of the Coulomb potential. 
The subscript s in $H_{\rm s}$ emphasis that this interaction is symmetric with respect to any unitary rotations of the pseudospin $\psi_{\vec{k},\alpha'}=\sum_{\alpha}U_{\alpha'\alpha}\psi_{\vec{k},\alpha}$. The effect of sublattice and valley dependent short-range interaction is discussed in Section~\ref{sec:g}.
Given that most multilayer graphene devices are encapsulated by hBN and controlled with electrical gates (refer to Section I), small-$q$ screening effects become significant. 
For the dual-gate configuration, electrostatic screening from metallic gates leads to the modified interaction potential 
\begin{align}
	V_q = \frac{2\pi e^2 \tanh(qd)}{\epsilon q \mathcal{A}},
\end{align}
and for the single-gate configuration, the  modified potential is 
\begin{align}
	V_q = \frac{2\pi e^2 (1-e^{-qd}) }{\epsilon q \mathcal{A}}.
\end{align}

Since $H_{\rm s}$ is pseudospin-independent, the exchange interaction is minimized when all the pseudospins are polarized along the same directions. 
This polarization allows the spatial components of the wavefunction [cf.~\cref{eq:wavefunction}] to avoid overlapping through exchange holes, effectively reducing the expectation value of the Coulomb Hamiltonian, akin to the conventional electron gas problem. 
However, such distribution of pseudospin polarization clearly presents a conflict with the band Hamiltonian $H_0$  which tends to favor changes in sublattice and layer pseudospin across momentum space, particularly near singularities. 
Consequently, the initial approach to tackling the interaction problem in multilayer graphene, which involves both $H_0$
and $H_{\rm s}$, is to determine the optimal distribution function of electrons and pseudospin orientations in momentum space. 
This can be achieved using the self-consistent Hartree-Fock theory. 
Early Hartree-Fock studies on multilayer graphene, such as those in Ref.~\cite{macdonald2012pseudospin,zhang2010spontaneous,min2008pseudospin}, predominantly focused on neutrality and utilized a simplified band structure, often omitting effects like trigonal warping. 
Notably, Ref.~\cite{macdonald2012pseudospin} emphasizes the conceptual framework for understanding exchange fields in multilayer graphene electronic systems and draws parallels to itinerant electron magnetism, enhancing our comprehension of these complex interactions.
Following the experimental findings by Zhou et al., more detailed mean-field studies have been developed. These studies \cite{szabo2022metals,koh2024correlated,xie2023flavor,chatterjee2022inter,huang2023spin} are guided by high-precision magnetic oscillation data and they delve deeper into the complexities of electronic interactions and bandstructure in multilayer graphene.

In the self-consistent Hartree-Fock theory, to minimize the combined effects of both the exchange energy and band Hamiltonian, the spinor orientations must satisfy the Hartree-Fock eigenvalue equation
\begin{align}\label{eq:HF_eigen}
	\hat{H}_{\vec{k}}^{\text{MF}}\chi_{n\vec{k}} =\epsilon_{nk} \chi_{n\vec{k}} 
\end{align}
with the mean-field Hamiltonian 
\begin{subequations}
\begin{align} \label{eq:MF_ham}
	\hat{H}^{\text{MF}}_{\vec{k}} = \hat{T}_{\vec{k}} + \hat{\Sigma}_{\vec{k}}.
\end{align}
The self-energy matrix $\hat{\Sigma}_{\vec{k}}$ and the density matrix $\hat{\rho}_{\vec{k}}$ are 
\begin{align} \label{eq:Fock}
	\hat{\Sigma}_{\vec{k}} & = -\sum_{q} V_{q} \,\hat{\rho}_{\vec{k}+\vec{q}},                        \\
	\hat{\rho}_{\vec{k}}     & = \sum_{n} n_F(\epsilon_{n\vec{k}}-\mu) \chi_{n\vec{k}} \chi_{n\vec{k}}^\dagger,
\end{align}
\end{subequations}
where $\mu$ is the chemical potential and  $n_F(\epsilon)=1/(e^{\beta \epsilon}+1)$ is the Fermi-Dirac distribution at inverse temperature 
$\beta=1/k_BT$. The eigenvalue equation \eqref{eq:HF_eigen} must be solved self-consistently, with the chemical potential $\mu$ determined by the electron density $n_e$ through the relation 
\begin{align} \label{eq:chemical_pot_root}
	n_e - \frac{1}{\mathcal{A}}\sum_{n\vec{k}} n_F(\epsilon_{n\vec{k}}-\mu) = 0, 
\end{align}
where the sample area is related to the momentum discretization in the simulation through $\mathcal{A}=(2\pi/dk)^2$. 
Numerically, the bisection method is well-suited to efficiently solve \cref{eq:chemical_pot_root} due to the monotonically decreasing nature of $n_F(\epsilon)$. 

Solving \cref{eq:HF_eigen} requires self-consistent iteration, which generally involves variants of the following procedure:   
\begin{enumerate}[label=(\roman*)]
\item First step: Guess an initial self-energy matrix $\hat{\Sigma}_{\vec{k}}^{(0)}$.
\item Step ${i\to i+1}$: Given $\hat{\Sigma}_{\vec{k}}^{(i)}$, solve \cref{eq:HF_eigen} for $\epsilon_{nk}^{(i)}$, $\chi_{n\vec{k}}^{(i)}$. Then solve \cref{eq:chemical_pot_root} for $\mu^{(i)}$ and use \cref{eq:Fock} to construct $\hat{\Sigma}_{\vec{k}}^{(i+1)}$ (optionally with Anderson mixing).  
\item Compare $\hat{\Sigma}_{\vec{k}}^{(i)}$ and $\hat{\Sigma}_{\vec{k}}^{(i+1)}$, repeat (ii) if not converged. 
\end{enumerate}
Different initial guesses can lead to different converged  solutions. Finding the solution with the lowest energy per particle 
\begin{align}\label{eq:hf_energy}
    E_{\rm HF} = \frac{1}{|n_e|} \left(\frac{1}{\mathcal{A}} \sum_{\bm{k}} \Tr[\left(\hat{T}_{\vec{k}} + \frac{1}{2} \hat{\Sigma}_{\vec{k}} \right)\hat\rho_{\bm{k}}] - \epsilon_{\rm HF}^{n_e=0}\right)
\end{align}
requires a systematic exploration that seeds multiple initial states. Note that for continuum models of graphene multilayers, some care is required to define the total energy relative to the paramagnetic state at charge neutrality by subtracting $\epsilon_{\rm HF}^{n_e=0}$. 

The most time-consuming step in the iteration is the construction of the Fock-self energy in \cref{eq:Fock} because $\rho_{\vec{k}+\vec{q}}$ can extend beyond the simulation cell boundary. 
A fast and effective method to construct the Fock-self energy is to use the convolution theorem: 
by taking the Fast Fourier Transforms (FFTs)  $\mathcal{F}[\hat\Sigma_{\vec{k}}]=\hat\Sigma_{\vec{r}}$, 
$\mathcal{F}[V_{\vec{q}}]=V_{\vec{r}}$ and 
$\mathcal{F}[\hat\rho_{\vec{k}}]=\hat\rho_{\vec{r}}$, the self-energy matrix elements in real space become simple element-wise products  $\hat\Sigma_{\vec{r}}=V_{\vec{r}}\hat\rho_{\vec{r}} $. The inverse FFT then yields $\hat\Sigma_{\vec{k}}=\mathcal{F}^{-1}[\hat\Sigma_{\vec{r}}]$.

\begin{figure*}[ht]
\centering
\includegraphics[width=1.0\linewidth]{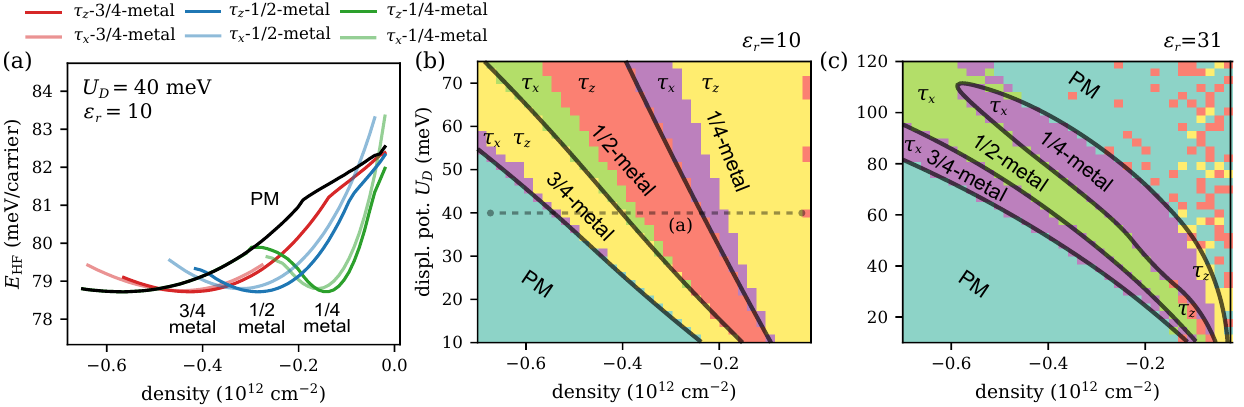}
\caption{%
Hartree-Fock phase diagram of hole-doped bernal bilayer graphene in a displacement potential $U_D$ at electron density $n_e<0$. 
(a)~Total energy per carrier $E_{\rm HF}$ of the symmetric state (i.e., the paramagnet) and the symmetry-broken flavor-polarized states as $n_e$ is varied and $U_D$ constant. The symmetry-broken states can be classified as $3/4$-metal, $1/2$-metal, and $1/4$-metal states, and whether they have valley-polarization $\tau_z$ or intervalley-coherence $\tau_x$. 
(b--c)~Two phase diagrams showing the lowest-energy Hartree-Fock state at given density $n_e$ and potential $U_D$, with different dielectric constant $\epsilon_r$.  When Coulomb interactions are stronger ($\epsilon_r=10$), the competition between $\tau_z$ and $\tau_x$ flavor-polarization is closer, with multiple first-order transitions. When Coulomb interactions are weaker ($\epsilon_r=31$), intervalley-coherent states dominate the phase diagram. The dashed line in panel (b) indicates the linecuts taken in panels (a). 
%
}
\label{fig:phase_diagram_bilayer}
\end{figure*}

\subsection{Nature of the Order Parameters}
In multilayer graphene, the order parameter that encapsulates both spin and valley degree's of freedom can require up to 15 parameters for a complete specification, unlike the usual magnetization order parameter which only needs three numbers to specify the magnetization direction. 
These $15$ numbers correspond to the generators of the $\mathrm{SU}(4)$ group, encompassing both spin and valley space, also referred to as flavor degrees of freedom.  
The order parameters of some simpler symmetry-broken states can be classified based on their low-frequency dynamics into three categories: Ising-like, XY-like, or Heisenberg-like.

\begin{enumerate}
\item \textbf{Spin Polarization ($\vec{s}=(s_x,s_y,s_z)$}) --- This order parameter is Heisenberg-like when the temperature exceeds the spin-orbit coupling strength. However, with modern experimental capabilities reaching temperatures as low as approximately $20$~mK, even intrinsic spin-orbit couplings, estimated at around $50\mu\text{eV}$ (approximately $500$~mK), are sufficiently strong to sustain spin order.

\item \textbf{Valley Polarization ($\tau_z$)} --- The valley-polarization, defined as $\tau_z = (n_K - n_{K'})/(n_K + n_{K'})$ quantifies the imbalance of carriers between opposite valleys and is characterized by Ising-like behavior. It can be stabilized at finite temperatures even in the absence of spin-orbit coupling, as the Hamiltonians for the opposite valleys are different.

\item \textbf{Intervalley Coherence ($\tau_{x,y}$)} ---  Intervalley coherent order emerges when the wavefunction becomes a superposition of states from opposite valleys, characterized by a phase $z_{n\vec{k},K}/z_{n\vec{k},K'} = e^{i(\phi_{\vec{k}}+\phi)}$.
      The collective rotation of $\phi_{\vec{k}}$ is an XY-like order parameter. When states from opposite valleys mix, the effective area of the Brillouin zone is reduced, corresponding to a tripling of the unit cell area in real space. The intrinsic microscopic mechanisms that pin the phase degree of freedom is an active area of research \cite{an2024magnetic}.
\end{enumerate}

The spin-polarized phase can be identified by examining the movement of phase boundaries with an in-plane magnetic field, particularly useful when adjacent phases lack net spin polarization or have low spin susceptibility \cite{zhou_half_2021,zhou_isospin_2021}. 
The valley-polarized phase, which breaks orbital time-reversal symmetry, results in a pronounced anomalous Hall effect due to orbital moments. The relationship between orbital magnetization and valley polarization is complex. Ref.~\cite{das2023unconventional} reports that orbital magnetization can undergo two sign changes and display non-analyticities while the ground state order parameters for valley and spin polarization remain unchanged.

Detecting the intervalley coherent phase is more challenging compared to the other phases; however, its effect of tripling the graphene unit cell can be revealed through atomic-scale scanning tunneling microscopy \cite{li2019STM,liu2022visualizing,coissard2022imaging}.

Even in the absence of electric-displacement-field-induced layer polarization, a dilute electronic system can spontaneously break layer-inversion symmetry, as observed in rhombohedral pentalayer graphene \cite{han2023orbital}. This spontaneous layer polarization behaves similarly to an Ising-like order parameter \cite{das2024superpolarized}. Ising domain walls and the defects associated with XY order parameters have been extensively explored in the literature, including in monolayer graphene under a strong magnetic field \cite{zhang2013valley, huang2021current, liu2022visualizing}. Ising domain walls have been studied theoretically \cite{zhang2013valley, huang2021current} and directly imaged through nano-SQUID \cite{arp2023intervalley}.
Topological defects associated with valley XY-phase have been reported in monolayer graphene under a strong magnetic field \cite{liu2022visualizing}. These findings illustrate the richness of the order parameters and their manifestations across different conditions and graphene configurations.

\subsection{Bernal Bilayer Graphene}

We now discuss the energy competition of symmetry-broken states and the resulting phase diagram of Bernal bilayer graphene obtained from self-consistent Hartree-Fock theory as described in \cref{sec:hf_meanfield}.  
A central finding from these calculations is that non-local exchange interactions drive Fermi-surface reconstructions with flavor polarization and momentum condensation when single-particle bands have a large density of states. 
The significant increase in degrees of freedom, coupled with the long-range nature of Coulomb interactions, leads to a plethora of symmetry-broken states that appear as local minima of the Hartree-Fock energy functional. 
To thoroughly explore these states numerically and identify the one with lowest energy, it is essential to start the self-consistent Hartree-Fock iterations with a variety of initial guesses, including different (spin-valley) magnetic states and random states. 
When the symmetry-broken state is unstable, the initial configurations (the seed) evolve towards a stable configuration during the iteration. 
Across much of the phase space, many seed states evolve into locally-stable excited states, indicating a rich landscape of magnetic order. 

\Cref{fig:phase_diagram_bilayer}a shows how the Hartree-Fock energy per particle [cf.~\cref{eq:hf_energy}] evolves with density $n_e$ for various competing states, featuring numerous energy crossings that correspond to first-order phase transitions.  
The energy curves in \cref{fig:phase_diagram_bilayer}a can be classified according to the imbalance in the occupation of spin--valley flavors: we have a paramagnetic state (PM, black line), $3/4$ metals (red), $1/2$ metals (blue) and $1/4$ metals (green). Further, each of these fractional metals has competing states with either valley polarization ($\tau_z$) or intervalley coherence ($\tau_x$). 
We see that the carriers in bilayer graphene show a characteristic Hartree-Fock symmetry-breaking pattern of increasing flavor polarization as the hole density decreases from $n_e = -6.5 \times 10^{-11} \text{cm}^{-2}$ towards charge neutrality: from a paramagnetic with four Fermi surfaces and no symmetry breaking, to the $3/4$ metal, to the $1/2$ metal, to the $1/4$ metal. The energy curves of these states appear \textit{horizontally shifted} towards lower hole densities relative to the paramagnetic state, and increase in steepness with decreasing flavors. Notably, this symmetry-breaking pattern and density-dependence is similar to the $4$-component two-dimensional electron gas with parabolic dispersion in the Hartree-Fock approximation \cite{huang2024competition}. 

A unique characteristic of multilayer graphene is that the band Hamiltonians in opposite valleys are not identical but are related by time reversal symmetry, i.e., 
${H^{\tau=+1}_{\vec{k}}= \left[H^{\tau=-1}_{-\vec{k}}\right]^*}$. 
This subtle difference causes the energy curves for these reduced Fermi-surface magnetic states to display two closely degenerate states, which we term valley doublets. 
These doublets correspond to the energy level of the valley-polarized Ising-like state ($\tau_z$) and the intervalley-coherent XY-like state ($\tau_{x,y}$), indicated by dark and light line colors in \cref{fig:phase_diagram_bilayer}a.
Consequently, the determination of whether $\tau_z$ or $\tau_{x,y}$ represents the lower energy state of the doublet is highly sensitive to small energy scales, a topic we will explore further in subsequent sections.
Generally, the band energy tends to favor the
$\tau_{x,y}$ as the lower energy state within the doublet, while exchange energy tends to favor $\tau_{z}$. 
As we explore different parts of the $n_e$--$U_D$ parameter space, the relative significance of band energy versus exchange energy varies, potentially altering the order of the doublets. In  \cref{fig:phase_diagram_bilayer}a,  the valley-polarized energy curve is observed to shift rightward relative to the intervalley-energy curve in all three-quarter, half-metal and quarater metal regime, indicating that the system transitions into the XY state prior to  the Ising state as the density decreases.

\Cref{fig:phase_diagram_bilayer}b and c display the Hartree-Fock phase diagrams in the $n_e-U$ parameter space for dielectric screening constants $\epsilon_r=10$ and $\epsilon_r=30$, respectively, at zero temperature. 
With $\epsilon_r=10$, increasing the interlayer potential $U_D$ expands the magnetic phases, pushing the paramagnetic phase to larger hole densities. 
Conversely, at smaller $U_D$, the magnetic transitions occur at lower carrier densities, favoring valley-polarized states, while intervalley-coherent states are suppressed. 
This suppression occurs because at lower carrier densities, exchange interactions (which favor the Ising over XY order) dominate over the band energy. 
For larger screening ($\epsilon_r=30$), the paramagnetic phase becomes more energetically favorable than the magnetic states, leaving the intervalley-coherent state as the only magnetic ground state, with the valley-Ising state only appearing at small displacement potentials $U_D$. 
Even with significant screening, the intervalley-coherent state persists due to the strongly enhanced density of states at the line of van-Hove singularities within the $n_e$--$U_D$ space -- resulting in the banana-like shape as shown in \cref{fig:bands_params_bilayer}c. 
The noise observed at very low densities in \cref{fig:phase_diagram_bilayer}b and c because the magnetic solutions approach the paramagnetic state while spin and valley polarizations fall below our numerical accuracy. 
Comparing this Hartree-Fock phase diagram to the non-interacting bandstructure calculations of the SWMc model, we observe that magnetic metals appear at the low-density side of the van-Hove singularities. 




\begin{figure}[t]
\centering
\includegraphics[width=8.6cm]{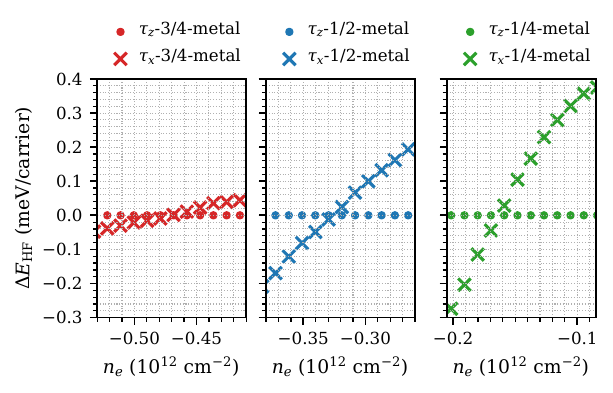}
\caption{
The energy-difference between the valley-doubles ($E_{\rm XY}-E_{\rm Ising}$) for the $3/4$-metal, $1/2$-metal, and $1/4$-metal occurs within a density window of
$\delta n=1.2\times10^{11}$cm$^{-2}$. The data indicate that the chemical potential jump is most pronounced for the $1/4$-metal, followed by $1/2$-metal and then $3/4$-metal. Consequently, the valley-order is softest in the $3/4$-metal, followed by the $1/2$-meta, and then $1/4$-metal. Here, $U_D=40$ meV and $\epsilon_r=10$ as in \cref{fig:phase_diagram_bilayer}a. 
}
\label{fig:bilayer_IVC_VP_competition}
\end{figure}

\subsection{Valley-Doublet Energy Crossing}

The energy difference between the valley-Ising and valley-XY doublets allows for a simple estimate of the susceptibility to rotate in the valley space. \cref{fig:bilayer_IVC_VP_competition} shows the energy levels of these doublets across the $3/4$-metals, $1/2$-metals, and $1/4$-metals. 
Notably, the energy splitting in the $3/4$-metal regime, which occurs at the highest density, is an order of magnitude smaller than in the other two cases. 
This smaller splitting indicates that the valley anisotropic energy is minimal, making it difficult to maintain long-range valley-order. 
Generally, when one of the valley doublets is observed in experiments, the other is energetically close, thus easily excitable, particularly at higher temperatures. 
The results in \cref{fig:bilayer_IVC_VP_competition} show that the magnetic anisotropic transition between Ising-like valley order to XY-like valley order leads to a discontinuity in thermodynamic incompressibility, and this discontinuity is most prominent at the low density, quarter metal region.


\subsection{Rhombohedral Trilayer Graphene}

\begin{figure*}[t]
\centering
\includegraphics[width=1.0\linewidth]{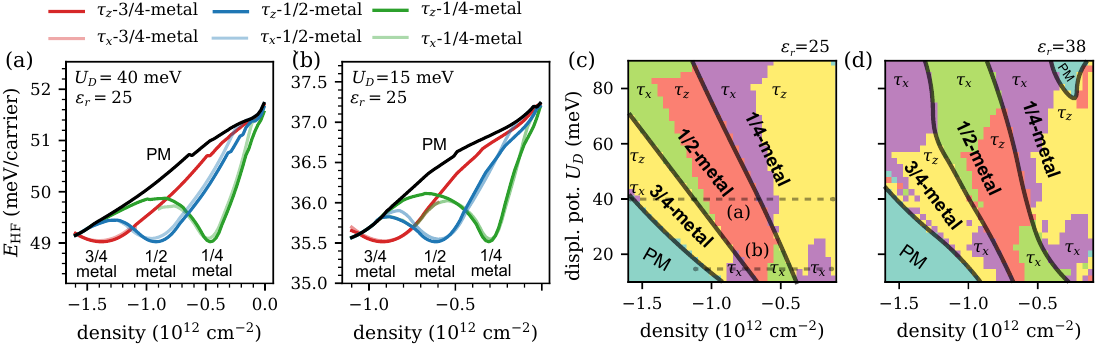}
\caption{%
Hartree-Fock phase diagram of hole-doped rhombohedral trilayer graphene in a displacement potential $U_D$ at electron density $n_e<0$ [cf.~\cref{fig:phase_diagram_bilayer}]. 
(a--b)~Total energy per carrier $E_{\rm HF}$ of the paramagnet and the symmetry-broken flavor-polarized states as $n_e$ is varied, for two different displacement potentials $U_D$. The symmetry-broken states are classified as $3/4$-metal, $1/2$-metal, and $1/4$-metal states, and according to valley-polarization $\tau_z$ or intervalley-coherence $\tau_x$. The displacement potential $U_D$ affects the sequence of first-order phase transitions $\tau_x\to\tau_z$ in (a) compared with  $\tau_z\to\tau_x$ in (b) as $n_e$ increases within each fractional-metal phase. 
(c--d)~Two phase diagrams showing the lowest-energy Hartree-Fock state at given density $n_e$ and potential $U_D$, with different dielectric constant $\epsilon_r$.  When Coulomb interactions are weaker ($\epsilon_r=38$),  intervalley-coherent phases take up more of the parameter space. The dashed lines in panel (c) indicate the linecuts taken in panels (a) and (b), respectively. 
}
\label{fig:phase_diagram_trilayer}
\end{figure*}

In this section, we explore the Hartree-Fock phase diagram of rhombohedral trilayer graphene.
As discussed in Sec.~\ref{sec:SWMc},  a significant difference compared to Bernal bilayer graphene is the large density range featuring an annular Fermi sea, which expands with increasing displacement field. In what follows, we will see that this characteristic significantly affects the mean-field phase diagram.

\Cref{fig:phase_diagram_trilayer}a shows the density dependence of competing states at displacement potential $U_D=40$~meV, including the paramagnetic (PM) symmetric state, magnetic $3/4$-metals, magnetic $1/2$-metals, and magnetic $1/4$-metals, all exhibiting closely degenerate valley-doublets. As seen in Bernal bilayer graphene, as the number of occupied flavors decreases, the energy curves shift horizontally towards lower densities relative to the paramagnetic state and become steeper. Among the doublets, the energy curve for the valley-polarized state ($\tau_z$) is consistently shifted to lower density compared to the intervalley coherent order ($\tau_x$), resulting in the intervalley coherent order appearing before the valley polarized order across all fractional-metal phases. 

However, an opposite trend is observed at lower displacement potentials, as shown in \cref{fig:phase_diagram_trilayer}b for $U_D=15$~meV: the energy curve for the valley-polarized state is shifted to higher density relative to the intervalley-coherent state. 
Consequently, for all fractional metals, the intervalley-coherent $\tau_x$ order is favored after the valley-polarized $\tau_z$ order -- meaning that the sequence of valley order  transitions are reversed. As shown in the $n_e$--$U_D$ phase diagram in \cref{fig:phase_diagram_trilayer}c and d, this reversal is a general feature at low displacement fields. 
The unexpected reversal in the transition sequence, where XY occurs at higher density compared to the Ising, has recently been reported in experimental studies in Ref.~\cite{arp2023intervalley} and explored theoretically in Ref.~\cite{das2023quarter}, observing that it results from subtle details: 
At low electric displacement fields, the Ising order parameter per particle saturates at $\tau_z=1$ and the XY order parameter also approaches saturation at $\tau_x\sim1$, and they both become relatively density-independent. Thus, they do not play an important role in the energy competition.  However, because the intervalley coherent state has a rounder $C_6$ Fermi surface compared to the warped $C_3$ Fermi surface, combined with the density-dependence of the layer polarization, the intervalley coherent state is not suppressed at the low-density limit in trilayer graphene.
Generally, the intervalley-coherent order is predicted to be more easily observed in graphene with more layers \cite{das2023quarter}: for thicker graphene stacks, the phase transition occurs at higher density, which increases the homogeneity of the valley order parameter enclosed by the Fermi surface. Consequently, there is a larger exchange energy per particle, enabling this order to manifest more readily in graphene with multiple layers. 

Aside from the transition sequence reversal at small $U_D$, the phase diagrams in \cref{fig:phase_diagram_trilayer}c and d have the same qualitative features as the Hartree-Fock phase diagrams of bernal bilayer graphene in \cref{fig:phase_diagram_bilayer}. In particular, we observe that intervalley coherent phases expand and become increasingly favored over valley polarization when the screening constant $\epsilon_r$ is reduced, as a result of the band energy becoming increasingly dominant.

\subsection{Hartree-Fock Phase Diagrams versus Experiments} 

After exploring some fundamental aspects of the magnetic phases in the Hartree-Fock phase diagram, it is now crucial to scrutinize the mean-field results against experimental data. In what follows, we highlight two important points.

The first point is that experimental studies on quantum oscillation frequencies \cite{zhou_half_2021,zhou_superconductivity_2021} have primarily identified magnetic metals corresponding to $1/2$-metal and $1/4$-metal phases. They are characterized by dominant fundamental frequencies close to $1/2$ and $1$, when normalized against carrier density. The $3/4$-metal phase, which was predicted by Hartree-Fock theory, was not observed in the intial studies of Zhou et.al~\cite{zhou_half_2021,zhou_superconductivity_2021}. 

Several factors might explain this discrepancy: First, in \cref{fig:phase_diagram_bilayer,fig:phase_diagram_trilayer}, the $3/4$-metal exists within a narrow density range, flanked by the paramagnetic and $1/2$-metal states. 
This characteristic is also observed in the simpler $4$-component parabolic-dispersion electron gas. Thus, when correlation energies suppress ordered states, the three-quarter metal is likely the first to be affected because it is the symmetry-broken state at the largest density. 
Second, as discussed in the theory section and supplemental materials of Ref.~\cite{zhou_half_2021}, the so-called Hund's coupling term $H_{\text{Hund}}= J \int \vec{s}_{K}(r)\cdot \vec{s}_{K'}(r)d^2r$ can affect the energy competition. Whether $J>0$ or $J<0$ can favor the $1/2$-metal over the $3/4$-metal and these interactions potentially eliminate the latter entirely when $|J|$ is large enough. Importantly, $J$ is one of the many symmetry-allowed terms that can influence the mean-field phase diagram. These interactions may arise from both purely electronic mechanisms and electron-phonon coupling, and are discussed in more detail in \cref{sec:g}. In summary, the combination of the above factors likely contributes to the absence of the $3/4$-metal as a ground state in multilayer graphene systems. We note that this 3/4 metallic phase has been recently detected in rhombohedral trilayer graphene when proximitized to WSe$_2$ \cite{patterson2024superconductivity,yang2024diverse}.

The second point is that the Hartree-Fock phase diagrams were computed with large dielectric screening constants of
$\epsilon=10$ and $\epsilon=30$. Neither of them agrees well to the experimental phase diagram. Naturally, with a larger screening $\epsilon=30$, the phase space for the paramagnetic state expands, while a lower  $\epsilon=10$, results in a reduced paramagnetic phase space.  Both fail to capture the re-entrance of the paramagnetic state observed in bilayer graphene \cite{zhou_superconductivity_2021}. 
We have applied the Bohm-Pines random-phase approximation to incorporate correlation energy. However, this approach significantly overestimates the importance of correlation energy, leading to a prediction of only a single phase transition from the paramagnetic to the quarter metal state. This prediction contradicts experimental data, which clearly indicate the presence of a half-metal phase as well.
Calculating effective interactions for metals is inherently challenging, as evidenced by the difficulties encountered in relatively simple systems like Helium 3 \cite{anderson2018theory}. Nevertheless, employing a reliable form of effective density interaction, such as those developed in Ref.~\cite{kukkonen1979electron}, might shed light on why screening constants are so large and elucidate the nature of paramagnetic re-entrance. 


\section{Spin-valley Paragmagons and Density-Wave Instabilitites} \label{sec:TDHF}
Until now, our discussion has focused on symmetry broken states where spin and valley polarization are spatially uniform. However, the electron gas model also supports a magnetic ground state characterized by oscillatory magnetization patterns rather than uniform distributions. This state, known as a spin-density wave, was first introduced by Overhauser in his seminal work \cite{overhauser1962spin}, and its properties are extensively discussed in Ref.~\cite{herring1967exchange}.  In the context of multilayer graphene, the inclusion of additional valley and layer degrees of freedom enriches the behavior of the spin-density wave. This state is characterized by a wavefunction that exhibits a spatially varying spinor component:
\begin{equation}
	\psi_{n\vec{k}}(\vec{r}) = e^{i\vec{k}\cdot\vec{r}}\chi_{n\vec{k}}(\vec{r}).
\end{equation}
A characteristic feature of this generalized pseudospin-density wave state is that it satisfy a  ``generalized'' Bloch's theorem. Here, the wavefunction $ \psi_{n\vec{k}}(\vec{r})$ picks up the usual Bloch-phase factor after a combined spatial translation $T_{a}$ \textit{and} rotation in the layer-sublattice-valley-spin space $\mathcal{R}$:
\begin{equation}
	\mathcal{R} T_{a}\psi_{n\vec{k}}(\vec{r}) =  \mathcal{R} \psi_{n\vec{k}}(\vec{r}+\vec{a})= e^{-i\vec{k}\cdot \vec{a}}\psi_{n\vec{k}}(\vec{r}).
\end{equation}
Readers are encouraged to compare this wavefunction with the simpler form presented in Eq.~\eqref{eq:wavefunction}
to appreciate the additional complexities introduced by the spatially varying spinor components.
This state represents a balance between the paramagnetic state, which lacks any order, and uniform magnetic states, which exhibit long-range magnetic order.  It achieves finite local spin and valley order, effectively minimizing the most important component of the exchange energy. This local ordering allows for significant reductions in exchange energy without imposing global spin and valley order, thereby optimizing kinetic energy as well.
Implementing this state is challenging due to the complexities of the Fermi surface and the large degrees of freedom.
Nonetheless, inspired by experimentally reported non-linear  $I-V$ transport characteristics shown in \cref{fig:nonlinear-IV} it might be worth explore this type of magnetic behavior further.  Recent theoretical work by Vituri et al. \cite{vituri2024incommensurate}, which employs self-consistent Hartree-Fock calculations with a band-projected Hamiltonian, suggests that states exhibiting oscillatory magnetic order—such as valley spirals and valley-crystal states—could be energetically more favorable than the uniform ground state. Moreover, they suggest the soft-fluctuations within these states could mediate effective interactions that lead to superconductivity. To this end, we will employ time-dependent Hartree-Fock (TDHF) theory to investigate this type of instability and to explore collective modes in the paramagnetic state, i.e.~paramagnons.

Following the approach in Ref.~\cite{thouless2014quantum}, we provide a concise introduction to time-dependent Hartree-Fock theory. This theory studies the dynamics of the Dirac density matrix $\rho$ generated by both the band Hamitonian~$T$ and the time-dependent self-energy~$\Sigma$:
\begin{align}
	i \hbar\partial_{t}\rho = [T+\Sigma[\rho],\rho].
\end{align}
In the above equation, we write the self-energy as $\Sigma[\rho]$ to emphasize that it is a functional of the yet-to-be-determined density matrix $\rho$. The mean-field density matrix $\rho^{\text{MF}}$ solves the above equation in the stationary limit $\partial_t\rho=0$, with the conditions $\rho^2=\rho$ and $\mathrm{Tr}(\rho)$ equals number of occupied states. This means $[\rho^{\text{MF}},H^{\text{MF}}]=0$, where
$T+\Sigma[\rho^{\text{MF}}]=H^{\text{MF}}$. To analyze fluctuations around this mean-field solution, we linearize the density matrix as $\rho(t)=\rho^{\text{MF}}+\delta \rho(t)$:
\begin{align} \label{eq:linearized_tdhf}
	i \hbar \partial_t \delta \rho = [H^{\text{MF}}, \delta \rho ] +  [\Sigma[\delta\rho], \rho^{\text{MF}} ].
\end{align}
Given that the Hartree-Fock self-energy depends linearly on the density matrix ($\Sigma[\delta \rho]\propto \delta \rho$), the resulting differential equation is homogeneous. This allows for straightforward solution in the eigenbasis of $H^{\text{MF}}$. From here and what follows, we adopt the notation from Ref.~\cite{thouless2014quantum}, where $m,n$ label the mean-field particle states with energies $\epsilon_m,\epsilon_n >\epsilon_F$ and $i,j$ label the mean-field hole states with energies $\epsilon_i,\epsilon_j <\epsilon_F$. Here $\epsilon_F$ is the Fermi energy of the mean-field groundstate. A key aspect of the linearized time-dependent Hartree-Fock equations is the coupling between the matrix elements $\delta \rho_{mi}(t)$ and their complex conjugates $\delta \rho_{im}(t)$, necessitating a normal modes expansion that includes both forward and backward time components:
\begin{align} \label{eq:tdmf_delta_rho}
	\langle m|\delta \rho(t)|i\rangle \equiv \delta \rho_{mi}(t)= X_{mi} e^{-i\omega t}+ Y_{mi}^* e^{i\omega t}.
\end{align}
This expansion is similar to the expansion of position operator in terms of both boson creation and annihilation operators: $x_{\vec{q}}(t)=\sqrt{\frac{\hbar}{2m\omega_q}}(a_{\vec{q}}e^{-i\omega_qt}+ a_{-\vec{q}}^\dagger e^{i\omega_qt})$. In simpler TDHF problems with real scattering matrix elements on the right-hand side of \cref{eq:linearized_tdhf}, there is a direct analogy in which the real and imaginary parts of $\delta \rho$ act as dynamical conjugate variables, i.e., $\partial_t \mathrm{Re}(\delta \rho)\propto \mathrm{Im} \rho$ and  $\partial_t \mathrm{Im}(\delta \rho)\propto -\mathrm{Re} \rho$. This is identical to the Hamilton equations of motion for a simple harmonic oscillator. An illustrative example is the spin-reversal excitation in a ferromagnet with ground-state spin polarization~$s^z_0$. In this case, $\mathrm{Re}(\delta \rho) = s_x$ and
$ \mathrm{Im}(\delta \rho) = s_y$ form dynamical conjugate variables that lead to the linearized Landau-Lifshitz equation. 

\begin{figure*}[t]
\centering
\includegraphics[width=1.0\linewidth]{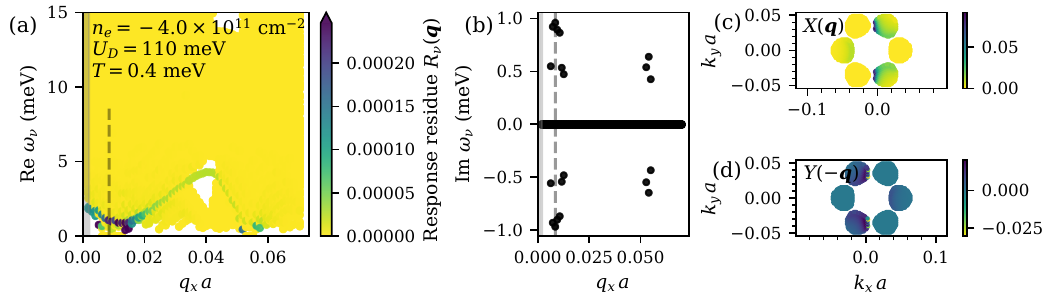}
\caption{%
Valley-flip particle-hole excitation spectrum derived from the generalized random-phase-approximation that accounts for the direct and exchange scattering.
(a) Real energy eigenvalue spectrum showing an incoherent particle-hole continuum (light yellow) and (damped) collective mode dispersion (dark blue).
The color scale corresponds to the valley-flip response residue $R_{\nu\bm{q}}$  in the $\chi_{\tau_+,\tau_-}$ response function [cf.~\cref{eq:TDHF_response_residue_valleyflip}] which highlights the enhanced susceptibility coming from valley paramagnons.
(b) Imaginary part of the energy eigenvalues indicates a finite $q$ valley-wave instabilities.
(c-d) Show the collective mode wavefunction for the critical mode at the wavevector $q$ marked by a dashed line in panels (a) and (b).  The particle-hole pairs predominantly arise from the flat regions of the Fermi surfaces in opposite valleys. 
}
\label{fig:intervalley}
\end{figure*}

Using \cref{eq:tdmf_delta_rho}, \cref{eq:linearized_tdhf} leads to the compact eigenvalue equation 
\begin{align}
\begin{bmatrix}
    A    & B    \\
    -B^* & -A^*
\end{bmatrix}
\begin{bmatrix}
    X \\
    Y
\end{bmatrix}
= \omega
\begin{bmatrix}
    X \\
    Y
\end{bmatrix} .
\end{align}
Here, $A$ and $B$ represent matrices in the particle-hole space, with matrix elements between particle-hole pair $mi$ and $nj$ given by
\begin{align}
	A_{mi,nj} & = (\epsilon_m -\epsilon_i)\delta_{mn}\delta_{ij} + V_{mj,in}- V_{mj,ni}, \\
	B_{mi,nj} & =  V_{mn,ij}- V_{mn,ji},
\end{align}
where $V_{mn,ij}$ represents the matrix elements of the Coulomb interaction within the mean-field eigenbasis. 
These expressions are commonly encountered in the literature on nuclear matter
\cite{rowe1968equations,schuck2021equation}. 
However, unlike in nuclear matter where the mean-field eigenbasis describes self-contained quantum fluids with localized wavefunctions, extended systems in condensed matter theory often involve extended wavefunctions. 
In such extended systems, it is feasible and often necessary to categorize excitations based on their wave-vector $\vec{q}$ \cite{wolf2023quasi}.
Additionally, for scenarios where the many-body Hamiltonian exhibits additional symmetries, these excitations should also be classified according to their corresponding quantum numbers. We emphasize that it is important to consider particle-hole excitations with wave-vectors $\vec{q}$ and $-\vec{q}$ in extended systems because of \cref{eq:tdmf_delta_rho}. To illustrate this, let us define the notation for particle and hole indices more clearly:


\begin{widetext}
\begin{align}
(m,i) \mapsto \begin{cases}
                  (n_p \vec{k}+\vec{q},\, n_h \vec{k}) & (+) \\
                  (n_p \vec{k}-\vec{q},\, n_h \vec{k}) & (-)
              \end{cases},
\qquad
(n,j) \mapsto \begin{cases}
                  (n_p' \vec{k}'+\vec{q},\, n_h' \vec{k}') & (+) \\
                  (n_p' \vec{k}'-\vec{q},\, n_h' \vec{k}') & (-)
              \end{cases}.
\end{align}
where $n_p$ and $n_h$ denote the bands of the particles and holes, respectively. The $\pm$ sign represents the excitations has momentum $\pm \hbar \vec{q}$. 
Then the matrix elements for $A$ and $B$ are 
\begin{align} \label{eq:TDMF_matrixelements_mapping}
A_{mi,nj} & \mapsto \begin{pmatrix}
                        A_{(n_p \vec{k}+\vec{q},\, n_h \vec{k}),(n_p' \vec{k}'+\vec{q},\, n_h' \vec{k}')} & 0                                                                                 \\
                        0                                                                                 & A_{(n_p \vec{k}-\vec{q},\, n_h \vec{k}),(n_p' \vec{k}'-\vec{q},\, n_h' \vec{k}')}
                    \end{pmatrix}
\equiv
\begin{pmatrix}
    (\mathcal{A}_{\vec{q}})_{\vec{k},\vec{k'}} & 0                                        \\
    0                                       & (\mathcal{A}_{-\vec{q}})_{\vec{k},\vec{k'}}
\end{pmatrix},                                                                                                                                                                                                        \\
B_{mi,nj} & \mapsto \begin{pmatrix}
                        0                                                                                 & B_{(n_p \vec{k}+\vec{q},\, n_h \vec{k}),(n_p' \vec{k}'-\vec{q},\, n_h' \vec{k}')} \\
                        B_{(n_p \vec{k}-\vec{q},\, n_h \vec{k}),(n_p' \vec{k}'+\vec{q},\, n_h' \vec{k}')} & 0
                    \end{pmatrix}
\equiv
\begin{pmatrix}
    0                                        & (\mathcal{B}_{\vec{q}})_{\vec{k},\vec{k'}} \\
    (\mathcal{B}_{-\vec{q}})_{\vec{k},\vec{k'}} & 0
\end{pmatrix}.
\end{align}
In the last line, we absorbed the band indices into the labels $\vec{k}$ and $\vec{k'}$. Consequently, the eigenvalue equation simplifies to
\begin{align}
\begin{bmatrix} \label{eq:gRPA}
    A    & B    \\
    -B^* & -A^*
\end{bmatrix}
\begin{bmatrix}
    X \\
    Y
\end{bmatrix}
= \omega
\begin{bmatrix}
    X \\
    Y
\end{bmatrix}
\;\; \mapsto \;\; \sum_{\vec{k}'}
\begin{pmatrix}
    (\mathcal{A}_{\vec{q}})_{\vec{k},\vec{k'}}      & (\mathcal{B}_{\vec{q}})_{\vec{k},\vec{k'}}      \\
    (-\mathcal{B}_{-\vec{q}}^*) _{\vec{k},\vec{k'}} & (-\mathcal{A}_{-\vec{q}}^*) _{\vec{k},\vec{k'}}
\end{pmatrix}
\begin{bmatrix}
    (X_{\nu\vec{q}})_{\vec{k'}} \\
    (Y_{\nu\vec{q}})_{\vec{k'}}
\end{bmatrix}
= \omega_{\nu\vec{q}}
\begin{bmatrix}
    (X_{\nu\vec{q}})_{\vec{k}} \\
    (Y_{\nu\vec{q}})_{\vec{k}}
\end{bmatrix}.
\end{align}
\end{widetext}
%

This eigenvalue equation is sometimes called the generalized RPA (gRPA) eigenvalue problem. It represents the solution to the homogeneous equation derived from a geometric series of Feynman diagrams \cite{thouless1964green}.
The primary focus of our study are the eigenvalues $\omega_{\nu\vec{q}}$ of Eq.~\eqref{eq:gRPA}.
Here $\nu$ enumerates the number of possible particle-hole excitations at a given $\vec{q}$. An instability arises when $\omega$ becomes imaginary signaling a second order phase transition. In practical computations, the size of the gRPA matrix is constrained by the $k$-mesh resolution and must remain manageable because this matrix is non-Hermitian, complicating the application of many standard diagonalization accelerations. These errors may appear as zero-energy solutions with slight imaginary components; however, they typically represent stable, uncorrelated particle-hole pair excitations, not genuine instabilities.
To determine whether an eigenmode represents a real instability or a numerical artifact, one can examine the mode's distribution. If the eigenmode is strongly peaked at a few specific $k$-value, it is just an uncorrelated particle-hole pairs, whereas a broad superposition across the basis suggests a collective mode. The collective behavior of each eigenmode can also be quantitatively assessed through the transition density 
\begin{align}\label{eq:TDHF_response_residue}
R_{\nu\vec{q}} &=\frac{1}{N}\left|\sum_{k} \langle \nu \vec{q}|c_{\vec{k+q}}^\dagger c_{\vec{k}}|\text{HF}\rangle\right|^2 \nonumber \\
 &
 =\frac{1}{N}\left|\sum_{\vec{k}}(X_{\nu\vec{q}})_{\vec{k}}+(Y_{\nu\vec{q}})_{\vec{k}}\right|^2,
\end{align}
where $N=\mathrm{Tr}(\rho)$ is the number of occupied states, $|\text{HF}\rangle$ is the groundstate and $|\nu \vec{q}\rangle$ is the eigenvector of the gRPA equation. 
For uncorrelated particle hole pairs, where $X$ and $Y$ only peak at a few $\vec{k}$ points, $R_{\nu\vec{q}}$ vanishes in the thermodynamic limit as $1/N$. Conversely, for collective modes,  $R_{\nu\vec{q}}$ remains an order of unity in the thermodynamic limit. For a simple electron gas, this quantity, which appears as the residue in the Lehman representation of the density-density response function, reflects the strength of each mode in the spectrum.

This concludes our brief introduction to TDHF theory. We will now explore the application of TDHF theory to examine the excitations of the so-called symmetric-$12$ state in Bernal bilayer graphene, see Table.~\ref{table:terminologies}. This state features three spin-degenerate Fermi pockets in each valley, with the corresponding pockets in the opposite valley rotated by $180$ degrees. This fully-symmetric state has been experimentally confirmed through Shubnikov-de Haas oscillations, is the ground state at very low densities, and appears in between the almost-half-metal ferromagnetism (PIP$_2$) and almost-quarter-metal ferromagnetism (PIP$_1$). In the latter symmetric-$12$ state, introducing holes causes the three pockets within each spin-valley flavor to expand and converge near a saddle-point van Hove singularity where the Fermi sea topology changes quickly to a simply connected Fermi surface. Close to this singularity, the resistance increases as shown in \cref{fig:nonlinear-IV}. This enhanced resistive state appears to be related to superconductivity, as shown in  \cref{fig:n_D_BBG_in_plane_field}.

Our analysis includes two channels: the inter-valley channel, where particle and hole pairs are drawn from opposite valleys [$(n_p=K, n_h=K')$ in Eq.~\eqref{eq:TDMF_matrixelements_mapping}], and the intra-valley channel  [$(n_p=K, n_h=K)$ in Eq.~\eqref{eq:TDMF_matrixelements_mapping}], where particle and hole pairs originate from the same valley. Although the electron and hole pairs are from opposite valleys, contributing to a large momentum of $2\vec{K}$, they can propagate through multilayer graphene electron gas at very low frequency.  For the intra-valley channel, we specifically consider excitations where particles and holes have opposite spins, as these pairs do not undergo direct scattering and exhibit a greater tendency towards divergence compared to the charge channel.

\begin{figure*}[t]
\centering
\includegraphics[width=1.0\linewidth]{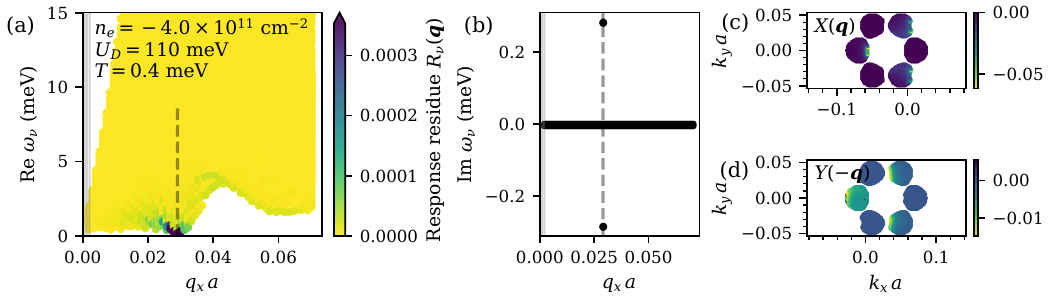}
\caption{%
Spin-flip particle--hole spectrum derived from the generalized random phase approximation that accounts for both direct and exchange scattering.
(a) Real energy eigenvalue spectrum displaying the typical incoherent particle-hole continuum (light yellow). Each mode is colored with $R_{\nu\vec{q}}$ which corresponds to the residue of the spin-flip susceptibility $\chi_{s_+,s_-}$, c.f. Eq.\eqref{eq:spin_flip_residue}. There is a strong enhancement at wavevector $q_{x}^*=2k_{F,x}$, corresponding to the largest Fermi diameter projected in the 
$x$ direction.
(b) Imaginary part of the gRPA energy eigenvalues indicates that this mode is unstable and leads to the formation of a spin-density wave.
(c-d) The wavefunction at $q_{x}^*$ indicates most particle hole pairs that made up this collective mode comes from both end of the Fermi surface. 
}
\label{fig:intravalley}
\end{figure*}

\subsection{Valley Paramagnon}

Paramagnon is a concept used to describe enhanced spin fluctuations that manifest as a broad peak in the magnetic susceptibility ($\text{Im} \chi$ vs $\omega$) in a paramagnetic metal near a ferromagnetic critical point \cite{berk1966effect} or close to Mott-insulating transitions \cite{vollhardt1984normal}.  Early studies introduced a phenomenological parameter ($I$ in Ref.~\cite{brinkman1968spin,berk1966effect,anderson1973anisotropic}) to provide an elegant theoretical-minimal description of enhanced spin susceptibility in a Fermi gas. They study the effects of this enhanced susceptibility on the heat capacity of Helium-3 near its melting curve \cite{brinkman1968spin}, its role in generating spin-singlet repulsion in late transition metals \cite{berk1966effect}, and its role in stabilizing the anisotropic superfluid phase in Helium-3 \cite{anderson1973anisotropic}.

In this subsection, we show that TDHF theory provides an efficient framework for studying paramagnon dispersion without the need for phenomenological parameters. By analyzing the magnetic fluctuation spectral weight of each mode, we effectively distinguish collective modes from uncorrelated particle-hole excitations, even when they are not spectrally isolated. We observe that valley-paramagnons are low-lying excitations that soften at finite wavevectors. When the 12 Fermi pockets are bigger than a critical volume, this softening leads to valley-density wave instability. The critical wavevector where the valley-wave instability occurs corresponds to the distance between the faces of the two Fermi pockets from opposite valleys. Our results suggest that the symmetric-12 paramagnetic metal is more likely to undergo a second-order transition into an intervalley coherent state at finite $q$ rather than at $q=0$, due to the effects of trigonal warping.


We sourced the particle and hole from opposite valleys and diagonalize the gRPA matrix to obtain the eigenvalue spectrum shown in \cref{fig:intervalley}. We account for both the direct and exchange scattering generated by the long-range Coulomb interaction. \cref{fig:intervalley}a) shows the eigenvalues $\omega_{\nu\vec{q}}$ v.s.~$q_xa$. $a=0.246$nm is the lattice constant. Since there are many extended Fermi surfaces in the sym-12 state, there is a continuum of excitations. To distinguish uncorrelated particle-hole pair excitations from collective modes, we use the eigenvector $\ket{\nu \vec{q}}$ of the gRPA equation and compute 
\begin{align}\label{eq:TDHF_response_residue_valleyflip}
    R_{\nu\vec{q}}^{\text{VF}}=\left|\braket{\nu \vec{q}|\tau_{-}(\vec{q})| \text{sym}_{12}}\right|^2 ,
\end{align}
where $\ket{\text{sym}_{12}}$ is the Hartree-Fock ground-state of the symmetric-12 metal and
\begin{align}
    &\braket{\nu \vec{q}|\tau_{-}(\vec{q})| \text{sym}_{12}}
    = \frac{1}{\sqrt{\mathcal{A}}}\sum_{\vec{k} n_h n_p}  \bigg( 
    \braket{n_h \vec{k}|n_p \vec{k}+\vec{q}}
     X^{\nu}_{n_p \vec{k}+\vec{q},n_h \vec{k}} \nonumber \\
    &\quad + \braket{n_p \vec{k}-\vec{q}|n_h \vec{k}}
    Y^{\nu}_{n_p \vec{k}-\vec{q},n_h \vec{k}}
    \bigg).
\end{align}
Here, the sum over $n_h$ includes all hole bands (i.e., below the Fermi energy) in one valley and $n_p$ includes particle bands in the opposite valley (i.e., above the Fermi energy). This quantity measures the collectivity of the eigvector $\ket{\nu \vec{q}}$. For uncorrelated particle-hole pair, $R_{\nu\vec{q}}^{\text{VF}}$ vanishes in the thermodynamic limit as the area of the system
$\mathcal{A}\rightarrow\infty$.
In contrast, for collective modes, $R_{\nu\vec{q}}^{\text{VF}}$ remains finite as $N$ increases because the non-vanishing $X$ and $Y$ components increase with the basis size.  As shown in \cref{fig:intervalley}a), there exists a mode at each $q$ where $R_{\nu\vec{q}}^{\text{VF}}$ is of order one, corresponding to a highly collective mode. We termed this collective mode valley paramagnon. The yellow region in \cref{fig:intervalley}a) corresponds to the valley-flip continuum.

Due to trigonal warping in the SWMC band Hamiltonian, the valley-paramagnon at $q=0$ has a finite gap. However, the paramagnon softens as $q$ increases and reaches a minimum at a critical wave vector corresponding to the distance between the faces of the two Fermi pockets in opposite valleys. Beyond this critical wave vector, the frequency of the paramagnon increases again, passing through a region devoid of the valley-flip continuum, the white region around $(q,\omega)=(0.04, 4)$.  In this window, the paramagnon is spectrally isolated from the continuum due to Pauli blocking influenced by the shape of the Fermi pockets. The valley paramagnon becomes well-defined collective mode in this window.

The quantity $R_{\nu\vec{q}}^{\text{VF}}$ actually represent to the residue function of the valley-flip susceptibility 
\begin{align} \label{eq:chi_valley_flip}
	\chi_{\tau_+,\tau_-}(\vec{q},\omega) & = \sum_{\nu}\mathrm{sgn}(\omega_{\nu\vec{q}})
	\frac{R_{\nu\vec{q}}^{\text{VF}}}{\omega-\omega_{\nu\vec{q}}+i\delta}.
\end{align}
The paramagnon contribute most significantly to the valley-flip susceptibility. However, because the paramagnon mode exists within the valley-flip continuum in most of the $\omega-q$ space, it does not lead to a sharp Lorentzian peak in $\chi_{\tau_+,\tau_-}$ like a ferromagnon would. Instead, it leads to a rather broad-peak that sometimes refered to as incoherent magnetic-fluctuations. Numerically, this braod-peak appears when we set the smearing parameter $\delta$ in Eq.~\eqref{eq:chi_valley_flip} to be approximately the typical spacing within the continuum. Because the low frequency susceptibility is strongly enhanced by the paramagnon, fermionic quasiparticles experience intense electron-quasiboson scattering even when they are close to the Fermi surface. This could potentially amplifying the non-analytic components of the thermodynamic response function considerably \cite{berk1966effect}.

 For the results shown in \cref{fig:intervalley}, we selected a density at which the valley paramagnon is already unstable. The imaginary parts of the eigenvalues are shown \cref{fig:intervalley}b), and the wavefunctions are shown in \cref{fig:intervalley}c) and d). It is evident that most of the wavefunctions are concentrated at the faces of the Fermi pockets. We emphasize that this valley-density wave instability corresponds to a line in the $n_e-D$ space, rather than a specific point. Most importantly, our results show that a second-order transition into intervalley coherent order is more likely at finite $q$ than at $q=0$ especially when trigonal warping is strong.

\subsection{Spin Paramagnon}

In this subsection, we discuss neutral spin-1 excitations of the symmetric-12 paramagnetic metal. We source electron and holes from the same valley but with opposite spin projections and diagonalize the gRPA eigenvalue equation. Because they have opposite spin, they only undergo exchange scattering. The results are shown in \cref{fig:intravalley}. To differentiate spin paramagnon from the continuum, we used the eigenvector of the gRPA equation ($\ket{\nu \vec{q}}$) to compute the transition density from the  Hartree-Fock ground-state $\ket{\text{sym}_{12}}$ to the excited state via spin-flip operator $s_{-}(\vec{q})$:
\begin{align} \label{eq:spin_flip_residue}
R_{\nu\vec{q}}^{\text{SF}}=\left|\braket{\nu \vec{q}|s_{-}(\vec{q})| \text{sym}_{12}}\right|^2, 
\end{align}
where
\begin{align}\label{eq:TDHF_response_residue_spinflip}
    &\braket{\nu \vec{q}|s_{-}(\vec{q})| \text{sym}_{12}}
    = \frac{1}{\sqrt{\mathcal{A}}}\sum_{\vec{k} n_h n_p}  \bigg( 
    \braket{n_h \vec{k}|n_p \vec{k}+\vec{q}}
    \; X^{\nu}_{n_p \vec{k}+\vec{q},n_h \vec{k}} \nonumber \\
    &\quad + \braket{n_p \vec{k}-\vec{q}|n_h \vec{k}}
    Y^{\nu}_{n_p \vec{k}-\vec{q},n_h \vec{k}}
    \bigg).
\end{align}
Here, the summation over $n_h$ includes all hole bands of one spin projection and $n_p$ includes all particle bands with opposite spin projection.  Unlike the valley-paramagnon, where a single mode at each $q$ dominates the residue, the spin paramagnon appears more dispersed; at a given $q$, several modes may exhibit significant residues, giving rise to the dark blue regions in \cref{fig:intravalley}a). These regions represent enhanced paramagnetic spin susceptibility.

Due to long-range Coulomb interactions, the paramagnon softens at a wavevector corresponding to the diameter of the Fermi surface, leading to finite $q=2k_F$ spin susceptibility stronger than $q=0$. This phenomenon is known to lead to spin-density wave instability in a 3D electron gas \cite{kranz1984exchange}. Given the presence of multiple Fermi pockets and the consideration of excitations projected along the 
$q_x$ direction, the projected Fermi diameters of different pockets vary because of trigonal warping, resulting in the appearance of multiple spin paramagnons. At this density, the spin paramagnon at the $q=2k_F$ also becomes unstable, as shown in the \cref{fig:intravalley}b). 
The corresponding wavefunctions, shown in \cref{fig:intravalley} c) and d), primarily consist of particle-hole pairs concentrated at both ends of the Fermi surface. This spin paramagnon instability appears to be weaker than the valley paramagnon instability. Thus, according to TDHF theory, as the Fermi pockets of the symmetric-12 paramagnetic metal approach each other with increasing $|n_e|$, the metal is more likely to transition into a valley-density wave rather than a spin-density wave. However, this conclusion could be influenced by the specifics of the interactions.

\subsection{Discussions}

From these calculations, we learned that the paramagnetic metal in multilayer graphene exhibits a strongly enhanced spin and orbital magnetic susceptibility compared to that of a non-interacting Fermi gas derived from SWMC Hamiltonian. Due to the long-range Coulomb interactions and the specific geometry of the Fermi sea, the enhanced magnetic susceptibility is strongest at finite wavevector.
Consequently, the symmetric-12 paramagnetic metal can be regarded as an almost spin-valley polarized Fermi liquid. These paramagnons can lead to increased resistance as the paramagnetic state transitions into a symmetry-broken phase.

Let us now briefly discuss the influence of these spectrum under an in-plane magnetic field. With an in-plane Zeeman field, the spin flip spectrum shifts from $\omega_{\nu\vec{q}}\rightarrow \omega_{\nu\vec{q}}+\omega_{L}$ where $\omega_L$ is the Larmor frequency. This shift suppresses the spin-density wave instability. In the intervalley channel,  a weak orbital field induced by the vector potential $\vec{A}(z)=z(\vec{B}_{\parallel}\times \hat{z})$,
imparts momentum to plane waves in opposite layers in opposite directions, creating a band energy difference between opposite valleys. This mechanism also serves to suppress valley-density wave instability.  If we associate the non-ohmic IV behavior with density-wave formation, as discussed in Ref.~\cite{zhou_superconductivity_2021}, the isotropic suppression of non-ohmic IV behavior with magnetic field direction suggests that spin density waves are more likely to be the ground state than valley density waves.

Looking ahead, using this enhanced susceptibility to compute the thermodynamic free energy will be beneficial. This analysis will allow comparison with the temperature-dependent compressibility measurements reported in recent experiment \cite{holleis2024isospin}. It is important to understand how these paramagnons influence electron-electron interactions and their role in facilitating superconductivity.  This issue has been recently addressed in Ref.~\cite{dong2023superconductivity}, where interactions—including the long-range Coulomb repulsion— are modeled using delta-function approximations. An estimate of the strength of the sublattice-valley dependent delta-function-like interactions is provided in the next section.


\section{The Intriguing Magnetic Properties of $1/4$-Metal} 

\begin{figure}[t]
    \centering
\includegraphics[width=\columnwidth]
{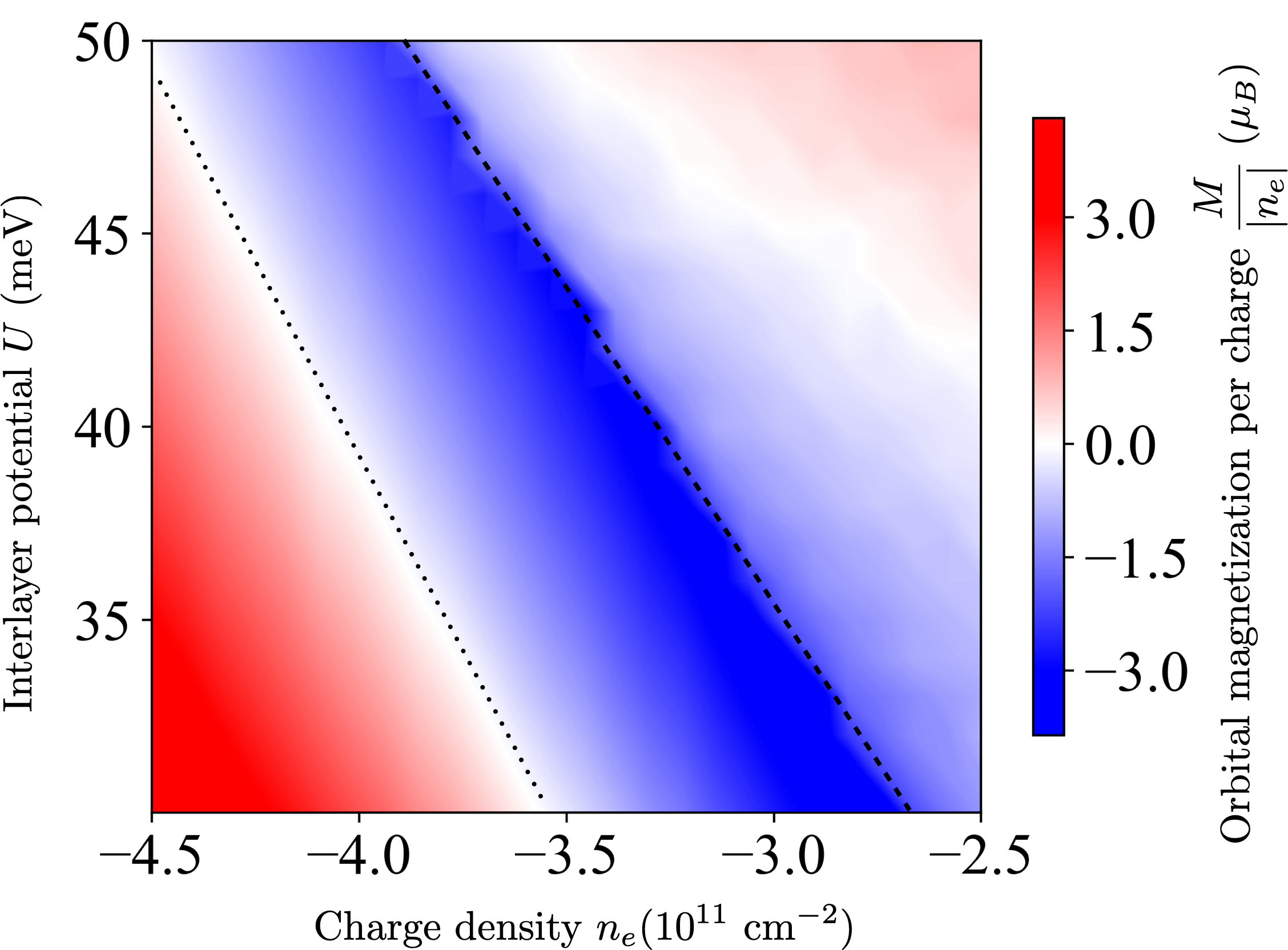}
    \caption{
    The orbital magnetization of the rhombohedral trilayer graphene quarter metal displays a rich landscape across the $n_e-U$ parameter space, even as the ground state valley polarization remains unchanged. In the white region where $M$ vanishes, the groundstate is unresponsive to weak orbital magnetic field. Along the blue diagonal line, where $M$ reaches a minimum, it exhibits non-analytic behavior due to a annular Lifshitz transition. 
    }
    \label{fig:OM_RTG}
\end{figure}

In this section, we explore one novel aspects of magnetism in multilayer graphene: itinerant electron orbital magnetism.  Traditionally, the magnetism in ferromagnetic metals such as Ni, Fe, and Co has been primarily understood through the Pauli exclusion principle which adjusts the the momentum distributions of electrons based on their spin orientations. In these materials, magnetization has generally been attributed to the accumulation of spin moments, with the contributions of orbital moments—induced by weak spin-orbit coupling—considered secondary or negligible However, this conventional perspective changes in the context of lightly-doped multilayer graphene, where the electrons' discrete valley degree of freedom ($\tau=\pm1$) is intrinsically related to orbital moments. This enables a clear separation of well-defined orbital and spin moments, even in the presence of intrinsic spin-orbit coupling (of the Kane-Mele type). In this section, we first discuss the anomalous behavior of orbital magnetism \cite{das2024unconventional}, which can exhibit sign changes and show non-analyticities while the spin and valley polarizations remain unchanged. We then examine the low-energy ferromagnon excitations in this system. Lastly, we highlight some counterintuitive properties of these magnetic states.

The so-called modern theory of orbital magnetization can be  addressed through various methods\cite{hanke2016role,shi2007quantum,thonhauser2011theory,xiao2021adiabatically,PhysRevLett.95.137205,PhysRevLett.110.087202,ceresoli2010first,ceresoli2006orbital,resta2010electrical,raoux2015orbital,piechon2016geometric,PhysRevB.102.184404,PhysRevB.100.054408,PhysRevB.105.195426}. 
They all aim to compute the orbital magnetization from a generic Bloch Hamiltonians defined in the absence of an external magnetic field. A robust quantum mechanical approach is to follow the methodology in Ref.~\cite{shi2007quantum}, which involves expanding the expectation values of the minimal-coupled Bloch Hamiltonian to linear order in the perturbation
$V_B=e\sum_{i=x,y}\{\hat{v}_i,A^i\}/2$ and equate it to the magnetization energy $-MB$. Here, $\hat{v}_i=\partial_{k_i} \hat{H}$ is the velocity operator and $A^i$ is the vector-potential field.  Three distinct contributions affect the energy shift of the Slater determinants: the ground state expectation value of $V_B$, the change in Bloch-wave energy to linear order in $V_B$, and the change in the Bloch wavefunction to the same order. Crucially, only changes in the wavefunction contribute to the orbital magnetization: $-MB=\text{Tr}(\delta\rho H_{0})$ where
$\delta \rho =\sum_{n\vec{k}} n_{F}(\epsilon_{n\vec{k}})\left( |\delta\psi_{n\vec{k}}\rangle\langle \psi_{n\vec{k}}|+\text{h.c.}\right),
$ and
\begin{equation} \label{eq:del_psi}
    |\delta \psi_{n,\vec{k}}\rangle = \sum_{n'\neq n,\vec{k}\neq\vec{k'}}\frac{\langle\psi_{n',\vec{k'}}|\hat{V}_B|\psi_{n,\vec{k}}\rangle}{\left(\epsilon_{n,\vec{k}}-\epsilon_{n',\vec{k'}}\right)} |\psi_{n',\vec{k'}}\rangle.
\end{equation}
As a result, $M$ is sensitive not only to the properties at the Fermi surface but also to the entire eigenvalue spectrum.  For further discussion and a deeper theoretical background, we refer the readers to Ref.~\cite{supmat,shi2007quantum}. Below, we present the final expression for zero-temperature valley-projected orbital magnetization per area ($M^\tau$):
\begin{align} \label{eq:OM}
 M^\tau =& \; \frac{e\mu}{\hbar}  \sum_{n=1}^{M} \int \frac{d^2k}{(2\pi)^2} \bigg[
n_F(\epsilon_{n\vec{k}}^{\tau})\Omega_{n\vec{k}}^{\tau}  \bigg] \nonumber \\
 +& \frac{e }{2 \hbar  } 
 \sum_{\substack{
   n=1}}^{M}
  \int \frac{d^2k}{(2\pi)^2}   n_{F}(\epsilon_{n\vec{k}}^{\tau})\sum_{\substack{n'=1 \\ n' \neq n}}^{M} \bigg[
(\epsilon_{n,\vec{k}}^{\tau}+\epsilon_{n',\vec{k}}^{\tau})  \nonumber \\
  \; & \times 
 \frac{ \epsilon_{ij} \operatorname{Im} \left( \langle \psi_{n,\vec{k}}^{\tau}| \hat{v}_i |\psi_{n',\vec{k}}^{\tau}\rangle  \langle \psi_{n',\vec{k}}^{\tau} |\hat{v}_j| \psi_{n,\vec{k}}^{\tau} \rangle \right)}{(\epsilon_{n,\vec{k}}^{\tau}-\epsilon_{n',\vec{k}}^{\tau})^2} \bigg].
\end{align}
Here $\epsilon_{ij}=-\epsilon_{ji}$ is the antisymmetric tensor and $\Omega^\tau$ is the valley-projected Berry-curvature
\begin{equation}
    \Omega_{n\vec{k}}^{\tau}=-\sum_{\substack{
   n'=1 \\
   n' \neq n
  }}^{M}  
 \frac{ \epsilon_{ij} \operatorname{Im} \left( \langle \psi_{n,\vec{k}}^{\tau}| \hat{v}_i |\psi_{n',\vec{k}}^{\tau}\rangle  \langle \psi_{n',\vec{k}}^{\tau} |\hat{v}_j| \psi_{n,\vec{k}}^{\tau} \rangle \right)}{(\epsilon_{n,\vec{k}}^{\tau}-\epsilon_{n',\vec{k}}^{\tau})^2},
\end{equation}
which integrates to $2\pi$ times the Chern number $C^{\tau } = \pm 3$ when the valley-projected valence bands are completely occupied. Here $M=2Ng_s$ where $2N$ is the number of sublattices in $N$-layer graphene and $g_s=2$ is related to the spin degrees of freedom.

\begin{figure*}[t]
\centering
\includegraphics[width=0.99\linewidth]{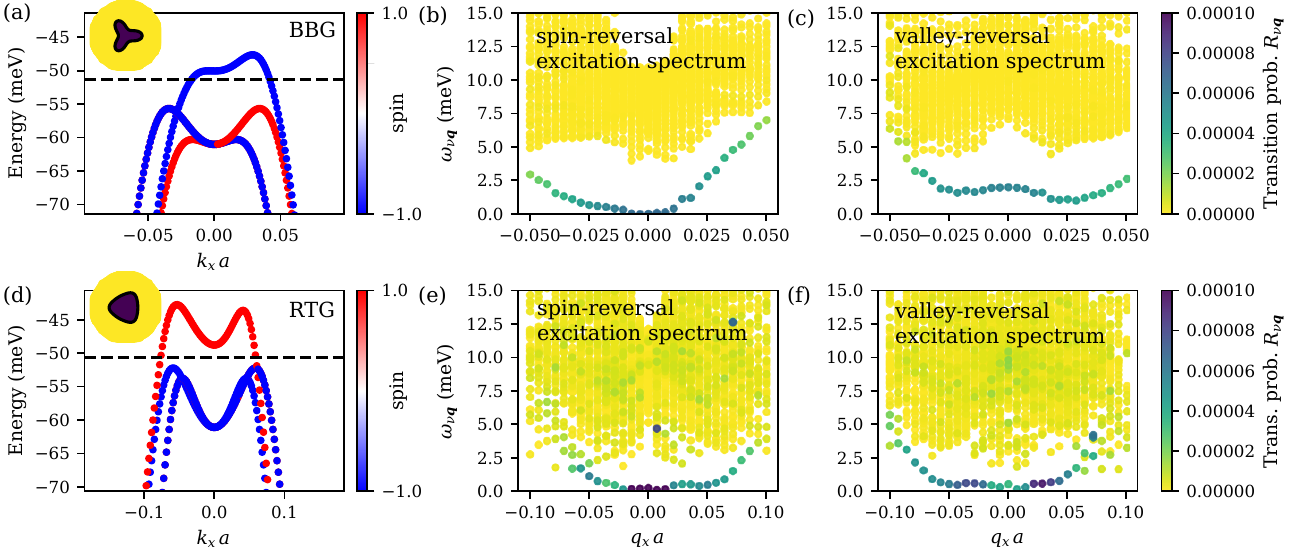}
\caption{%
Particle--hole excitation spectrum derived from the generalized random phase approximation that accounts for both direct and exchange scattering for $1/4$-metal states in Bernal bilayer graphene (a--c) and rhombohedral trilayer graphene (d--f). 
(a,d) Hartree-Fock electronic bands of the fully spin--valley-polarized $1/4$-metal. The dashed line indicates the chemical potential and the inset shows the shape of the Fermi sea. 
(b--c,e--f) Energy eigenvalue spectra displaying the typical incoherent particle-hole continuum (light yellow) for spin-reversal and valley-reversal excitations, respectively. Each mode is colored with $R_{\nu\vec{q}}$ which corresponds to the residue of the susceptibility [cf.~\cref{eq:spin_flip_residue}] of the respective channel. 
We used (a--c) $n_e=-1.1 \times 10^{11} \mathrm{~cm}^{-2}$, $U_D=60\mathrm{~meV}$, $\epsilon=6.3$ and (d--f) $n_e=-5.8 \times 10^{11} \mathrm{~cm}^{-2}$, $U_D=60\mathrm{~meV}$, $\epsilon=14.5$, respectively.
}
\label{fig:quartermetal_collectivemodes}
\end{figure*}

Fig.~\ref{fig:OM_RTG} shows the total orbital magnetization $M=M^{\tau=+1}+M^{\tau=-1}$ in a window of hole density and displacement field where the groundstate is a valley and spin polarized quarter metal. 
To obtain this result, we solve the Hartree-Fock equation self-consistently for a given $n_e$ and $D$ and obtained the converged $\psi_{nk},\epsilon_{nk}$, and the chemical potential and substitute them into Eq.~\eqref{eq:OM}.  Importantly, while the degree of  valley-polarization remains fixed in this $n_e-D$ window, $M$ shows sign-changes and features a singularities. This independent relation between the spin and valley-polarization parameter and magnetization starkly deviates from conventional ``spin'' ferromagnets, where magnetization arises primarily from the alignment of spin moments and thus is directly proportional to the groundstate spin polarization via the gyromagnetic ratio. The non-analyticity of orbital magnetization coincides with the annular Lifshitz transition lines because, at these transitions, the inverse electronic compressibility becomes singular, and thus, through thermodynamic Maxwell relations, orbital magnetization also manifests non-analytic behavior \cite{das2023unconventional}.

The unconventional sign changes arise because the ``edge contributions''—defined as the first term in Eq.~\eqref{eq:OM} and the rest of the term, called the "bulk" contributions, have oppopsite sign. The edge contribution, which dominates close to the band edges, leads to a sign change in $M$ \cite{das2023unconventional}. 
This effect was first identified by Ref.\cite{zhu2020voltage} in the context of insulating states, illustrating the unique magnetic properties of an orbital Chern insulator.   This unconventional sign change of orbital magnetization was also recently observed experimentally using the nano SQUID-on-tip technique, discussed briefly in Section~\ref{sec:Sample Characterization}. The nano-SQUID can detect the magnetic field generated by the electric current flow in 2D metals.


Given the non-trivial functional dependence of magnetization to charge density, it is interesting to study the associated small amplitude magnetic fluctuations. We employed the time-dependent Hartree-Fock method, as outlined in the previous section, to study the excitations corresponding to single spin-reversal, single valley-reversal, and combined spin-and-valley reversal. Our analysis in each channel revealed a single low-lying collective mode, spectrally isolated from the Stoner continuum, indicating that for each electron removed from the valence band, there are roughly three low-lying charge neutral bosonic excitations.
As shown in Fig.~\ref{fig:quartermetal_collectivemodes}, the spin-reversal and valley-reversal excitations show characteristics distinct from typical itinerant electron ferromagnets. The Stoner continuum has a gap that does not close at any finite wavevector due to strong ferromagnetism (100 percent polarization) in these dilute electron systems. Furthermore, the collective modes have relatively flat dispersions and extend across a wide range of wavevectors 
$q$ comparable to the Fermi surface diameter. Due to the shape of the bandstructure, the valley-reversal collective modes softens at intermediate $q$ vector, similar to the paramagnon dispersion we discussed in previous paragraph. Additionally, the valley-reversal excitation at $q=0$ has a finite energy gap because the Hamiltonian in opposite valleys is different. This gap is more pronounced with stronger trigonal warping as shown in Fig.~\ref{fig:quartermetal_collectivemodes}.
We note that the mode reversal that reverse both spin and valley at the same time has  identical dispersion to the valley-reversal spectrum due to the absence of spin-orbit coupling and short-range lattice constants.

These findings suggest that these well-defined neutral excitations can significantly alter thermodynamic properties \cite{kasner1996thermodynamics} and quasiparticle behaviors \cite{kasner2000quasiparticle} in these strong ferromagnets. Indeed, Ref.~\cite{holleis_isospin_2024} recently reports a counterintuitive increase in resistance observed when the paramagnetic state transitions into the ferromagnetic quarter metal phase.


\section{Sublattice and Valley dependent Lattice-Scale Interactions}\label{sec:g}
In this section, we discuss sublattice and valley-dependent interactions, initially focusing on monolayer graphene before expanding our discussion to multilayer graphene. A straightforward symmetry analysis reveals eight non-vanishing coupling constants that serve as inputs to the long-wavelength theory. We then describe how to compute these coupling constants by expanding the sublattice-projected Bloch-functions as linear combination of spatially localized wavefunctions at the two sublattices. Lastly, we discuss how the relative strength of these coupling constants is expected to change under the influence of a strong displacement field in multilayer graphene.

In the $\bm k\cdot \bm p$ expansion,  the three-dimensional wavefunction of multilayer graphene is decomposed around the two valleys, $\pm K$  into a fast-moving part and a slowly varying part. The slowly varying component is described in Eq.~\eqref{eq:wavefunction}. Until now, our focus has primarily been on interactions between the slowly varying components of the wavefunctions which dominate in the dilute doping limit. Recall in the experiments, the filling fraction per carbon atom is approximately $10^{-4}$. This low filling fraction implies that electrons rarely approach each other at the atomic scale. Nevertheless, Coulomb interactions must have weak sublattice, layer and valley dependencies at very short distances. At these short distances, optical phonons can also become relevant. Although a crude approximation might find these interactions to be much weaker than the long-range Coulomb potential, the many-body renormalization effects from the high density of states in the $\pi$ and $\pi^*$ bands significantly boost their magnitudes, thereby making them important in the low-energy limit.

In their seminal work, Aleiner, Kharzeev, and Tsvelik \cite{aleiner2007spontaneous}, first proposed modeling these lattice-scale interactions using delta-function interactions in the continuum model, described by
\begin{align}\label{eq:hasym}
	H_{\rm a}= \frac{1}{2}\sum_{\substack{\alpha,\mu =\\0,x,y,z}}g_{\alpha\mu}^{0}\int :\left[\psi^{\dagger}(\vec{r})\tau^{\alpha}\tilde{\sigma}^{\mu}\psi(\vec{r})\right]^2: \; d^2 r.
\end{align}
In this section, we use their basis \cite{aleiner2007spontaneous}
\begin{align} \label{eq:aliner_basis}
	\psi(\vec{r})=[\psi_{KA}(\vec{r}),\psi_{KB}(\vec{r}),\psi_{K'B}(\vec{r}),-\psi_{K'A}(\vec{r})],
\end{align}
so that the Dirac Hamiltonian in opposite valleys are identical. Here $\tilde{\sigma}$ and $\tau$ act on the sublattice and valley-degrees of freedom as defined in Eq.~\eqref{eq:aliner_basis}. The tilde in $\tilde{\sigma}$ emphasizes the unconventional order of the basis.  This model, which incorporates the Dirac Hamiltonian, long-range Coulomb interactions, and short-range interactions, can be considered "graphene’s standard model."
It is characterized by a finite set of parameters, $g_{\alpha \mu},v_F,\epsilon$ that can be used to model any electronic properties in multilayer graphene.

To compute $g_{\alpha\mu}^{0}$, it is necessary to first solve the three-dimensional Schrödinger equation for particles moving in the periodic potential generated by graphene's honeycomb lattice. The solutions to this equation are Bloch wavefunctions. Specifically, we need two Bloch wavefunctions with momenta $\tau=\pm K$. These wavefunctions must then be transformed into two wavefunctions that have a large weight on the  $\sigma=A,B$ sublattices of the honeycomb structure. A convenient representation for these wavefunctions is
\begin{align} \label{eq:3d_wavefunction}
	\Psi_{\tau \sigma}(x,y,z)\equiv \Psi_{\tau \sigma}(\vec{r})=\frac{1}{\sqrt{N}}\sum_{\vec{R}_\sigma}e^{i\tau\vec{K}\cdot\vec{R}_\sigma}\phi_{\sigma}(\vec{R}_\sigma-\vec{r}).
\end{align}
We emphasis here $\vec{r}=(x,y,z)$ denotes a three-dimensional coordinate vector while $\vec{K}$ represents the two-dimensional wavevector at the Brillouin zone corner. The summation over $\vec{R}_\sigma=m\vec{a}_1+n\vec{a}_2+\vec{\ell}_{\sigma}$ includes $N$ lattice sites, with $\bm a_{1}$ and $\bm a_2$ as the two-dimensional primitive lattice vectors. $\vec{\ell}_{\sigma}$ is the relative distance between sublattice $A$ and $B$. For the localized functions $\phi$ in Eq.~\eqref{eq:3d_wavefunction}, we used $2p_z$ atomic orbitals with an effective nuclear charge
$\tilde{Z}$ in Ref.~\cite{wei2024landau}:
\begin{align}
	\phi(r,\theta,\phi)=\frac{2\tilde{Z}^2\sqrt{\tilde{Z}}}{\sqrt{96a_B^5}}re^{-\frac{\tilde{Z}r}{2a_B}}Y_1^0(\theta,\phi),
\end{align}
where $\int |\phi(\vec{r})|^2 d^3r=1$.
Here, $a_B$ is the Bohr radius. For a more accurate representation of localized wavefunctions, we recommend Wannierizing the $\pi$ and $\pi^*$ bands based on first-principles calculations.

Given these functions, the coupling constants  $g_{\alpha\mu}^{0}$ can be simply derived from the appropriate sum of the matrix elements $V_{ij,kl}$ as follows:
\begin{align}\label{eq:g0}
	g_{\alpha\mu}^{0} & =\frac{\mathcal{A}}{4}
	\sum_{ijkl} V_{ik,jl} (\tau^{\alpha}\tilde{\sigma}^{\mu})_{ij}   (\tau^{\alpha}\tilde{\sigma}^{\mu})_{kl},
\end{align}
where $V_{ij,kl}$ are calculated by solving the six-dimensional spatial integrals 
\begin{align} \label{eq:Vc_mat_ele}
	V_{ik,jl} = \int \int \rho_{ij}(\vec{r}_1)\;V(\vec{r}_{12})\;\rho_{kl}(\vec{r}_1)\;d^3 r_1d^3r_2.
\end{align}
Here  $V(\vec{r}_{12})$ is the mutual repulsion between the electrons and
\begin{align}
	\rho_{ij}(\vec{r}) \equiv \Psi^{*}_i(\bm r)\Psi_j(\bm r)
\end{align}
represents the overlap charges between two wavefunctions $i$ and $j$.
The subscript $i,j,k,l$ in $\Psi$ represents the four possibility $KA,KB,K'B,K'A$. The concept of overlap charges is important for determining certain generic properties of the exchange integral \cite{anderson1997re}. Notably, one property that can be rigorously demonstrated is the positive definiteness of these matrix elements, as detailed in the supplementary materials of Ref.~\cite{wei2024landau}.
We note that $V_{ik,jl}\rightarrow 0$ vanishes in the thermodynamic limit as the sample area $\mathcal{A}$ approaches infinity, similar to scaling of Coulomb matrix elements between plane waves. However, by multilplying a factor of $\mathcal{A}$ in Eq.~\eqref{eq:g0}, we ensure that  $g_{\alpha\mu}^{0}$  — now having the dimensions of energy times area - remains finite as the sample area $\mathcal{A}$ approaches infinity.


The number of independent coupling constants can be reduced through symmetry analysis of graphene's honeycomb lattice. These symmetries include a 120 degree rotation ($C_3$), a 180 degree rotation ($C_2$), two mirror planes - one bisecting the (real-space) honeycomb vertices ($M_y$), and the other perpendicular to the $AB$ bond ($M_x$) \footnote{We note there is an error in  Eq.~2.5c of Ref.~\cite{aleiner2007spontaneous}. The mirror symmetry that interchanges the valley should not interchange the sublattice. This discrepancy appears to be a typographical error and does not seem to impact the overall analysis and conclusions presented in the paper}. These symmetry operators transform the wavefunction $\Psi_{\tau\sigma}(\vec{r}) $ as follows:
\begin{subequations}
\begin{align}
C_3         & :\quad  \Psi_{\tau\sigma}(\vec{r})  \mapsto e^{\frac{2\pi i}{3}  \tau\sigma}\Psi_{\tau\sigma}(R_{2\pi/3}\vec{r}). \\
C_2         & :\quad  \Psi_{\tau\sigma}(x,y,z)  \mapsto \Psi_{\bar{\tau}\bar{\sigma}}(-x,-y,z),                                 \\
M_y         & :\quad  \Psi_{\tau\sigma}(x,y,z)  \mapsto \Psi_{\bar{\tau}\sigma}(-x,y,z),                                        \\
M_x         & :\quad  \Psi_{\tau\sigma}(x,y,z)  \mapsto \Psi_{\tau\bar{\sigma}}(x,-y,z)                                         \\
T_{\vec{a}} & : \quad\Psi_{\tau\sigma}(x,y,z)  \mapsto e^{2\pi i \tau/3}\Psi_{\tau\sigma}(x,y,z)
\end{align}
\end{subequations}
In the last line, we also included the phase factor that the wavefunction acquires upon translation by a lattice constant $\vec{a}$. 
Using $T_{\vec{a}}$ and $C_3$ symmetries, we can deduce that certain matrix elements vanish due to the phase relationship. For example, under $C_3$ symmetry, the matrix elements map as 
\begin{equation}
	V_{KB, KB\;,\; KA, KA} \rightarrow e^{2\pi i/3} V_{KB,KB\;,\; KA, KA},
\end{equation}
and therefore $V_{KB,KB\;,\; KA, KA}=0$. From $C_3$ symmetry, we find
\begin{subequations}
\begin{align}
	g_{xx} & = g_{xy}, \\
	g_{zx} & = g_{zy}, \\
	g_{0y} & = g_{0x}, \\
	g_{yx} & = g_{yy}.
\end{align}
\end{subequations}
Using $T_{\vec{a}}$, we find
\begin{subequations}
\begin{align}
	g_{xz} & = g_{yz}, \\
	g_{x0} & = g_{y0}, \\
	g_{yx} & = g_{xx}.
\end{align}
\end{subequations}
These $7$ equations reduce the number of independent coupling constants from 16 to 9. After neglecting the spin-valley independent component ($g_{00}^0$), since it is accounted by the long-range Coulomb interactions, the number of independent coupling constants consistent with the space group symmetry of graphene is reduced to $8$:
\begin{align}
	g_{0z}^{0}\,,\,g_{0\perp}^{0}\,,\,g_{z0}^{0}\,,\,g_{z\perp}^{0}\,,\,g_{z z}^{0}\,,\,g_{\perp 0}^{0}\,,\,g_{\perp \perp}^{0}\,,\,g_{\perp z}^{0}.
\end{align}
Here $g_{\alpha x}^{0}=g_{\alpha y}^{0}\equiv g_{\alpha \perp}^{0}$, $g_{x\mu}^{0}=g_{y\mu}^{0}\equiv g_{\perp\mu}^{0}$.
Upon further application of time-reversal symmetry,
\begin{align}
	\text{TR} & :\quad  \Psi_{\tau\sigma}(\vec{r})  \mapsto \Psi^{*}_{\bar{\tau}\sigma}(\vec{r}),
\end{align}
we find $g_{0\perp}=g_{0z}=g_{\perp0}=g_{z0}=0$ and there are only four non-vanishing coupling constants. They are: 
\begin{widetext}
\begin{subequations}
\begin{align} \label{eq:g_zz}
	g_{zz}^0          & = \frac{\mathcal{A}}{4}\sum _{\tau,\sigma,\tau^\prime,\sigma^\prime}\mathrm{sgn}(\sigma)\mathrm{sgn}(\sigma^\prime)\int \int \left[ \rho_{\tau\sigma,\tau\sigma}(\vec{r}_1)V(\vec{r}_{12})\rho_{\tau^\prime\sigma^\prime, \tau^\prime\sigma^\prime}(\vec{r}_2) \right]
	d^3\vec{r}_2 d^3\vec{r}_1\;,\; \text{sublattice-antisymmetric}
	\\
	g_{\perp \perp}^0 & = \frac{\mathcal{A}}{4}\sum_{\tau \sigma}\int \int \left[ \rho_{\tau\sigma,\bar{\tau}\sigma}(\vec{r}_1) V(\vec{r}_{12}) \rho_{\bar{\tau}\sigma, \tau\sigma}(\vec{r}_2) \right] d^3\vec{r}_1 d^3\vec{r}_2
	\;,\;  \text{AAAA, inter-valley} \label{eq:g_xx}
	\\
	g_{z\perp}^0      & = \frac{ \mathcal{A}}{4}\sum_{\tau\sigma}2\int  \int \left[\rho_{\tau\bar{\sigma},\tau\sigma}(\vec{r}_1)
		V(\vec{r}_{12})
		\rho_{\tau\sigma, \tau\bar{\sigma}}(\vec{r}_2)
		\right] d^3\vec{r}_1 d^3\vec{r}_2
	\;,\;  \text{inter-sublattice, $KKKK$}
	\\
	g_{\perp z}^0     & =\frac{\mathcal{A}}{4}\sum_{\tau\sigma}2\int \int \left[ \rho_{\tau\sigma,\bar{\tau}\bar{\sigma}}(\vec{r}_1) V(\vec{r}_{12})
		\rho_{\bar{\tau}\bar{\sigma}, \tau\sigma}(\vec{r}_2)
		\right]
	d^3\vec{r}_1 d^3\vec{r}_2 \;,\;  \text{inter-valley, inter-sublattice}
\end{align}
\end{subequations}
\end{widetext}

Next, we discuss the estimates of these 4 coupling constants. $V(\vec{r}_{12})$ is the mutual repulsion between the electrons, typically modeled by the Coulomb potential $e^2/(\epsilon|r_1-r_2|)$ with $\epsilon$ being the dielectric constant.  Given the slow $1/r$ decay, it becomes essential to compute these integrals in Fourier space.
We note that the Fourier components of $V(\vec{r}_{12})$ involved in the computation of $g_{\alpha\mu}^{0}$ necessarily involve very large \textit{in-plane} momentum transfers. These components include $\vec{G}(\neq 0)$ for intra-valley coupling constants ($g_{z\perp}$ and $g_{zz}$) and $2|\vec{K}|+\vec{G}$ for inter-valley coupling constants ($g_{\perp z}$ and $g_{\perp \perp}$), where $\vec{G}$ is the reciprocal lattice vector of graphene. The screening of the bare Coulomb potential at these large momentum vectors markedly differs from the many-body screening effects generated by the dilute electron gas in doped graphene, which generally influence interactions at much longer length scales compared to the carbon-carbon distance.


\begin{table}[t]
\begin{ruledtabular}
\begin{tabular}{c|cccccc}
$\text{meV nm}^2$   & $\tilde{Z}=1$ & $\tilde{Z}=1.5$ & $\tilde{Z}=2$ & $\tilde{Z}=2.5$ & $\tilde{Z}=3$ & $\tilde{Z}=4$ \\
\hline
$g_{\perp \perp}^0$ & 8.88          & 61.5            & 156.4         & 268.9           & 384.3         & 600.6         \\
$g_{zz}^0$          & 0.55          & 12.55           & 67.46         & 184.03          & 353.7         & 781.15        \\
$g_{z \perp}^0$     & 0.234         & 4.41            & 15.84         & 25.11           & 25.38         & 12.42         \\
$g_{\perp z}^0$     & 5.94          & 29.7            & 46.08         & 42.93           & 30.42         & 9.72          \\
\end{tabular}
\end{ruledtabular}
\caption{%
The electronic contributions to lattice scale interaction. $\tilde{Z}$ is the effective nucleus charge.
}
\label{tab:lattice_scale_interaction_g}
\end{table}


In Ref.~\cite{wei2024landau}, we neglect the screening effect at these large momentum transfers, setting the dielectric constant $\epsilon$ to 1, leading to the values presented in Table.~\ref{tab:lattice_scale_interaction_g}.
There are a few important messages. First,  the intersublattice coupling constants ($g_{\perp z}^0$ and $g_{z\perp}^0$)  are almost an order of magnitude smaller than the intra-sublattice  coupling constants ($g_{\perp \perp}^0$ and $g_{zz}^0$) for $2<\tilde{Z}<4$.
This is due to the inherently smaller overlap charge for inter-sublattice processes $\Psi^*_{A\tau_1}\Psi_{B\tau_2}$ compared to intra-sublattice overlap charge $\Psi^*_{A\tau_1}\Psi_{A\tau_2}$.
Second, there is a pronounced dependence on the effective nucleus charge $\tilde{Z}$ in these calculations. The estimated $\tilde{Z}$ for an isolated carbon atom, based on Hartree-Fock calculation, is approximately $\tilde{Z}=3$ \cite{clementi1963atomic}. Given this sensitivity, employing density-functional theory becomes a more appropriate choice as it can more accurately account for  variations in wavefunctions at the atomic scale.
Let us  point out that the estimates presented in Ref.~\cite{knothe2020quartet} appear to be inaccurate, particularly their conclusion that $g_{\perp \perp}^0<0$. This discrepancy might stem from their approach of performing the summation in real space instead of in reciprocal space.

While the intersublattice coupling consstants $g_{\perp z}^0$ and $g_{z\perp}^0$ show minimal electronic contributions, they receive significant contributions from interactions mediated by optical phonons. Optical phonons couple more strongly to Dirac electrons than other phonon modes due to the Kohn anomaly — a non-analytic phonon dispersion observed at the $\Gamma$ and $K$ points \cite{piscanec2004kohn}. 
Neglecting retardation effects, estimates for their contributions \cite{wu2018theory,piscanec2004kohn} are 
\begin{align}\label{eq:g0_estimate2}
	(g_{\perp z}^0,\, g_{z\perp}^0)_{\text{phonon}} = (-69, -52)\text{meV} \cdot \text{nm}^2
\end{align}
By setting the effective nucleus charge $\tilde{Z}=2.5$, a value slightly reduced from the $\tilde{Z}=3$ typically associated with an isolated carbon atom \cite{clementi1963atomic}, and factoring in the electron-phonon contributions, we derived the estimates 
\begin{align}\label{eq:g0_estimate}
	(g_{\perp \perp}^0,g_{zz}^0 , g_{\perp z}^{0},\, g_{z\perp}^{0})
	= (269 ,  184 , -26, -27)\text{meV}\cdot\text{nm}^2.
\end{align}

To provide context, we compare these values with the long-range Coulomb interaction, represented as
\begin{align}
	V_c = \frac{e^2}{\epsilon_r k_F} = \frac{2 a_B}{\epsilon_r k_F} \times 13.6\text{eV} ,
\end{align}
where $a_B$ is the Borh radius and $\epsilon_r$ is the dielectric constant.
Assuming an electron density $n_e=10^{11}$cm$^{-2}$ in a paramagnetic state with four Fermi surfaces, $n_e\sim4\times \frac{\pi k_F^2}{(2\pi)^2}$, and using a dielectric constant $\epsilon_r=4$  typical for a Boron Nitride substrate, we find 
\begin{align}
	V_c = 6430 \,\text{meV}\cdot\text{nm}^2.
\end{align}
This shows that the long-range Coulomb potential is at least 20 times stronger than the largest values in \cref{eq:g0_estimate}.



Let us now extend the symmetry analysis from monolayer to $N$-layer graphene. Notably, in $N$-layer graphene, the spinors of the first conduction and valence bands are predominantly localized to the $A$ sublattice of the 1st layer and the opposite sublattice ($B_N$) of the last layer, $N$.  From a top view,  these two sublattices, $A_1$
and $B_N$ sites, form a honeycomb lattice configuration, despite being separated vertically by the interlayer distance. This spatial arrangement allows us to directly apply the symmetry considerations used in monolayer graphene to the $A_1$ and $B_N$ sublattices in multilayer graphene. In the case of Bernal bilayer graphene, assuming the Bloch-function are made up of linear-combination of $2p_z$ atomic orbitals, we follow the same procedure and derive the layer- and valley-dependent coupling constants values
\begin{align}\label{eq:g0_estimate_bilayer}
	(g_{\perp \perp}^0,g_{zz}^0 , g_{\perp z}^{0},\, g_{z\perp}^{0})
	= (269\,,\, 3853\,,\, 5.1\,,\, 0.3)\,\text{meV}\cdot\text{nm}^2,
\end{align}
assuming an effective nuclear charge $\tilde{Z}=2.5$. We note that the inter-layer coupling constants $g_{\perp z}^{0}$ and $g_{z\perp}^{0}$ are suppressed due to the exponential decrease in overlap caused by vertical displacement between the low-energy sublattices (or layers). Next, we note that the inter-valley scattering process $g_{\perp \perp}^0$ for the bilayer remains identical to that of monolayer graphene. This is because the inter-valley scattering only occurs for electrons that are projected to the same sublattice sites, i.e. that are all within the same layer, c.f.~\cref{eq:g_xx}.

Most significantly, the layer-antisymmetric coupling constant $g_{zz}^0$ is greatly enhanced because electrons from opposite layers can align directly above each other in the $x-y$ plane, and they experience a strong repulsion arising from the Fourier components $V(\vec{G}=0,q_z)$.  It is important to note that the layer-antisymmetric part of the interactions should not be modeled as a delta function in the continuum $K\cdot p$ approximation because electrons in different layers can still experience layer antisymmetric interactions even when they are widely separated within the 2D material plane. Therefore, it is more accurate to model the layer-antisymmetric interaction with a long-range potential 
\begin{align} \label{eq:layer_dependent}
	V_{l,l'}(x,y) = \frac{e^2}{\epsilon_r \sqrt{x^2+y^2+(l-l')^2a_c^2}}, 
\end{align}
where $a_c$ is the inter-layer separartion in multilayer graphene.  This layer-anisotropic repulsive interaction tends to localize identical fermions within the same layer, minimizing their exchange energy. In undoped multilayer graphene that is not encapsulated by hBN (small $\epsilon_r$) and without external fields, this anisotropic interaction lead to a so-called layer-antiferromagnet where electrons with opposite spin projections are localized in opposite layers. This leads to an insulating behavior instead of a semi-metallic behavior predicted by SWMc model band-calculations. At high displacement fields, valence band electrons near the Brillouin zone corner are mostly polarized to the lower energetic layers, regardless of their flavors, making this anisotropic term less important.

In summary, the most dominant pseudospin-dependent interaction stems from the layer-antisymmetric long-range Coulomb interaction. The next most significant interaction is the layer-projected inter-valley scattering process  $g_{\perp \perp}^0$ , whose value is the same as monolayer case in the simplest  (linear-combination of $2p_z$ atomic orbitals) approximation. We can usually neglect the coupling constants that involve electrons changing layers at a vertex ($g_{\perp z}^{0}$ and $g_{z\perp}^{0}$).

For electron-phonon mediated interactions, we can approximate that they predominantly originate from intra-layer optical phonons. Notably, these phonons scatter electrons from the low-energy $A1$ sublattice to the dimer site $B1$. Since the first conduction and valence bands have very little weight on the dimer-site sublattices, we can neglect them at first approximations. Their contributions to $g_{\perp z}^{0}$ and $g_{z\perp}^{0}$ will also be very small. In summary, taking into account the lowest-order electronic and phonon contributions to pseudospin-dependent interactions in multilayer graphene, we are left primarily with the sublattice projected inter-valley scattering term $g_{\perp\perp}^0$. This term is solely influenced by electron-electron interactions, without any contributions from electron-phonon interactions, making it inherently repulsive $g_{\perp\perp}^0=269\,\text{meV}\cdot\text{nm}^2>0$. This finding provides a simple explanation for the repulsive nature of the inter-valley Hund's coupling identified in Ref.~\cite{zhou_half_2021}. The repulsive nature of these interactions promotes ferromagnetic alignment of spins in opposite valleys of the half-metal and
explain the suppression of three-quarter metal states.



\section{Influence of Proximity-Induced Spin-Orbit Coupling on Magnetism and Superconductivity}
\label{sec:SOC}
%
\begin{figure}
    \centering
    \includegraphics[width=1\linewidth]{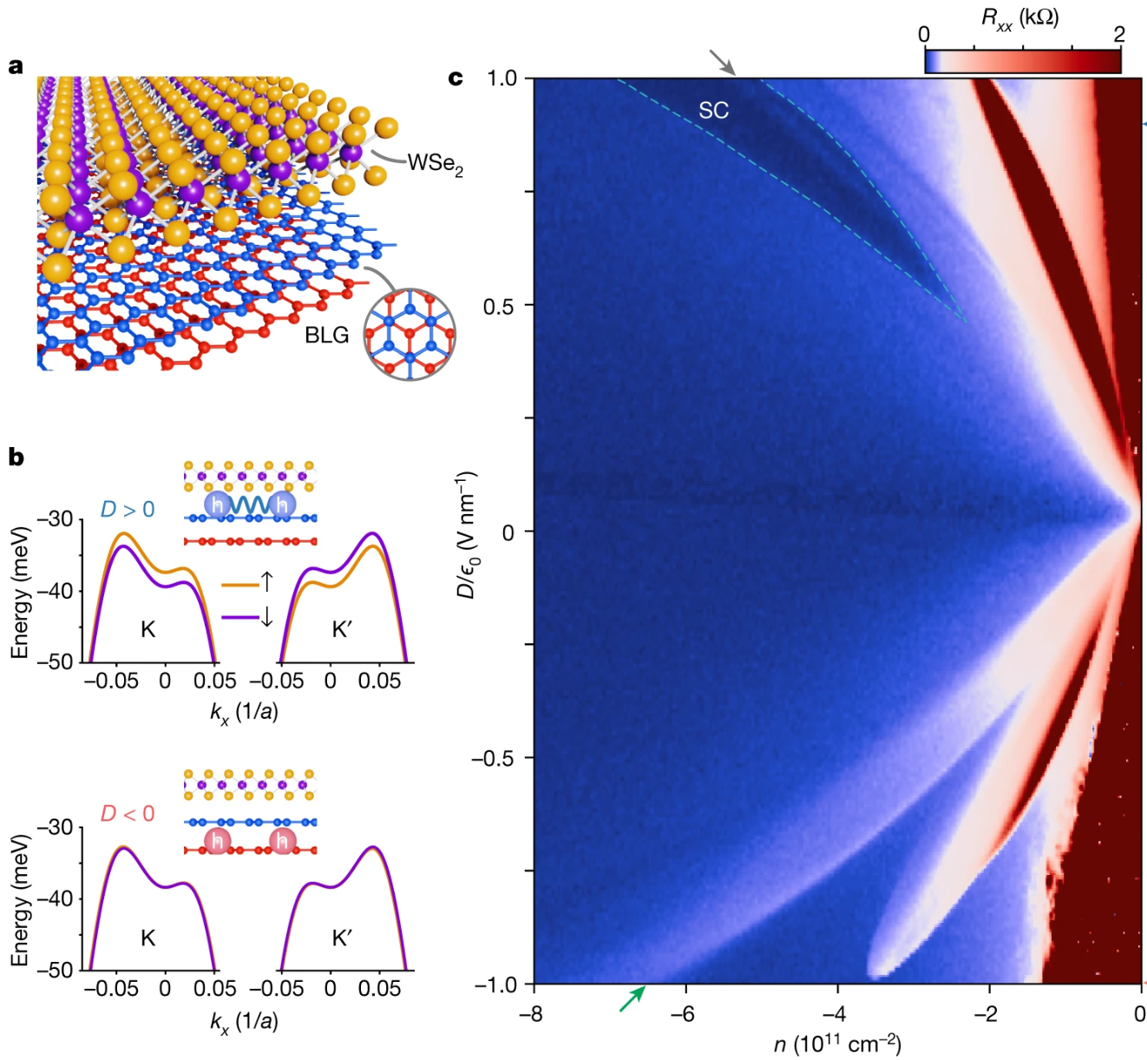}
    \caption{Phase diagram of BBG/monolayer WSe$_2$ at zero magnetic field. a. A schematic of the crystal lattice of BLG (blue and red) with a WSe$_2$ monolayer (yellow and purple) on top. b. The non-interacting valence bands near the $K$ and $K'$ points of the Brillouin zone for $D/\epsilon_0 = 1$\si{\volt\per\nano\metre} (top) and $-1$\si{\volt\per\nano\metre} (bottom) are calculated by including an Ising spin-orbit coupling ($\lambda_I = 1$\si{meV}) on the top layer. Here, $a = 0.246$\si{nm} is the graphene lattice constant. The schematics illustrate that electronic wavefunctions in the valence bands of BBG are polarized towards the top layer for $D > 0$ and towards the bottom layer for $D < 0$. c. The plot of $R_{xx}$ versus doping density $n$ and displacement field $D$ measured at zero magnetic field shows that flavor-polarized states exhibit strong asymmetry with respect to the sign of the $D$ field. The negative $D$ field part of the phase diagram is reminiscent of pristine BBG including a resistive state with non-Ohmic transport behavior appears at the region marked by the bottom grey arrow. Superconductivity, indicated by a dashed line, spans a wide range of doping and $D$ values at positive $D$ fields, where wavefunctions are strongly polarized towards the WSe$_2$. A resistive region appears in the middle of the superconducting region, as marked by the top grey arrow. This figure is adapted from Ref.~\cite{zhang2023enhanced}.}
    \label{fig:wse2}
\end{figure}
Transition metal dichalcogenides (TMD) can be used as high-quality insulating dielectrics for graphene-based devices and importantly, introduce spin-orbit coupling (SOC) that could modify the electronic properties of graphene. There have been accumulating experiments showing that proximitizing TMD such as WSe$_2$ to graphene can enrich strong correlation phenomena in graphene multilayers  enrich the phenomenology in graphene multilayers \cite{arora2020superconductivity,lin2022spin,su2023superconductivity}. Ref.~\cite{zhang2023enhanced} first explored the low-temperature phase diagram of BBG in proximity with a monolayer WSe$_2$ in the strong displacement field $D$, see \cref{fig:wse2}a and b for illustration, and discovered a remarkable enhancement of superconductivity (SC) when carriers are polarized towards the interface of the BBG and WSe$_2$.
\Cref{fig:wse2}c depicts $R_{xx}$ as a function of carrier density $n_e$ and displacment field $D$ when WSe$_2$ is stacked on one side of BBG. The result indicates that the phase diagram of BBG depends on not only the magnitude but also the directions of the $D$ field. On the hole-doped regime, the negative $D$ field repels holes from the WSe$_2$ side and the resulting phase diagram resembles the hBN-encapsulated BBG in the zero in-plane magnetic field, where several spontaneous spin-valley symmetry broken phases emerge at low-density and large-$D$ regime, a resistive state with non-Ohmic tranpsort behavior appear at similar values of $n_e$ and $D$, and no superconductivity is observed. For a positive $D$ field, the holes are pushed towards the WSe$_2$ layer and, remarkably, superconductivity becomes stable in zero magnetic field. Compared to the superconductivity triggered by a finite out-of-plane magnetic field in hBN-encapsulated BBG with similar hole densities, this superconductivity occupies a much larger region of the $n_e-D$ phase diagram and has an optimal critical temperature reaching approximately 300 \si{\milli\kelvin}, representing an almost tenfold increase. The critical in-plane magnetic field $B_{c\scriptstyle\parallel}$ of the SC phase induced by WSe$_2$ in BBG has an unconventional doping dependence. The ratio of $B_{c\scriptstyle\parallel}$ and the Pauli-limit field $B_P$ ($B_{P}=\frac{\Delta_0}{\sqrt{2}\mu_B}$ estimated from the weak-coupling BCS theory with $g-$factor $g=2$ ) can be as large as 6 at the low-density end of the SC dome and gradually reduces to about 1 at the large-density side of the dome. The Pauli-limiting field is defined by equating the Zeeman energy of the normal state ($N_F\mu_B^2 B^2$ ) to the energy of the superconducting state at zero temperature ($\frac{N_F\Delta_0^2}{2}$) and the zero-temperature superconducting gap $\Delta_0$ can be estimated from the critical temperature $T_c$. The critical temperature of superconducting phase shows an abrupt drop within a narrow range of density inside the superconducting dome, which is close to the doping range to observe the resistive states in both the $D<0$ side and hBN-encapsulated samples. Besides the main SC phase, there is evidence for additional and more fragile SC phases at different carrier densities \cite{holleis2023ising}, including a SC phase on the electron-doped side when the displacement field polarizes the conduction band electrons to the layer adjacent to WSe$_2$ reported in Ref.~\cite{li2024tunable}.

Shubnikov–de Haas (SdH) oscillation measurements provide further insights into the effects of WSe$_2$ on electronic structures. When the holes are polarized to the layer away from WSe$_2$ by a large $D$ field, the dominant oscillation frequencies are overall consistent with the hBN-encapsulated BBG. However, when the holes are polarized to the layer next to WSe$_2$, even for the symmetric phase at large hole densities, the nomralized quantum oscillation frequencies change from a single peak at $1/4$ to two slightly different frequencies centered at $1/4$ \cite{holleis2023ising} due to single-particle band splitting generated by the proximity-induced spin-orbit coupling (SOC), as shown in the band structure schematics in \cref{fig:wse2}b. As the hole density lowers, careful analyses of quantum oscillation data reveal that the system enters a nematic PIP phase \cite{holleis2023ising,zhang2023enhanced} where majority of holes occupy two spin-valley flavors equally and two minority bands have tiny occupation numbers. 
Holleis et.~al.~\cite{holleis2023ising} found that  to match the quantum oscillation frequency with the Luttinger's theorem, each minority band in this PIP phase must have twofold degenerate Fermi surfaces. Additionally, in relatively large out-of-plane magnetic field, magnetic breakdown between the small Fermi pockets doubles the oscillation frequency. These observations are incompatible with $C_{3z}$ symmetry of BBG and signal nematicity in the PIP phase of WSe$_2$-proximitized BBG. This nematic PIP phase and the almost-half-metallic ferromagnet in hBN-encapsulated BBG appear in similar regions in the $n_e-D$ phase diagrams and they both consist of large and tiny Fermi surfaces. However, the PIP phase could represent either a spin-orbit field-polarized paramagnet or a ferromagnet — a distinction that cannot be made solely based on their Luttinger volumes. Further reduction in hole density transitions the system to a paramagnetic state with three small Fermi surfaces per spin-valley flavor, akin to the Sym-12 phase in hBN-encapsulated BBG but with a slight SOC-induced Fermi surface splitting. This paramagnetic phase is separated from the nematic PIP phase by a presumably first-order phase transition, as indicated by a relatively sharp dip in inverse compressibility. 

Despite the uncertainties on the roles of WSe$_2$ in the enhancement of SC in BBG, the experimental data show clearly that the proximity-induced SOC can modify the electronic structures and order parameters of correlated phases in BBG. In contrast to the exceptionally small intrinsic SOC of pristine graphene $\sim 50$\si{\micro eV} due to the small atomic number of carbon \cite{sichau2019resonance,banszerus2020observation,banszerus2021spin}, TMD possess a strong SOC arising from heavy transition metal elements and inversion-asymmetric lattice structures. Weak hybridization between the spin-split Bloch bands of TMD and low energy states in graphene inside the TMD band gap
effectively introduces a SOC Hamiltonian to the graphene layer adjacent to TMD \cite{wang2015strong},
\begin{equation}\label{eq:h_soc}
    \hat{H}_{\rm soc} = \frac{\lambda_{I}}{2}\tau s_z + \frac{\lambda_R}{2} (\tau\sigma_x s_y - \sigma_y s_x),
\end{equation}
where $\tau=1(-1)$ labels $K(K')$ valley, and $\sigma_i$ and $s_i$ are Pauli matrices for sublattice pseudospin and spin, respectively. $\lambda_I$ and $\lambda_R$ are parameters of the Ising and Rashba types of SOC, respectively. The precise values of these SOC parameters vary with experimental conditions such as the twist angle between TMD and the adjacent graphene layer \cite{li2019twist,david2019induced}. Theoretical predictions of these values are typically on the order of $1$\si{meV}, an order of magnitude or more stronger than the intrinsic SOC in graphene. 

With only one layer (\textit{e.g.}, layer 1) in proximity with WSe$_2$, the non-interacting Hamiltonian of BBG reads
\begin{equation}\label{eq:h_soc_bbg}
    \hat{H}_{\bm k}^{\rm AB} = \hat{T}_{\bm k}^{\rm AB} + 
    \begin{pmatrix}
        \hat{H}_{\rm soc} & 0\\
        0 & 0
    \end{pmatrix},
\end{equation}
where $\hat{T}_{\bm k}^{\rm AB}$ 
represents the kinetic term consists of SWMC hopping parameters. 
Using standard perturbation theory o project the four-dimensional Hamiltonian onto the low-energy basis of non-dimer sites $(A1, B2)$, and by neglecting $\gamma_{3,4}$,  we arrive at the following:
%
\begin{equation}\label{eq:h_soc_2band}
\hat{h}_{\bm k}=
\begin{pmatrix}
 \frac{\lambda_I}{2}\tau s_z + U_1 & -\frac{v_0^2}{\gamma_1}(\pi^{\dagger})^2-\frac{\lambda_Rv_0}{2\gamma_1}\pi^{\dagger}\hat{s}^{\dagger}\\
 -\frac{v_0^2}{\gamma_1}\pi^2-\frac{\lambda_R v_0 }{2\gamma_1}\pi \hat{s} & U_2\\
\end{pmatrix},
\end{equation}
with $\pi = \tau k_x + i k_y$ and $\hat{s} = \tau s_y - i s_x$. Here, Rashba SOC is layer(sublattice) off-diagonal and is effectively reduced by a factor of $v_0 k/\gamma_1$ compared to the Rashba SOC in $\hat{H}_{\rm soc}$ for a single layer graphene because in $\hat{H}_{\rm soc}$ the Rashba SOC couples the low-energy non-dimer carbon atom sites only to the high-energy dimer sites of BBG which has a large energy separation $\sim \gamma_1$ from the low energy states. In the regime $|U_1-U_2|, v_0 k\ll \gamma_1$ where the two-band approximation is valid, effects of Rashba SOC in Eq.~\eqref{eq:h_soc_2band} are further suppressed by a large interlayer potential difference $|U_1-U_2|$.

The exciting experimental progresses have stimulated theoretic studies on zero-temperature phase diagrams of BBG with proximity induced SOC in strong displacement field \cite{xie2023flavor}. These works consider only Ising SOC, supposing the large displacement field suppresses the spin splitting generated by Rashba SOC. On one hand, because SOC is weaker than the long range Coulomb interaction energy scale and the Fermi energy, the Hartree-Fock phase diagrams share many qualitative features with the one without SOC, as discussed in Sec.~\ref{sec:hf_meanfield}, including cascade phase transitions from the paramagnetic metal to three-quarter metals to half metals and then to quarter metals, as. On the other hand, Ising SOC, which lowers the spin symmetry from SU(2) to U(1),  is significantly larger than the magnetic anisotropic energies between nearly degenerate spin-valley orders without SOC, see Fig.\ref{fig:bilayer_IVC_VP_competition}. This substantial SOC presence is thus effective in tipping the balance between competing ferromagnetic phases. For instance, Ising SOC tends to polarize electrons in two valleys to opposite spins and favors states with spontaneous valley-spin locking ($\tau^z s^z$ order parameter) over the spin or valley paramagnetic states \cite{xie2023flavor}. 
This effect could lead to an expansion of phases with finite spin-valley polarization in the $n_e-D$ phase diagram.
This effect might explain the expansion of the PIP phase with large spin-valley polarization to lower density before entering the paramagnetic state with 3 small Fermi pockets per flavor, compared to the hBN-encapsulated BBG. However, the precise nature of the spin-valley order in the PIP phase remains elusive.
Koh \textit{et.al.} \cite{koh2024symmetrybroken} revealed that competition between Ising SOC and short range ferromagnetic ``Hunds coupling" related to the valley exchange lattice scale interactions discussed in Sec.~\ref{sec:g} results in an intervalley coherent spin-canted ground state with both big and small Fermi pockets, which can be a candidate for the PIP phase as well as the parent state for spin-orbital-enhanced SC \cite{dong2024superconductivity}. In a broader context, Ref.~\cite{wang2024electrical} proposes to exploit the proximity induced SOC in TMD-encapsulated BBG to electrically and selectively switch isospin of the correlated phases with potential applications in spintronics and valleytronics. Ref.~\cite{xie2023gatedefined} put forward a gate-defined SC/IVC/SC planar Joshphson junction in BBG with proximity induced (Rashba) SOC to realize topological superconductivity. 

Bridging the theoretical modeling and experiment results will benefit from accurate input of the SOC parameters. These parameters play an important role in quantitative understanding of various properties of BBG such as the Pauli violation ratios of SC. They have been measured via different experimental techniques for BBG/WSe$_2$ heterostructures, see Table~\ref{tab:soc}. Below, we briefly review some of these experimental techniques.
%

One method to determine $\lambda_{I}$ in BBG is to probe Landau level (LL) crossing patterns in BBG under an out-of-plane magnetic field $B_{\perp}$ using resistance or capacitance measurements. Ref.~\cite{island2019spin} noticed that the spin-valley ordering of the quantum Hall state at filling factor $\nu=3(-3)$ is determined by single-particle anisotropy energies including Ising SOC, $B_{\perp}-$induced Zeeman splitting, and $D-$induced valley splitting because zeroth LL orbitals in two valleys are polarized in opposite layers in BBG \cite{mccann2013electronic}. The LL of the highest(lowest) single-particle energy among the octet zeroth LLs will be empty(occupied) in the $\nu=3(-3)$ ground state. Therefore, tuning $D$ or $B_{\perp}$ field leads to magnetic phase transitions at $\nu=3(-3)$, signaled by either transport or capacitance measurements. The critical $D$ field and $B_{\perp}$ field corresponds to when two single-particle LLs cross at Fermi level and can be used to fit $\lambda_I$. Table~\ref{tab:soc} shows that different experiments based on similar LL crossing schemes report similar magnitude of $\lambda_I\sim 1$\si{meV}. These estimates are also consistent with theories \cite{li2019twist}.
In contrast, layer-off-diagonal Rashba SOC in Eq.~\eqref{eq:h_soc_2band} vanishes after projection into the zeroth LL and therefore its effects on the spectra of the zeroth LL are suppressed in the strong magnetic field.  Thus, probing LL crossing provides a clean method to determine $\lambda_I$ but is less commonly used to extract $\lambda_R$ except in Ref.~\cite{wang2019quantum}. 

SdH conductance oscillation measurements probe the splitting of Fermi surfaces induced by the SOC, which give rises to characteristic beating pattern from which the sizes of the split Fermi surfaces can be assessed and fit with the band structure calculations to extract the SOC parameters \cite{wang2016origin}. Eq.~\eqref{eq:h_soc_2band} indicates that the importance of Ising SOC decreases with Fermi momentum $k_{F}$ and carrier density $n_e$ while the splitting induced by Rashba SOC increases with $n_e$. Therefore, the $n_e-$dependence of the oscillation frequency splitting could constrain values of $\lambda_{I}$ and $\lambda_R$ together.

Besides, there are experiments to estimate SOC parameters for samples in the diffusive transport regime based on the weak antilocalization effects (WAL) \cite{amann2022counterintuitive,yang2017strong,mccann2012z}, which requires both intervalley disorder scattering and spin relaxation in BBG. One can fit the spin relaxation rates, which are related to $\lambda_{I,R}$, to the WAL features in weak-field magnetoresistance data. However, this type of measurements often involve multiple fitting parameters and assume no disorder-induced spin relaxation. The inferred SOC parameters could be sensitive to disorder.

To summarize, Table~\ref{tab:soc} indicates that the magnitude of interfacially induced Ising SOC , $|\lambda_I|$, measured by different groups are consistently below $6$\si{meV} and agree with the theoretical predictions$\sim 1$\si{meV}, whereas the uncertainties on the strength of Rashba SOC remains relatively large and calls for a more precise measurement in future experiments. 
The twist-angle dependence of SOC parameters is even less understood. Only recently, Ref.\cite{zhang2024twist} systematically investigated Ising SOC and BBG phase diagrams across a series of devices fabricated from the same BBG and WSe$_2$ crystals, but with varying twist angles between the BBG and WSe$_2$. This experiment reveals that $|\lambda_I|$ decreases monotonically from $1.6$\si{meV} at $\theta=0$ to nearly zero at $\theta=30^{\circ}$. In future, it will be desirable to use instruments \cite{inbar2023quantum} with \textit{in-situ} tunability of the twist angle between graphene and TMD to map out the SOC parameters as a function of the twist angle and explore the evolution of BBG phase diagrams with continuously tunable SOC in a single device.

Currently, the nature of these SC phases and the mechanism for the enhancement of pairing in the WSe$_2$ proximitized BBG is yet to be understood. Existing theoretical proposals range from attractions mediated by virtual tunneling to WSe$_2$ \cite{chou2022enhanced} to enhanced magnon-mediated attractions in the partially spin-polarized normal state with spin orders canted by the proximity-induced Ising spin-orbit coupling \cite{dong2021superconductivity}. On the experimental side, the asymmetry in the displacement field data indicates that the enhancement of superconductivity is not merely due to the larger dielectric screening in WSe$_2$ than hBN, but is instead related to effects generated on the interface of WSe$_2$ and BBG. Besides, an important feature shared by the most robust hole-doped SC phase and the electron-doped SC phase \cite{li2024tunable} in WSe$_2$-proximitized BBG is that superconductivity emerge inside the PIP phase at low temperature. In hBN-encapsulated BBG, superconductivity occurs near the boundary between the PIP and Sym$_{12}$ phase but still stays on the PIP side \cite{holleis2023ising}. In our opinion, these observations hint that small Fermi pockets promotes superconductivity in BBG. 

An interesting and rewarding research direction is to extend these experimental and theoretical studies to RTG \cite{yang2024diverse,patterson2024superconductivity,koh2024correlated} and other grapheen multilayers \cite{choi2024electric} in proximity with WSe$_2$ or other TMDs. For example, Ref.~\cite{han2024large} has recently reported quantum anomalous Hall states with Chern number $C=\pm 5$ in WS$_2$-proximitized pentalayer graphene, further highlighting the rich phenomena that can arise in multilayer graphene/TMD heterostructures, driven by the interplay of electron correlations, band topology, and proximity-induced spin-orbit coupling.



\begin{table}
    \centering
    \caption{Experimental results for the magnitude of Ising ($\lambda_I$) and Rashba ($\lambda_R$) spin orbit coupling in BBG on the layer proximitized to WSe$_2$ and the experimental observations from which these estimates are extracted. We use ``1L'' to emphasize monolayer, but otherwise refer to multilayer or bulk WSe$_2$. We do not distinguish the signs of $\lambda_{I,R}$ (\textit{e.g.}, Ref.~\cite{wang2019quantum} reported negative $\lambda_I$), because the signs do not affect the energy spectrum of Hamiltonian Eq.~\eqref{eq:h_soc_bbg}. Experimental schemes to estimate $\lambda_{I,R}$ include 1) probing the frequency splitting of quantum oscillations (SdH); 2) measuring critical $D$ fields and $B_{\perp}$ fields for the Landau levels with opposite valleys and spins to cross at the Fermi level (LL crossing); 
    3) fitting the spin relaxation rates to the weak-field magnetoresistance data which exhibit weak antilocalization effects (WAL).}
    \begin{ruledtabular}
    \begin{tabular}{ccccc}
         Device &  $\lambda_I$/\si{meV}& $\lambda_R$/\si{meV} &  Observation & Ref.\\
    \hline
        hBN/BBG/1L-WSe$_2$               & $0.7$ & $0\sim 4$ & LL crossing, SdH &\cite{zhang2023enhanced} \\
                            &$1.6$ & - & LL crossing & \cite{holleis_isospin_2024}\\
                         & $1.7$ & - & LL crossing  & \cite{li2024tunable}   \\
        hBN/BBG/WSe$_2$& $1.7$ & - & LL crossing &\cite{island2019spin} \\
                       & $2.1\sim  2.3$ & $10 \sim 20$ & LL crossing &\cite{wang2019quantum} \\
        BBG/WSe$_2$                    & $0 \sim 6$ & $10 \sim  15$ & SdH &\cite{wang2016origin} \\
        SiO$_2$/BBG/WSe$_2$ & $0.8$ & $0.5$ & WAL &\cite{amann2022counterintuitive}\\
    \end{tabular}
    \end{ruledtabular}
    \label{tab:soc}
\end{table}

\section{Conclusions and Outlooks}

In this concluding section, we synthesize the key insights gained from studying generalized ferromagnetism and superconductivity in Bernal bilayer and rhombohedral trilayer graphene. We identify aspects that remain elusive and suggest possible interesting directions for future research in this rapidly developing field.

The strong-field magnetotransport data have always been instrumental in revealing the Fermi surface degeneracies of graphene metals. Recently, as demonstrated in Refs.~\cite{zhou_half_2021,zhou_isospin_2021,zhou_superconductivity_2021,holleis_isospin_2024,holleis2023ising}, high-quality weak-field quantum oscillation data—where the magnetic flux density is significantly smaller than the electron density— have also proven to be highly informative. In this semiclassical limit, the frequency of quantum oscillations not only identify magnetic transitions associated with stepwise reductions in Fermi surface degeneracies but also detects almost-half metal and almost-quarter metal ferromagnets characterized by tiny minority flavor Fermi pockets. 
Additionally, it can also identify the topology of the Fermi sea, indicating whether the paramagnetic state has a disconnected 12 pockets or an annular shape. Recent studies \cite{holleis2023ising} suggests that these oscillations can even detect momentum space condensation, a nematic transition that break graphene lattice's $C_3$ rotation symmetry. All of this valuable information is derived from frequency analysis. Looking forward, extracting physical insights from the amplitudes and linewidths of these oscillation frequencies could be highly informative. Specifically, analyzing how these quantities vary with changes in density and displacement fields could provide deeper insights into how electron-electron interactions affect quasiparticle renormalization and lifetimes. This analysis is particularly valuable as the observed changes are independent of phonons and impurities, given that the temperature remains constant and the variations in electron density result solely from gate-tuning, not from fabricating multiple devices.


While this review primarily focuses on the interaction effects in lightly-doped multi-layer graphene such as 
metallic ferromagnetism and superconductivity, it is important to also acknowledge the extensive literature on interaction effects in undoped multilayer graphene \cite{zhang2011spontaneous,weitz2010broken,geisenhof2021quantum,stepanov2019quantum}. The ground state of charge-neutral, undoped multilayer graphene varies with the electric displacement field. At large displacement fields, the ground state becomes a field-polarized layer insulator, where valence band electrons across all flavors are predominantly localized in the low-energy layer, resulting in states with opposite valleys having opposite Chern numbers. Upon doping this field-induced layer-polarized insulator the groundstate naturally evolves into a paramagnetic metal. This state agrees with the symmetric-12 paramagnetic metal discussed throughout this review.
At low displacement fields, the neutrality groundstate transitions into an interaction-induced layer antiferromagnet. In this state, electrons with opposite spin projections localize to the opposite layers ($A_1$ and $B_N$), optimizing the exchange energy from layer-anisotropic Coulomb interaction, c.f.~Eq.~\eqref{eq:layer_dependent}.
Without spin-orbit coupling, the direction of spin projection remains arbitrary. At these small displacement fields, the quantum anomalous Hall insulator is also energetically close by and manifests upon the application of a magnetic field. Exploring how these correlated insulators transition into ferromagnetic metals upon doping at small displacement fields could be an interesting research direction.


One of the most surprising findings is that superconductivity emerges when a dilute amount number of electrons are introduced to/removed from the $\pi$ band of multilayer graphene which is made up by carbon's simple $2p_z$ atomic orbitals. 
We summarize here a list of experimental conditions under which superconductivity has been observed. In rhombohedral trilayer and Bernal bilayer graphene without adjacent layers of transition metal dichalcogenides, superconductivity appears at phase transition boundaries.  For rhombohedral trilayer graphene, it manifests at the boundary between an almost half-metal ferromagnet (PIP$_2$) and the paramagnetic state with an annular Fermi sea, as well as between PIP$_2$ and an almost quarter-metal ferromagnet (PIP$_1$). In bilayer graphene, field-induced superconductivity emerges within a small but finite density window between the PIP$_2$ at low $|n_e|$ and the fully-symmetric paramagnetic state with 12-pockets at high $|n_e|$. A common theme among these occurrences of superconductivity is their proximity to almost half-metal ferromagnet.

When WSe$_2$ is introduced to the rhombohedral trilayer graphene and bilayer graphene stack, it leads to spin-valley splitting and lifts the 4-fold degenerate electronic density of states. Specifically, the simplest form of Ising-type spin-orbit coupling changes the shape of the density-of-states function and reduces its degeneracy to 2-fold, without affecting the Bloch wavefunctions. These changes naturally reduce the tendency towards ferromagnetism that breaks time-reversal symmetry and strongly modify the magnetic anisotropic energy landscape. Notably, Ising spin-orbit coupling is known to increase the energy of valley-XY (inter-valley coherent) state \cite{arp_intervalley_2023,das2023quarter}.
Under these modified conditions, the region exhibiting field-induced superconductivity in Bernal bilayer graphene \cite{holleis2023ising} is now stabilized even without an in-plane magnetic field and expands within the density-displacement parameter space. In rhombohedral trilayer graphene, two significant changes occur: the previously robust superconducting state near the almost-half-metal ferromagnet no longer exhibits superconductivity \cite{yang2024diverse,patterson2024superconductivity}. This trend  supports theories proposing that superconductivity is mediated by intervalley fluctuations \cite{chatterjee2022inter}, as Ising spin-orbit coupling not only raises the energy of the intervalley coherent (IVC) state but also reduces its order parameter\cite{arp_intervalley_2023,das2023quarter}, which was previously identified as the ground state for the almost half-metal ferromagnet. Additionally, new superconducting phases emerge in various regions of the $n_e-D$ phase diagram on both the electron and hole-doped sides, whose characteristics remain to be fully understood \cite{yang2024diverse,patterson2024superconductivity}.

Since superconductivity in multilayer graphene without WSe$_2$ typically appears near an almost half-metal ferromagnet, it is suggestive that some form of spin-valley magnetic susceptibility is strongly enhanced above the paramagnetic SWMC-band susceptibility, generating effective interactions that lead to Cooper pairing. If this superconducting mechanism also applies to cases where multilayer graphene is adjacent to WSe$_2$, then the spin-orbit coupling induced by WSe$_2$ must not diminish the enhanced magnetic susceptibility; otherwise, the free energy of the superconductor would increase. This scenario is reminiscent to how $^3$He's enhanced spin susceptibility near the first-order liquid-solid phase transition line favors the anisotropic superfluid A-phase over the isotropic B-phase because the A-phase does not reduce the spin susceptibility \cite{anderson1973anisotropic}. 
Therefore, exploring how spin-orbit coupling influences the stability of paramagnetism versus ferromagnetism, magnetic anisotropic energy, and spin-valley magnetic susceptibility could be an important direction for future research. 

\begin{acknowledgments}
\textit{Acknowledgments.}
We thank Anna Seiler for sharing her non-linear IV data on Bernal bilayer graphene. We are grateful for valuable discussions with Allan MacDonald, Trevor Arp,  Ludwig Holleis, Cyprian Lewadowski, Owen Sheekey, Thomas Weitz,  Andrea Young and Fan Zhang. We thank Guopeng Xu for providing estimates of the short-range coupling constants for bilayer graphene. T.M.R.W. acknowledges financial support from the NSF (Award No. DMR–2308817 and DMR–2308817).
\end{acknowledgments}

\bibliography{references}
\end{document}